\documentclass[14pt]{article}
\pdfoutput=1
\usepackage{amsmath,amssymb,amsfonts,amsthm,bm,bbm,cancel,wasysym}
\usepackage{epsfig,graphics,graphicx,epstopdf}
\graphicspath{{Charts/}}
\usepackage{array,booktabs,colortbl,colordvi,multirow}
\usepackage{colordvi,color,xcolor}
\usepackage{hyperref}
\usepackage{rotating}

\usepackage{verbatim}
\usepackage{cite}
\usepackage{subfig}
\usepackage{setspace}
\usepackage{url}
\usepackage[percent]{overpic}
\usepackage{slashed}
\usepackage{authblk}
\usepackage{xspace}
\usepackage{fullpage}
\usepackage{hyperref}

\def\ie{{\it i.e.}}
\def\eg{{\it e.g.}}
\def\etc{{\it etc}}

\def\to{\rightarrow}

\newskip\zatskip \zatskip=0pt plus0pt minus0pt
\def\matth{\mathsurround=0pt}
\def\lsim{\mathrel{\mathpalette\atversim<}}
\def\gsim{\mathrel{\mathpalette\atversim>}}
\def\atversim#1#2{\lower0.7ex\vbox{\baselineskip\zatskip\lineskip\zatskip
  \lineskiplimit 0pt\ialign{$\matth#1\hfil##\hfil$\crcr#2\crcr\sim\crcr}}}

\begin{document}


\begin{flushright}
SLAC-PUB-17532\\
\today
\end{flushright}
\vspace*{5mm}

\renewcommand{\thefootnote}{\fnsymbol{footnote}}
\setcounter{footnote}{1}

\begin{center}

{\Large {\bf Kinetic Mixing, Dark Photons and Extra Dimensions III:  ~~~Brane Localized Dark Matter}}\\

\vspace*{0.75cm}

{\bf Thomas G. Rizzo and George N. Wojcik}~\footnote{rizzo, gwojci03@slac.stanford.edu}

\vspace{0.5cm}

{SLAC National Accelerator Laboratory}\\
{2575 Sand Hill Rd., Menlo Park, CA, 94025 USA}

\end{center}
\vspace{.5cm}

\begin{abstract}
 
\noindent 

Extra dimensions have proven to be a very useful tool in constructing new physics models. In earlier work, we began investigating toy models for the 5-D analog of the kinetic 
mixing/vector portal scenario where the interactions of dark matter, taken to be, \eg, a complex scalar, with the brane-localized fields of the Standard Model (SM) are mediated by a 
massive $U(1)_D$ dark photon living in the bulk. These models were shown to have many novel features differentiating them from their 4-D analogs and which, in several cases, 
avoided some well-known 4-D model building constraints. However, these gains were obtained at the cost of the introduction of a fair amount of model complexity, \eg, dark matter 
Kaluza-Klein excitations. In the present paper, we consider an alternative setup wherein the dark matter and the dark Higgs, responsible for $U(1)_D$ breaking, are both 
localized to the `dark' brane at the opposite end of the 5-D interval from where the SM fields are located with only the dark photon now being a 5-D field. The phenomenology of such a 
setup is explored for both flat and warped extra dimensions and compared to the previous more complex models.

\end{abstract}

\renewcommand{\thefootnote}{\arabic{footnote}}
\setcounter{footnote}{0}
\thispagestyle{empty}
\vfill
\newpage
\setcounter{page}{1}



\section{Introduction}\label{Intro}

The nature of dark matter (DM) and its possible interactions with the fields of the Standard Model (SM) is an ever-growing mystery. Historically, Weakly Interacting Massive 
Particles(WIMPs)  \cite{Arcadi:2017kky}, which are thermal relics in the $\sim$ few GeV to $\sim$ 100 TeV mass range with roughly weak strength couplings to the SM, and axions \cite{Kawasaki:2013ae,Graham:2015ouw} were considered 
to be the leading candidates for DM as they naturally appeared in scenarios of physics beyond the Standard Model (BSM) that were constructed to address other issues. Important 
searches for these new states are continuing to probe ever deeper into the remaining allowed parameter spaces of these respective frameworks. However, the null results so far 
have prompted a vast expansion in the set of possible scenarios \cite{Alexander:2016aln,Battaglieri:2017aum} which span a huge range in both DM masses and couplings. In almost all of this model space, new forces 
and, hence, new force carriers must also 
exist to mediate the interactions of the DM with the SM which are necessary to achieve the observed relic density \cite{Aghanim:2018eyx}. One way to classify such interactions is via ``portals" \cite{vectorportal} of 
various mass dimension that result from integrating out some set of heavy fields; at the renormalizable level, the set of such portals is known to be quite restricted \cite{vectorportal,KM}. In this paper, 
we will be concerned with the implications of the vector boson/kinetic mixing (KM) portal, which is perhaps most relevant for thermal DM with a mass in the range of a $\sim$ few MeV to 
$\sim$few GeV and which has gotten much attention in the recent literature \cite{KM}.  In the simplest of such models, a force carrier (the dark photon, a gauge field corresponding 
to a new gauge group, $U(1)_D$ under which SM fields are neutral) mediates the relevant DM-SM interaction. This very weak interaction is the result of the small KM between 
this $U(1)_D$  and the SM hypercharge group, $U(1)_Y$, which is generated at the one-(or two)-loop level by a set of BSM fields, called portal matter (PM)\cite{KM,PM}, which carry charges 
under both gauge groups. In the IR, the phenomenology of such models is well-described by suitably chosen combinations of only a few parameters: the DM and dark photon masses, 
$m_{DM,V}$, respectively, the $U(1)_D$ gauge coupling, $g_D$ and $\epsilon$, the small dimensionless parameter describing the strength of the KM, $\sim 10^{-(3-4)}$. Frequently, and 
in what follows below,  
this scenario is augmented to also include the dark Higgs boson, whose vacuum expectation value (vev) breaks $U(1)_D$ thus generating the dark photon mass. This introduces two 
additional parameters with phenomenological import: the dark Higgs mass itself and the necessarily (very) small mixing between the dark Higgs and the familiar Higgs of the SM. Successfully 
extending this scenario 
to a completion in the UV while avoiding any potential issues that can be encountered in the IR remains an interesting model-building problem.

Extra dimensions (ED) have proven themselves to be a very useful tool for building interesting models of new physics that can address outstanding issues that arise in 4-D\cite{ED}. 
In a previous pair of papers \cite{Rizzo:2018ntg,Rizzo:2018joy}, hereafter referred to as I and II respectively, we considered the implications of extending this familiar 4-D kinetic mixing picture into a (flat) 5-D scenario where it was assumed 
that the DM was either a complex scalar, a Dirac fermion or a pseudo-Dirac fermion with an $O(1)$ mass splitting. In all cases we found some unique features of the 5-D setup, \eg, the 
existence of strong destructive interference between the exchanges of Kaluza-Klein (KK) excitations of the dark photon allowing for light Dirac DM, which is excluded by CMB \cite{Ade:2015xua,Liu:2016cnk} constraints 
in 4-D, new couplings of the split pseudo-Dirac states to the dark photon that avoids co-annihilation suppression found in 4-D \cite{4dmaj}, or the freedom to choose appropriate 5-D wave function boundary 
conditions, \etc, all of which helped us to avoid some of the model building constraints from which the corresponding 4-D KM scenario can potentially suffer. 

The general structure of the model setups 
considered previously in I and II followed from some rather basic assumptions: ($i$) The 5-D space is a finite interval, $0\leq y\leq \pi R$, that is bounded by two branes, upon one of which 
the SM fields reside while the dark photon lives in the full 5-D bulk. This clearly implies that the $U(1)_D-U(1)_Y$ KM must solely occur on the SM brane. The (generalization of) the usual 
field redefinitions required to remove this KM in order to obtain canonically normalized fields then naturally leads to the existence of a very small, but {\it negative} brane localized kinetic 
term (BLKT) \cite{blkts} for the dark photon which itself then leads to a tachyon and/or ghost field in its Kaluza-Klein (KK) expansion.  
We are then led to the necessary conclusion that an $O(1)$ {\it positive} BLKT must already exist to remove this problem; the necessity of such a term was then later shown to be 
also very useful for other model building purposes. ($ii$) A simple way to avoid any significant mixing between SM Higgs, $H$, and the dark Higgs, $S$ which is employed in 4-D to generate the dark photon mass, is to eliminate the need for the dark Higgs to exist. This then removes the 
necessity of fine-tuning the parameter $\lambda_{HS}$ in the scalar potential describing the $\sim S^\dagger SH^\dagger H$ interaction in order to avoid a large branching fraction for the 
invisible width for the SM Higgs, $H$ \cite{HiggsMixing,inv}.\footnote{We note that this method of avoiding a dark Higgs while maintaining a massive dark photon is hardly unique. For example, a Stuckelberg mass may be introduced, as discussed in a different region of parameter space in the second reference of \cite{vectorportal}. However, the method employed in I and II (and here) affords far greater model-building flexibility than a Stuckelberg mechanism construction. We could, for instance, implement a non-Abelian dark gauge group using the extra-dimensional setup in I and II, which would be impossible if we assumed a Stuckelberg mass for the dark photon.} As is well known, one can employ appropriate (mixed) boundary conditions on the 5-D dark photon wave function on both branes to break 
the $U(1)_D$ symmetry and generate a mass for the lowest lying dark photon KK mode \cite{csakiEDs} without the presence of the dark Higgs. These boundary conditions 
generically have the form (in the absence of BLKTs or other dynamics on either brane) $v(y)|_1=0,~\partial v(y)|_2=0$ where $v$ is the 5-D dark photon wavefunction, $y$ describing the 
co-ordinate in the new extra dimension as above and $|_i$ implies 
the evaluation of the relevant quantity on the appropriate brane. Since the SM exists and the corresponding KM happens on one of these branes, it is obvious the $v$ cannot vanish there, 
so we identify the location of the SM with brane 2. It is then also obvious that the DM itself cannot live on brane 1 otherwise it would no longer interact with either the dark photon or, through 
it, with the SM, so the DM must {\it also} reside in the bulk, have its own set of KK excitations and, for phenomenological reasons, its own somewhat larger BLKT along with constrained boundary 
conditions. While allowing us to successfully circumvent some of the possible problems associated with analogous 4-D setup, this arrangement leads to a rather unwieldy structure. 
Can we do as well (or better) with a less cumbersome setup? This is the issue we address in the current paper. 

The complexity of the previously described structure follows directly from ($ii$), \ie,  employing boundary conditions to break $U(1)_D$ so that there is no dark Higgs-SM Higgs mixing 
(as there is no dark Higgs with which to mix). In the present analysis, we consider an alternative possibility which also naturally avoids this mixing in an obvious way, \ie, localizing the 
dark Higgs as well as the DM on the other, non-SM (\ie, dark) brane with only the dark photon now living in the full 5-D bulk to communicate their existence to us. Thus, keeping ($i$) but with the 
breaking of $U(1)_D$ on the dark brane via the dark Higgs vev, we eliminate the need for KK excitations of the DM field while also disallowing any tree-level dark Higgs-SM Higgs mixing, 
and thus significantly diminishing the phenomenological role of the dark Higgs itself as we will see below. In what follows, we will separately consider and contrast both flat as well as the 
warped \cite{Randall:1999ee,morrissey} versions of this setup in some detail assuming that the DM is a complex scalar field, $\phi$, which does not obtain a vev. This choice, 
corresponding to a dominantly p-wave annihilation via the spin-1 dark photon mediator to SM fermions, allows us to trivially avoid the constraints from the CMB while still recovering the observed relic density\cite{Steigman:2015hda,21club}.\footnote{While it was demonstrated in II that under certain conditions, the CMB constraints on $s$-channel fermionic DM annihilation might be avoided, these setups are dramatically more complicated than simply assuming that the DM is a complex scalar. To keep the construction here as simple as possible, we restrict our discussion to the scalar case and leave a detailed exploration of analogous fermionic DM models with an s-wave annihilation process to future work.} 
As we will see, in 
addition to the IR parameters noted above and suitably defined here in the 5-D context, only 2(3) additional parameters are present for the flat (warped) model version, 
these being the SM brane BLKT for the dark photon, $\tau$, and size of the mass term, $m_V$. In the warped version, as is usual, the curvature of the anti-deSitter space scaled by the compactification radius, $kR$, is also, in principle, 
a free parameter. Here, however, the value of this quantity is roughly set by the ratio of the weak scale, $\sim 250$ GeV, to that associated with the dark photon mass, $\sim 100$ MeV, \ie, $kR\sim 1.5-2$. 
In what follows, an $O(1)$ range of choices for the values of all of these quantities will be shown to directly lead to phenomenologically interesting results. Unlike in the previous 
setups, now the boundary conditions applied to wavefunction $v(y)$ will be significantly relaxed so that requirement of at least one of $v(y=0,\pi R)=0$ is no longer a 
problem, and the values of these wavefuctions will be determined by the values of the parameters $m_V (\tau)$ on the dark (SM) brane.

The outline of this paper is as follows: In Section \ref{Setup}, we present the construction of this model while remaining agnostic to the specific geometry of the extra dimension, while in Section \ref{FlatAnalysis} we specialize our discussion to the case in which the extra dimension is flat and present a detailed analysis of this scenario. In Section \ref{WarpedAnalysis}, we present an analogous discussion of the model in the case of a warped extra dimension, with appropriate comparison to the results from the flat case scenario. Section \ref{Conclusion} contains a summary and our conclusions.


\section{General Setup}\label{Setup}

Before beginning the analysis of the current setup, we will very briefly review the formalism from I that remains applicable, generalizing it slightly to incorporate either a flat or warped extra dimension. 

\subsection{Field Content and Kaluza-Klein Decomposition}
\begin{figure}[htbp]
\centerline{\includegraphics[width=5.0in,angle=0]{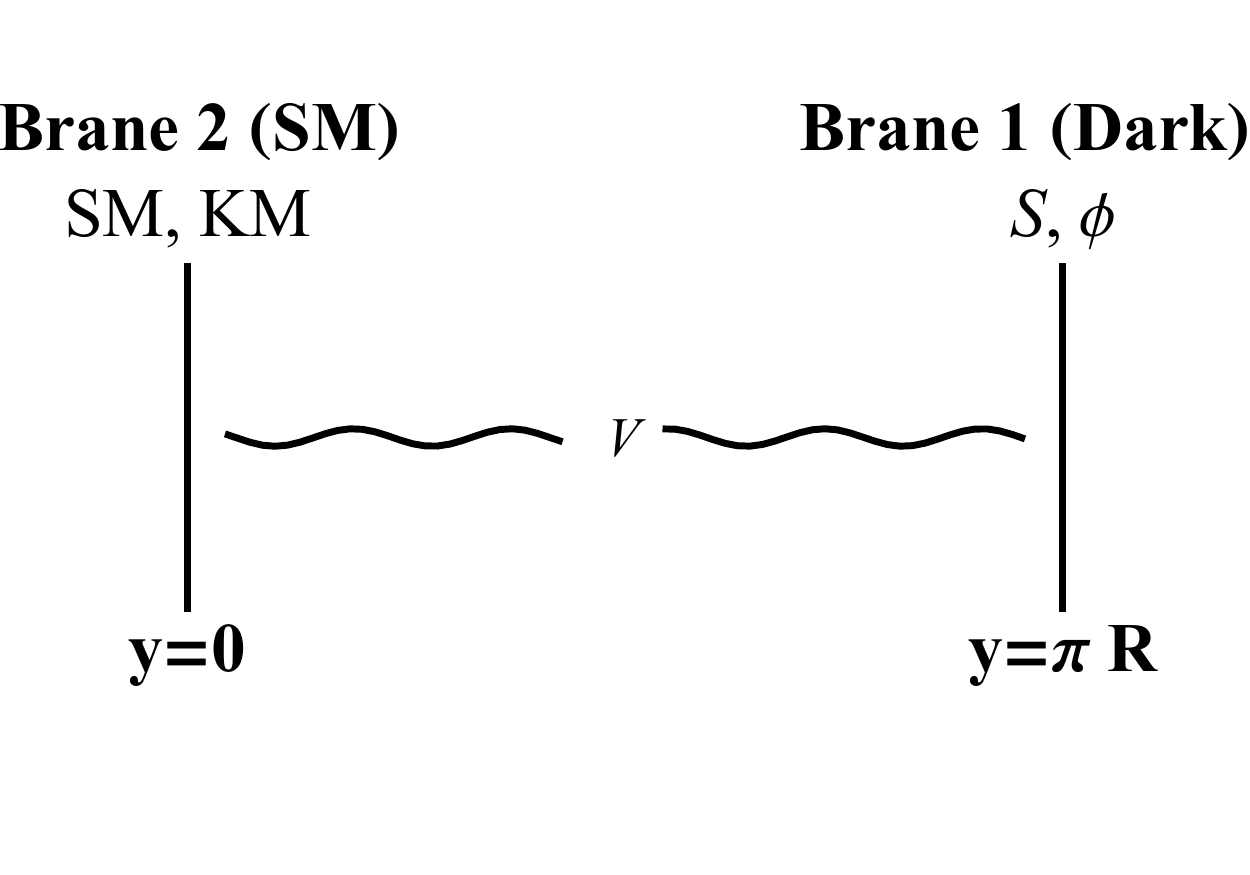}}
\caption{A simple diagram overviewing the construction of our model. The fields $\phi$ and $S$, representing the complex scalar dark matter (DM) and the dark Higgs, respectively, are localized on Brane 1 (the ``dark brane") at $y=\pi R$. The Standard Model (SM) and kinetic mixing (KM) terms are localized on Brane 2 (the ``SM brane") at $y=0$. The bulk contains only one field, the dark photon $V$.}
\label{figOverview}
\end{figure}

As noted in the Introduction, we consider the fifth dimension to be an interval $0\leq y \leq \pi R$ bounded by two branes; for definiteness we assume that the entirety of the SM is localized on the $y=0$ brane, while the dark matter (DM) field (which we shall refer to as $\phi$) and the dark Higgs (which we shall refer to as $S$) are localized on the opposite brane at $y= \pi R$. For a flat extra dimension, this assignment of branes is arbitrary, however we shall see that it has some physical motivation in the warped case. For clarity, we depict the localization of the different fields in the model in Figure \ref{figOverview}. It should be noted that although both $\phi$ and $S$ are complex scalars localized on the dark brane, they must be separate fields. This is in order to avoid potential pitfalls related to DM stability: If $\phi$ were to serve as both the dark matter and dark Higgs, then the physical DM particle could decay via a pair of virtual dark photons into SM particles, requiring draconian constraints on DM coupling parameters in order to maintain an appropriately long lifetime. This phenomenon is discussed in greater detail in the discussion of Model 2 in I, where the additional complexity added by making $\phi$ a bulk field allows for sufficient model-building freedom to circumvent these constraints. However, as our DM field $\phi$ is brane-localized in this construction, the methods outlined in that work are not applicable here, and we are forced to assume that $\phi$ acquires no vev and posit a separate dark Higgs, $S$, to break the dark gauge symmetry.

The metric of our 5-dimensional model is assumed to take the form,
\begin{equation}\label{genericMetric}
    ds^2 = f(y)^2 ~\eta_{\mu \nu} dx^{\mu} dx^{\nu}-dy^2,
\end{equation}
where $f(y)$ is simply some function of the bulk coordinate $y$: For a flat extra dimension, $f(y)=1$, while for a Randall-Sundrum setup \cite{Randall:1999ee}, $f(y)=e^{-ky}$, where $k$ is a curvature scale. The dark photon, described by a gauge field $\hat{V}_A(x,y)$ lies in the full 5-D bulk, and kinetically mixes with the 4-D SM hypercharge gauge field $\hat{B}_\mu (x)$ on the SM brane via a 5-D kinetic mixing (KM) parameter $\epsilon_{5D}$ as described (before symmetry breaking) by the action
\begin{align}\label{originalAction}
S=\int d^4x \int_{0}^{\pi R} dy &~\Big[-\frac{1}{4} \Big( \hat V_{\mu \nu} \hat V^{\mu \nu} -2 f(y)^{2}(\partial_\mu \hat{V}_y -\partial_y \hat{V}_\mu)(\partial^\mu \hat{V}^y -\partial^y \hat{V}^\mu) \Big)  \\
&~+\Big(-\frac{1}{4} \hat B_{\mu\nu} \hat B^{\mu\nu} +\frac{\epsilon_{5D}}{2c_w} \hat V_{\mu\nu} \hat B^{\mu\nu}  + L_{SM} \Big) ~\delta(y) \Big] \,,  \nonumber
\end{align}
where $c_w=\cos(\theta_w)$, the weak mixing angle, Greek indices denote only the 4-dimensional vector parts of the gauge field $\hat{V}$, and $\hat{V}_y$ denotes the fifth component of this field. Since spontaneous symmetry breaking takes place on the dark brane via the vev of the dark Higgs, $S$, we know \cite{5d,Casagrande:2008hr} that 
in the Kaluza-Klein (KK) decomposition the 5th component of $\hat V_A$ (which does not experience KM) and the imaginary part of $S$ combine to form the Goldstone bosons eaten by $\hat V$ to 
become the corresponding longitudinal modes. So, we are free in what follows to work in the $V_y=0$ gauge, at least for the flat and Randall-Sundrum-like geometries that we are considering here. Then the alluded-to relevant KK decomposition for the 4-D components 
of $\hat V$ is given by{\footnote{Note that $n=1$ labels the lowest lying excitation appearing in this sum.}}
\begin{equation}
\hat V^\mu (x,y) = \frac{1}{\sqrt{R}}\sum_{n=1}^\infty  ~ v_n (y) \hat V_n^\mu (x)\,,
\end{equation}
where we have factored out $R^{-1/2}$ in order to render $v_n(y)$ dimensionless. To produce a Kaluza-Klein tower, then we will require that the functions $v_n(y)$ must satisfy the equation of motion
\begin{align}\label{generalEOM}
    \partial_y [f(y)^2 ~\partial_y v_n(y)] = - m_n^2 v_n (y)
\end{align}
in the bulk, where here the $m_n$ are the physical masses of the various KK excitations.
Defining the KK-level dependent quantity $\epsilon_n = R^{-1/2} \epsilon_{5D} v_n (y=0)$, which we see explicitly depends on the values of the dark photon KK wavefunctions evaluated on the SM brane, we see that the 5-D KM becomes an infinite tower of 4-D KM terms given by 
\begin{equation}
\sum_n \frac{\epsilon_n}{2 c_w} \hat V_n^{\mu\nu} \hat B_{\mu\nu}\, .
\end{equation}
As discussed in I, the intuitive generalization of the usual kinetic mixing transformations, $\hat B^{\mu\nu} = B^{\mu\nu}+\sum_n \frac{\epsilon_n}{c_w} V^{\mu\nu}_n$, $\hat V^{\mu\nu} \to V^{\mu\nu}$, will be numerically valid in scenarios in which the infinite sum $\sum_{n} \epsilon_n^2/\epsilon_1^2$ is approximately $\lsim O(10)$, and $\epsilon_1 \ll 1$; in other words, $\epsilon_1$ is sufficiently small and $\epsilon_n$ shrinks sufficiently quickly with increasing $n$. Otherwise, terms of $O(\epsilon_1^2)$ (at least) become numerically significant and can't be ignored in the analysis, even if each individual $\epsilon_n$ remains small.

In both the cases of a warped and flat extra dimension, the sum $\sum_{n} \epsilon_n^2/\epsilon_1^2$ is within the acceptable range as long as there is a sufficiently large positive brane-localized kinetic term (BLKT) on the same brane as the SM-dark photon kinetic mixing, as was shown for flat space in I and will be demonstrated for warped space in Section \ref{WarpedAnalysis}. So, by selecting $\epsilon_1 \sim 10^{-(3-4)}$, as suggested by experiment, within our present analysis we can always work to leading 
order in the $\epsilon_n$'s, and thus the transformations $\hat B^{\mu\nu} = B^{\mu\nu}+\sum_n \frac{\epsilon_n}{c_w} V^{\mu\nu}_n$, $\hat V^{\mu\nu} \to V^{\mu\nu}$ will be sufficient for our purposes in removing the KM. 

It is interesting to note that we can see the requirement for a positive brane-localized kinetic term (BLKT) in our setup more immediately from the action of Eq.(\ref{originalAction}). In particular, as noted in I and the Introduction, making the usual substitution in the 5D theory to eliminate kinetic mixing (that is, $\hat{V}\rightarrow V$ and $\hat{B}\rightarrow B+\frac{\epsilon_{5D}}{c_w} \hat{V}$) produces the small negative 
BLKT $\sim -\frac{\epsilon_{5D}^2}{Rc_w^2}$ mentioned in the Introduction. In this 5-D treatment, the effective BLKT experienced by the $V$ on the SM brane would therefore be equal to whatever BLKT existed before mixing, \emph{shifted} by the mixing-induced negative brane term. This shift is highly suggestive of the necessity of introducing a positive BLKT to the model before mixing, in order to avoid the pitfalls associated with negative BLKT's (for example, in the case of a flat extra dimension, negative BLKT's such as this are well known to lead to tachyonic KK modes or ghost-like states); the BLKT before mixing is applied must be large enough that the effective term after mixing is non-negative.
In our explicit treatment of the model's kinetic mixing, because we only apply field shifts at the level of the effective 4-D theory, this negative brane term does not appear, but the requirement for a positive BLKT instead emerges as a condition to keep the kinetic mixing between the SM hypercharge boson and an infinite number of KK dark photons small. As will be seen, such considerations lead to a lower bound on the SM-brane BLKT.

Next, we note that the sum of the brane actions corresponding to the usual (positive) BLKT for $V$ on the SM brane, which we shall denote by $\tau$, and the corresponding dark 
Higgs generated mass term for $V$ on the dark brane, denoted by $m_V$, is given by 
\begin{equation}
S_{branes}=\int d^4x \int_{0}^{\pi R} dy ~\Big[-\frac{1}{4}  V_{\mu\nu}  V^{\mu\nu}  \cdot \tau R~\delta(y) +\frac{1}{2} m_V^2 R ~V_\mu V^\mu ~\delta(y-\pi R)\Big]\,,
\end{equation}
where factors of $R$ have been introduced to make $\tau$ dimensionless as usual and for $m_V$ to have the usual 4-D mass dimension. We note that one of the main advantages of our present setup is that the dark Higgs which generates the brane mass term $m_V$ is isolated from any mixing with the SM Higgs, and as a result, its phenomenological relevance in this construction is quite limited. Given that it is unstable (from decays via on- or off-shell dark photons, depending on the dark Higgs and dark photon masses), its most salient effect on any observables in the theory would be if a process such as $\phi \phi^\dagger \rightarrow V^{(n)*} \rightarrow V^{(m)} S$ were to dominate the calculation of the DM relic density. While this sort of construction may be of some interest (for example, \cite{Baek:2020owl} discusses a 4D model in a similar parameter space that may address the recent XENON1T electron recoil excess \cite{Aprile:2020tmw} in which a light dark Higgs plays such a role\footnote{It should be noted that without substantial modifications to our own setup, such as the addition of mass splitting between the two degrees of freedom of the complex scalar DM field \cite{Baek:2020owl} or additional slightly heavier DM scalars that facilitate the production of boosted $\phi$ pairs \cite{Jia:2020omh}, the XENON1T excess cannot be explained in the parameter space we are considering. A detailed discussion of how these or other mechanisms might be incorporated into a 5-D model like that discussed here is beyond the scope of this work.}), this effect can be easily suppressed by assuming that the dark Higgs (or rather, its real component after spontaneous symmetry breaking) has a sufficiently large mass (slightly greater than twice the DM mass, assuming cold dark matter), rendering this process kinematically forbidden. As such, for our analysis we can ignore this scalar and instead simply assume the existence of the brane-localized mass term $m_V$ without further complications. We will define the 4-D gauge coupling of the dark photon to be that between the DM and the lowest $V$ KK mode as evaluated on the dark brane. The action $S_{branes}$ supplies the boundary conditions, as well as the complete orthonormality condition, necessary for the complete solutions of the $v_n$. These are
\begin{equation}\label{generalBCs}
\Big (f(0)^2~\partial_y + m_n^2 \tau R\Big)v_n(0)=0, ~~~~~\Big( f(\pi R)^2 ~\partial_y+m_V^2 R\Big)v_n(\pi R)=0\,
\end{equation}
for the boundary conditions, and
\begin{equation}\label{orthoRelation}
    \frac{1}{R}\int_{0}^{\pi R} dy \; v_n(y) v_m(y) (1+ \tau R \delta(y)) = \delta_{n m}
\end{equation}
for the orthonormality condition. At this point, once the function $f(y)$ is specified, as we shall do in Sections \ref{FlatAnalysis} and \ref{WarpedAnalysis} for a flat and a warped extra dimension respectively, it is possible to uniquely determine the bulk wave functions $v_n(y)$ for all $n$ given the parameters $R$, $\tau$, $m_V^2$, and whatever additional parameters are necessary to uniquely specify $f(y)$.

Beyond discussing characteristics of individual KK modes, we shall find it convenient at times in our analysis to speak in terms of summations over exchanges of the entire dark photon KK tower. In particular, the sum
\begin{equation}\label{FDefinition}
    F(y,y',s) \equiv \sum_n \frac{v_n(y) v_n(y')}{s-m_n^2}
\end{equation}
shall appear repeatedly in our subsequent discussion, where for our purposes here $s$ is simply a positive number, but in our actual analysis shall denote the Mandelstam variable of the same name. To evaluate this sum, we can perform an analysis similar to that of \cite{Casagrande:2008hr,Hirn:2007bb}. First, we note that the orthonormality condition of the KK modes in Eq.(\ref{orthoRelation}) requires that
\begin{align}\label{generalOrthonormality}
    \frac{1}{R}\int_{0}^{\pi R} d y \; v_m (y)v_n(y)(1+\tau R \delta(y)) = \delta_{m n}\\
    \rightarrow \sum_{n} v_n(y) v_n(y')(1+\tau R \delta(y))=R \delta(y-y'), \nonumber
\end{align}
where the sum in the second line of this equation is over all KK modes $n$. We then note that the equation of motion Eq.(\ref{generalEOM}) and the $y=0$ boundary condition of Eq.(\ref{generalBCs}) can be recast in an integral form as
\begin{align}\label{generalIntegralEOM}
    v_n(y) =v_n(0)-m_n^2 \int_0^{y} dy_1 \; [f(y_1)]^{-2}\int_0^{y_1} dy_2 \; v_n (y_2)(1+\tau R \delta(y_2))
\end{align}
Using this integral form of the equation of motion for $v_n(y)$, we can now compute the sum $F(y,y',s)$. Combining Eqs.(\ref{generalOrthonormality}) and (\ref{generalIntegralEOM}), we can write the integral equation
\begin{align}\label{FIntegralEq}
    F(y,y',s) &= \int_0^{y} dy_1 \; f(y_1)^{-2}\int_0^{y_1} dy_2 \; [R\delta(y_2-y')-s F(y_2,y',s)(1+\tau R \delta(y_2))]+F(0,y',s).
\end{align}
Eq.(\ref{FIntegralEq}) can be straightforwardly rewritten as a differential equation,
\begin{align}\label{FDiffEq}
    \partial_y [f(y)^2 ~\partial_y F(y,y',s)] = R\delta(y-y')- s F(y,y',s), \nonumber\\
    \partial_y F(y,y',s)|_{y=0} = -s \tau R f(0)^{-2}F(0,y',s),\\
    \partial_y F(y,y',s)|_{y=\pi R} = - m_V^2 R f(\pi R)^{-2} F(\pi R, y',s), \nonumber
\end{align}
where the $y=0$ boundary condition is explicitly in the integral equation Eq.(\ref{FIntegralEq}), while the second is easily derivable from the $y=\pi R$ boundary condition on $v_n(y)$ given in Eq. (\ref{generalBCs}). Once a function $f(y)$ (and therefore a metric) has been specified, the function $F(y,y',s)$ is then uniquely specified by Eq.(\ref{FDiffEq}).

\subsection{Dark Photon Couplings}

With equations of motion for the Kaluza-Klein (KK) modes' bulk profiles $v_n(y)$ and the summation $F(y,y',s)$ specified, it is now useful to discuss some general aspects of our construction's phenomenology before explicitly choosing a metric. First, we note that the effective couplings of the $n^{th}$ KK mode of the dark photon to the DM on the dark brane are given by $g^{DM}_n = g_{5D} v_n(y=\pi R)/\sqrt{R}$, where $g_{5D}$ is the 5-dimensional coupling constant appearing in the theory, while recalling that the effective kinetic mixing (KM) parameters $\epsilon_n$ are similarly given by $\epsilon_n = \epsilon_{5D} v_n(y=0)/\sqrt{R}$. In terms of the value of these parameters for the least massive KK mode, $g_D \equiv g^{DM}_1$ and $\epsilon_1$, we can then write
\begin{align}\label{gepsilonDefs}
    g^{DM}_n &= g_D \frac{v_n (y=\pi R)}{v_1 (y= \pi R)}, \\
    \epsilon_n &= \epsilon_1 \frac{v_n(y=0)}{v_1(y=0)}. \nonumber
\end{align}
Armed with these relationships, our subsequent analysis will treat $\epsilon_1$, $g_D$, and the mass of the least massive KK dark photon excitation, $m_{1}$ (which we trade for $R$), as free parameters and identify them with the corresponding quantities that appear in the conventional 4-D KM portal model.

With the dark photon coupling to DM given in Eq.(\ref{gepsilonDefs}), we now can remind the reader of the slightly more complex form of the dark photon coupling to SM fermions, previously derived in I. In particular, once the shift $B \rightarrow B + \sum_n \frac{\epsilon_n}{c_w} V_n$ is applied, the $Z$ boson undergoes a small degree of mixing with the $V_n$ fields. Once the mass matrix of the $Z$ boson and the $V_n$ modes is diagonalized again, then to leading order in the $\epsilon$'s the $V_n$ fields couple to the SM fermions as
\begin{align}\label{gVff}
    \frac{g}{c_w}t_w \epsilon_n \bigg[ T_{3L} \frac{m_n^2}{m_Z^2-m_n^2}+Q \frac{c_w^2 m_Z^2 - m_n^2}{m_Z^2-m_n^2} \bigg] \xrightarrow{m_n \ll m_Z} e Q \epsilon_n,
\end{align}
where $Q$ is the fermion's electric charge, $T_{3L}$ is the third component of its weak isospin, $m_Z$ is the mass of the $Z$ boson, $m_n$ is the mass of the dark photon KK mode $V_n$, $e$ is the electromagnetic coupling constant, and $c_w$ and $t_w$ represent the cosine and tangent of the Weinberg angle, respectively. As pointed out in I and in Eq.(\ref{gVff}), when $m_n \ll m_Z$, the coupling in Eq.(\ref{gVff}) simplifies dramatically; we shall find this approximation exceedingly useful in our subsequent analysis.

We also remind the reader that, as discussed in I, the kinetic and mass mixing of the dark photon fields with the $Z$ boson results in non-trivial modifications to the $Z$ boson and SM Higgs phenomenology. In particular, the $Z$ boson gains an $O(\epsilon)$ coupling to the DM, as well as an $O(\epsilon^2)$ correction to its mass. It was pointed out in I that the $\epsilon$ suppression of these effects keeps them far below present experimental bounds (for example, from precision electroweak observables for the mass correction and measurements of the invisible $Z$ decay width for the $Z$ coupling to DM), and given the fact that the $Z$ boson is roughly $10^2$ to $10^3$ times more massive than the lighter dark photon KK modes, this coupling also does not contribute significantly to the DM relic abundance calculation or direct detection scattering processes. As a result, we shall ignore these couplings in our subsequent analysis.

Meanwhile, the SM Higgs field $H$ gains two phenomenologically interesting new couplings from this mixing, which may contribute to experimentally constrained Higgs decays to either a pair of dark photon modes or to a single dark photon mode with a $Z$. First, new $HZV_n$ couplings emerge of the form,
\begin{align}\label{gHZV}
    K_{HZV_n} = \frac{2 m_Z^2}{v_H} \bigg[ \frac{t_w \epsilon_n m_n^2}{m_Z^2-m_n^2} \bigg],
\end{align}
where $v_H$ denotes the SM Higgs vev $\sim$ 246 GeV. Meanwhile, $H V_n V_m$ couplings emerge of the form,
\begin{align}\label{gHVV}
    K_{H V_n V_m} = \frac{2 m_Z^2}{v_H} \bigg[ \frac{t_w \epsilon_n m_n^2}{m_Z^2-m_n^2} \bigg] \bigg[ n \rightarrow m \bigg].
\end{align}
Notably, the couplings of the Higgs in Eq.(\ref{gHZV}) and (\ref{gHVV}) to a given dark photon field $V_n$ are both proportional to the ratio $m_n^2/(m_Z^2-m_n^2)$, which results in an approximate $m_n^2/m_Z^2$ suppression when $m_n \ll m_Z$. Given that we are considering the parameter space in which the lightest dark photon, $V_1$, has a mass of $\sim 100$ MeV, this term provides extremely strong suppression of the SM Higgs couplings to the lighter KK modes of $V$. Given that the $\epsilon_n$'s shrink as $n$ gets very large (as necessitated by the convergence of the sum $\sum_n \epsilon_n^2/\epsilon_1^2$), and hence couplings of the Higgs to more massive dark photon KK modes are highly suppressed by smaller $\epsilon_n$'s, the numerical effect of these couplings on the observable physics in the model is minute, even compared to other $\epsilon$-suppressed quantities. As a result, and as discussed in detail for the analogous system in I, the contributions of the couplings in Eqs.(\ref{gHZV}) and (\ref{gHVV}) are many orders of magnitude below the present constraints from Higgs branching fractions. For example, if we assume that the $\epsilon_1 = 10^{-3}$, $m_1=0.1-1$ GeV, our results for the Higgs decay width from processes $H \rightarrow Z V_n$ never exceed $O(1)$ eV for the model parameter space we discuss in either our flat or warped space constructions. Meanwhile, the sum of the $H \rightarrow V_n V_m$ widths, being doubly suppressed, never achieves a value of more than $O(10^{-4})$ eV. Both of these processes represent negligible contributions to the Higgs decay width, and therefore negligible branching fractions.

\subsection{Dark Matter Phenomenology}

To round out our general discussion of this model's setup and phenomenology, it is useful to give symbolic results for two quantities that are of particular phenomenological interest for dark matter within the mass range we are considering, and which can be expressed in a manner agnostic to the specific functional form of the bulk wavefunctions $v_n(y)$ and the sum $F(y,y',s)$. In particular, we shall give symbolic expressions for the DM-electron scattering cross section and the thermally averaged annihilation cross section of DM into SM particles.\footnote{The simple thermal freeze-out treatment here is valid for DM particles of mass $\gsim O(\textrm{MeV})$, as long as the force mediator controlling annihilation, in our case the dark photon, is of similarly small mass\cite{Boehm:2003hm}. In most of the parameter space this cross section, up to factors of $\sim$5-40\% as we will see below, is identical to that obtained in well-studied 4-D models.} We note that for the class of models we consider, these represent the dominant sources of constraints. For example, in \cite{Cho:2020mnc}, constraints from direct detection of DM particles boosted by cosmic rays are found to be much weaker than conventional DM-electron scattering constraints in this region of parameter space for a generic 4-D model of kinetic mixing/vector portal DM.

It should be noted that when performing the computations of the direct detection scattering and annihilation cross sections, we have made two significant simplifying assumptions: First, we have approximated the coupling of any Kaluza-Klein (KK) dark photon mode $V_n$ to a given SM fermion species as $\approx e Q \epsilon_n$, which, according to Eq.(\ref{gVff}), is only a valid approximation when $m_n \ll m_Z$, and therefore breaks down if we sum over the entire infinite tower of KK modes once we reach sufficiently large $n$. Second, we have assumed that the contribution of the $Z$ boson exchange to both of these processes is negligible compared to that of the exchange of KK tower bosons. In practice, both of these approximations amount to letting $m_Z \rightarrow \infty$, namely assuming that the $Z$ boson is much heavier than \emph{all} KK modes of the dark photon. Numerically, we find that the $m_Z \rightarrow \infty$ approximation has a negligibly small effect on our results: Because the lightest dark photon modes are approximately $10^2$ to $10^3$ times lighter than the $Z$ boson, and these light modes also have the numerically largest kinetic mixing, applying the precise dark photon coupling of Eq.(\ref{gVff}) and including the effects of $Z$ boson exchanges in these computations serves only to significantly complicate our symbolic expressions while altering the numerical result at well below the percent level. As such, for the purposes of these computations we confidently work in the limit where $m_Z \rightarrow \infty$.

First, we note that in the limit where the DM particle $\phi$'s mass $m_{DM}$ is far greater than the mass of an electron, we can approximate the DM-electron scattering cross section for direct detection as
\begin{align}\label{sigmae}
    \sigma_{\phi e} = 4 \alpha_{\textrm{em}} m_e^2 (g_D \epsilon_1)^2 \bigg\lvert \frac{F(0,\pi R,0)}{v_1(\pi R) v_1 (0)} \bigg\rvert^2.
\end{align}

To ensure that our DM candidate produces the correct relic abundance, we also must compute the thermally averaged annihilation cross section for DM into SM particles (which we shall denote by the symbol $\sigma$), weighted by the M$\o$ller velocity of the DM particle pair system $v_{M\o l}$ in the cosmic comoving frame \cite{Gondolo:1990dk}. We are careful to note that $\sigma$ here refers to the Lorentz-$\textit{invariant}$ cross section. To find this average, we must integrate $\sigma v_{\textrm{M\o l}}$ weighted by the two Bose-Einstein energy distributions, $f(E)$, of the complex DM fields in 
the initial state. As noted in \cite{Gondolo:1990dk}, if the freeze-out temperature, $T_F$ satisfies $x_F=m_{DM}/T_F\gsim 3-4$ as it will below, we can approximate these Bose-Einstein distributions with Maxwell-Boltzmann ones, and can employ the following formula to express the thermal average as a one-dimensional integral,
\begin{equation}\label{singleIntegralAvg}
    <\sigma v_{\textrm{M\o l}}>= \frac{2 x_F}{K_2^2(x_F)}\int_0^{\infty} d \varepsilon \; \varepsilon^{1/2}(1+2 \varepsilon)K_1(2 x_F \sqrt{1+\varepsilon})\sigma v_{lab},
\end{equation}
where $K_n(z)$ denotes the modified Bessel function of the second kind of order $n$, $v_{lab}$ is the relative velocity of the two DM particles in a frame in which one of them is at rest, and $\varepsilon \equiv (s-4m_{DM}^2)/(4m_{DM}^2)$, \ie,  the kinetic energy per unit mass in the aforementioned reference frame. This integral can be performed numerically; in our numerical evaluations here we will assume $x_F=20$ but note that other values in the 20-30 range give very similar results. We can proceed now by computing the cross-section for the annihilation of a DM particle-antiparticle pair into a pair of SM fermions of mass $m_f$ and electric charge $Q_f$, in which case we invoke the following expression for the cross-section of a $2\rightarrow 2$ process, 
\begin{align}
    \sigma v_{lab} = \frac{\sqrt{s(s-4 m_f^2)}}{s(s-2 m_{DM}^2)}\int \frac{d \Omega ~|\mathcal{M}|^2}{(64 \pi^2)},
\end{align}
where $s$ is the standard Mandelstam variable, $m_{DM}$ is the mass of the DM particle, $\Omega$ is the center-of-mass scattering angle and $\mathcal{M}$ is the matrix element for the annihilation process we are considering. When $s$ is far from any KK mode resonances, we arrive at the result
\begin{align}\label{sigmavrelNonRes}
    \sigma v_{lab} = \frac{1}{3}\frac{g_D^2 \epsilon_1^2 \alpha_{\textrm{em}}Q_f^2}{v_1(\pi R)^2 v_1 (0)^2 }\frac{(s+2m_f^2)(s-4 m_{DM}^2)\sqrt{s(s-4 m_f^2)}}{s(s-2 m_{DM}^2)} \lvert F(0,\pi R,s) \rvert^2.
\end{align}
%
%
%
In practice, for both of the specific cases we shall discuss in our analysis, we shall find it necessary to consider regions of parameter space such that DM annihilation through the first KK mode enjoys some resonant enhancement \cite{Feng:2017drg}. In order to accommodate this scenario, we have to modify Eq.(\ref{sigmavrelNonRes}) slightly, arriving at
\begin{align}\label{sigmavrel}
    \sigma v_{lab} &= \frac{1}{3}\frac{g_D^2 \epsilon_1^2 \alpha_{\textrm{em}}Q_f^2}{v_1(\pi R)^2 v_1 (0)^2 }\frac{(s+2m_f^2)(s-4 m_{DM}^2)\sqrt{s(s-4 m_f^2)}}{s(s-2 m_{DM}^2)}\\
    &\times\Bigg\lvert F(0,\pi R, s) -v_1(\pi R) v_1 (0)\Big( \frac{1}{s-m_1^2}-\frac{1}{s-m_1^2+i m_1 \Gamma_1}\Big)\Bigg\rvert^2, \nonumber
\end{align}
%
where $\Gamma_i$ is the total width of $V_i$ which we need to calculate as a function of $m_{i}$. We note that $V_1$ in particular will be very narrow as $\Gamma_1/m_1\simeq \alpha \epsilon_1^2/3 \sim 10^{-10}$ when decays to DM pairs are not kinematically allowed. 
Physically, we have simply subtracted the contribution of the lowest-lying KK mode from the sum $F(0,\pi R, s)$, where its propagator appears with its pole mass, and added this contribution again with the Breit-Wigner mass instead. Since the annihilation of two complex scalars into a pair of fermions through a vector gauge boson is $p$-wave process, and so is $v_{rel}^2$ suppressed at later times (\ie, at lower temperatures when the DM is moving slowly), we are safe from the previously mentioned strong constraints on DM annihilation during the CMB at $z \sim 10^3$ \cite{Steigman:2015hda}.
We further note that if $m_{DM}>m_{1}$, then we would expect the $s$-wave process $\phi \phi^\dagger \to 2V_1$ to be dominant for unsuppressed values of $g_D$. In order to avoid this possibility, 
we must then require that $m_{DM}<m_{1}$ and this will be reflected in our considerations below. We note that if $m_{1}>2m_{DM}$ then the $O(g_D^2)$ decay 
$V_1\to \phi \phi^\dagger$ will dominate, otherwise, $V_1$ will decay to SM fermions with, as noted above,  a suppressed $O(\alpha \epsilon_1^2)$ decay partial width.

%

\section{Flat Space Model Setup}\label{FlatAnalysis}

\begin{figure}[htbp]
\centerline{\includegraphics[width=5.0in,angle=0]{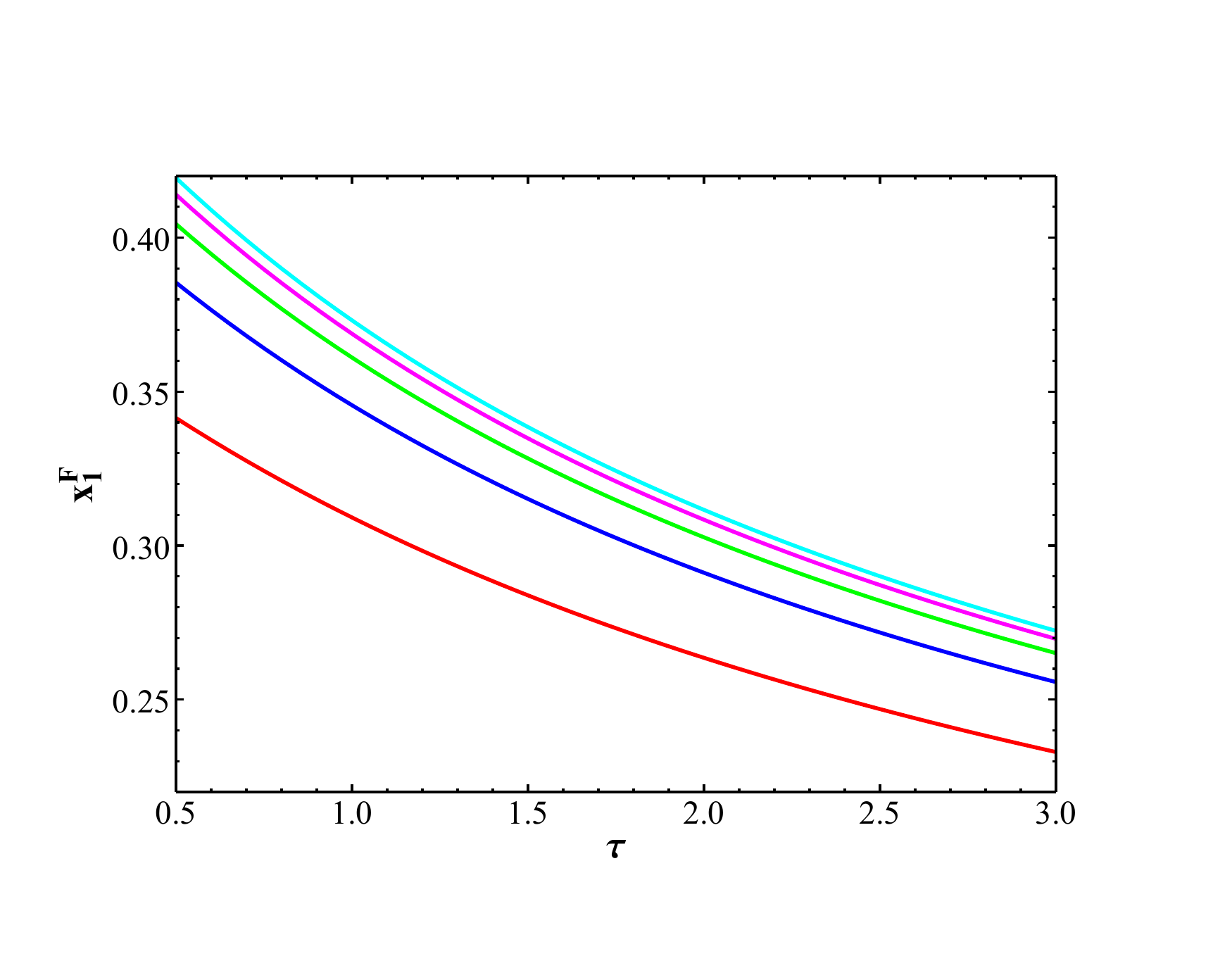}}
\vspace*{-2.0cm}
\centerline{\includegraphics[width=5.0in,angle=0]{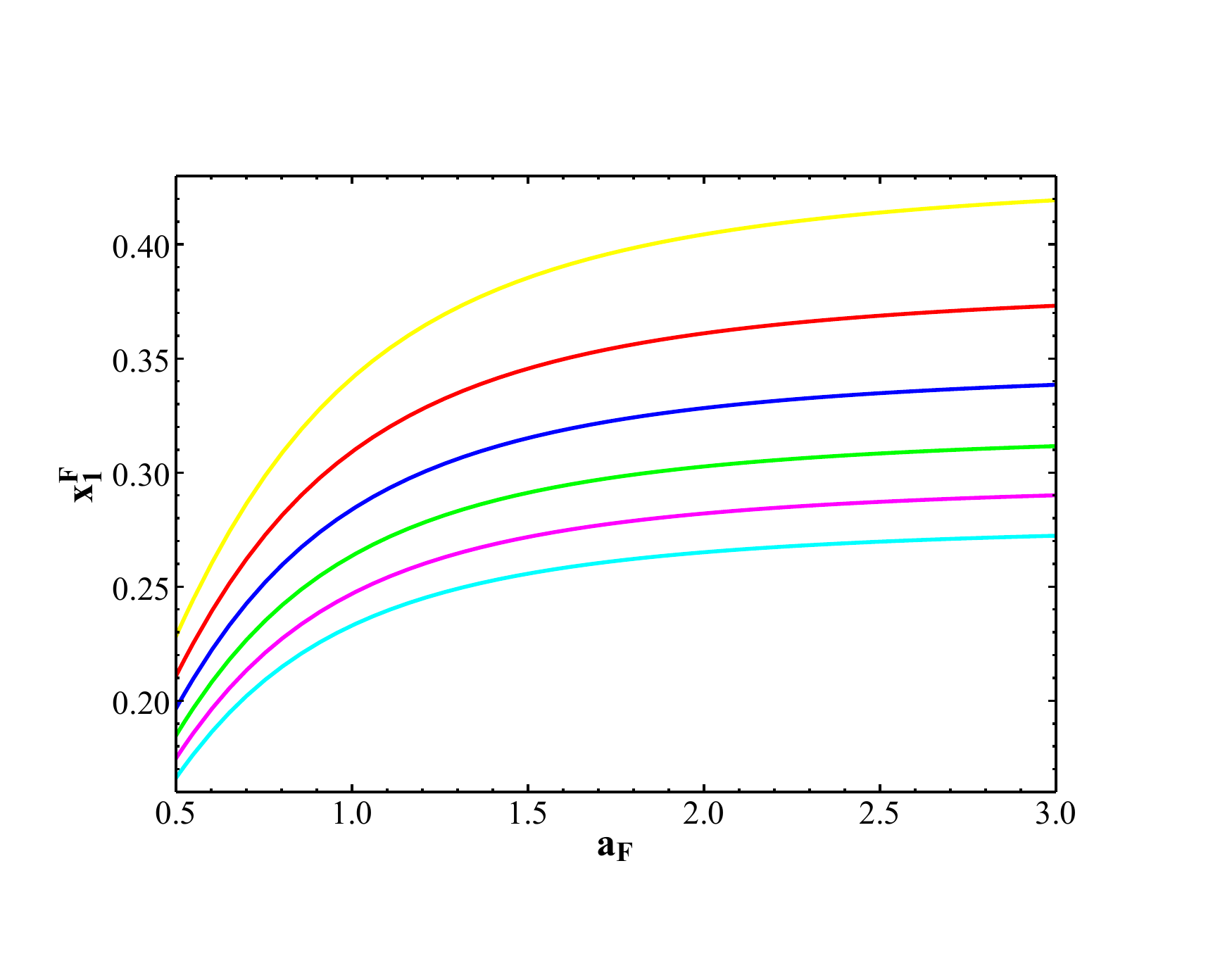}}
\vspace*{-1.0cm}
\caption{(Top) Value of the root $x^F_1$ as a function of $\tau$ for, from bottom to top, $a_F=$1, 3/2, 2, 5/2 and 3, respectively.
(Bottom) Value of the root $x^F_1$ as a function of $a_F$ for, from top to bottom, $\tau=$1/2, 1, 3/2, 2, 5/2 and 3, respectively. }
\label{fig1}
\end{figure}

In order to further explore the phenomenology of our construction, we must now specify the geometry of the extra dimension, namely by selecting a specific function $f(y)$ in Eq.(\ref{genericMetric}). With this determined, we can then straightforwardly find the spectrum of Kaluza-Klein (KK) gauge bosons $V_n$, their bulk wavefunctions $v_n(y)$, and concrete expressions for the cross sections of Eqs.(\ref{sigmae}) and (\ref{sigmavrel}). Initially, we shall consider the case of a flat extra dimension, \ie, $f(y)=1$. The equation of motion for the bulk profile $v_n(y)$ is then straightforward; from the generic case given in Eqs.(\ref{generalEOM}) and (\ref{generalBCs}), we quickly arrive at
\begin{align}\label{FlatEOM}
    \partial_y^2 v_n(y) = -m_n^2 v_n (y), \\
    (\partial_y + m_n^2 \tau R)v_n(y)|_{y=0} = 0, \nonumber \\
    (\partial_y + m_V^2 R)v_n(y)|_{y=\pi R} = 0. \nonumber
\end{align}
which when combined with the orthonormality condition Eq.(\ref{orthoRelation}) quickly yields the expressions 
\begin{align}\label{Flatvn}
    v_n(y) &= A_n (\cos(x^F_n (y/R))-\tau x^F_n \sin(x^F_n (y/R))), \\
    A_n &\equiv \sqrt{\frac{2}{\pi}}\Big( 1+(x^F_n \tau)^2 + (1-(x^F_n \tau)^2)\frac{\sin(2 \pi x^F_n)}{2 \pi x^F_n }+\frac{2 \tau}{\pi}\cos^2(\pi x^F_n)\Big)^{-\frac{1}{2}}, \nonumber \\
    x^F_n &\equiv m_n R, \;\;\;\; a_F \equiv m_V R, \nonumber
\end{align}
where we have defined the dimensionless quantities $x^F_n$ and $a_F$ from combinations of dimensionful parameters for the sake of later convenience{\footnote {The label ``F'' is used here 
to distinguish these flat space results from those of the warped case which we will discuss further below.}}. 
The allowed values of $x^F_n$ (and hence the mass spectrum of the KK tower) are given by the solutions to the equation,
\begin{align}\label{FlatEigenvalues}
    \tan(\pi x^F_n) = \frac{(a_F^2-(x^F_n)^2\tau)}{x^F_n (1+a_F^2 \tau)}.
\end{align}
Given the results of Eqs.(\ref{Flatvn}) and (\ref{FlatEigenvalues}), we can now examine the behavior of a number of phenomenologically relevant quantities. To begin, it is useful to get a feel for the numerics of $x^F_1$, the lowest-lying root of the mass eigenvalue equation Eq.(\ref{FlatEigenvalues}). Since we are free to choose $m_1$ within the $\sim 0.1-1$ GeV mass range of interest, the lowest root $x^F_1 = m_1 R$ tells us the value of the compactification radius $R$ within this setup, hence, the value of $x^F_1(a_F,\tau)$ is important to consider. In I, where boundary conditions were used to break $U(1)_D$, the parameter $a_F$ is, of course, absent. However, it was found that $x^F_1$ in that case was a decreasing function of $\tau$, as is typical for the effect of brane-localized kinetic terms (BLKTs), with $x^F_1(\tau=0)=1/2$. Here, on the other hand, it is the value of $a_F\neq 0$ that generates a mass for the lowest lying dark photon KK state so that we expect $x^F_1\to 0$ as $a_F\to 0$ and thus to grow with increasing $a_F$. The top and bottom panels of Fig.~\ref{fig1} show that, indeed, the values of $x^F_1$ follow this anticipated behavior. For a fixed value of $a_F$, $x^F_1$ decreases as $\tau$ increases and for a fixed value of $\tau$, $x^F_1$ increases with the value of $a_F$. 

\begin{figure}[htbp]
\centerline{\includegraphics[width=5.0in,angle=0]{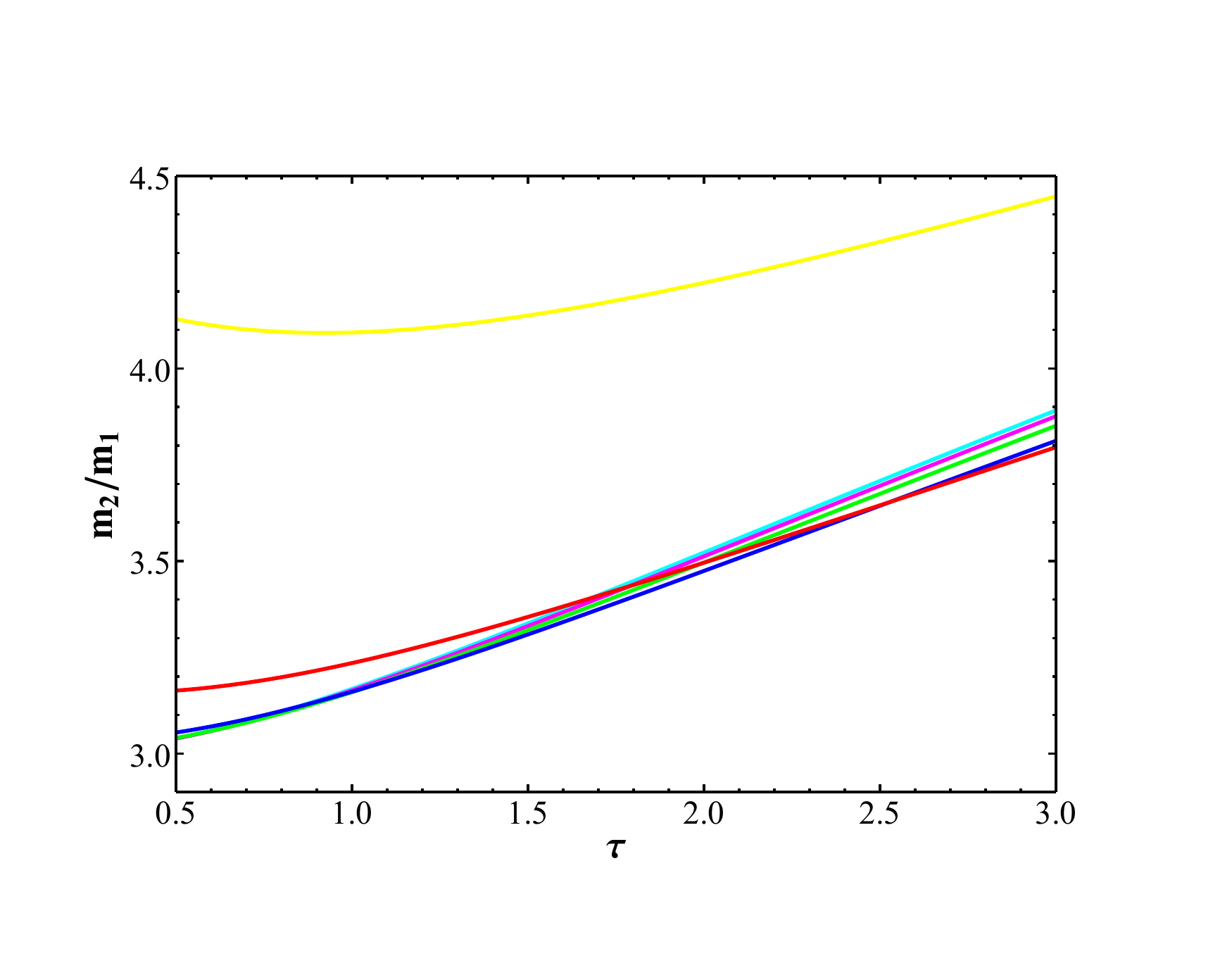}}
\vspace*{-2.0cm}
\centerline{\includegraphics[width=5.0in,angle=0]{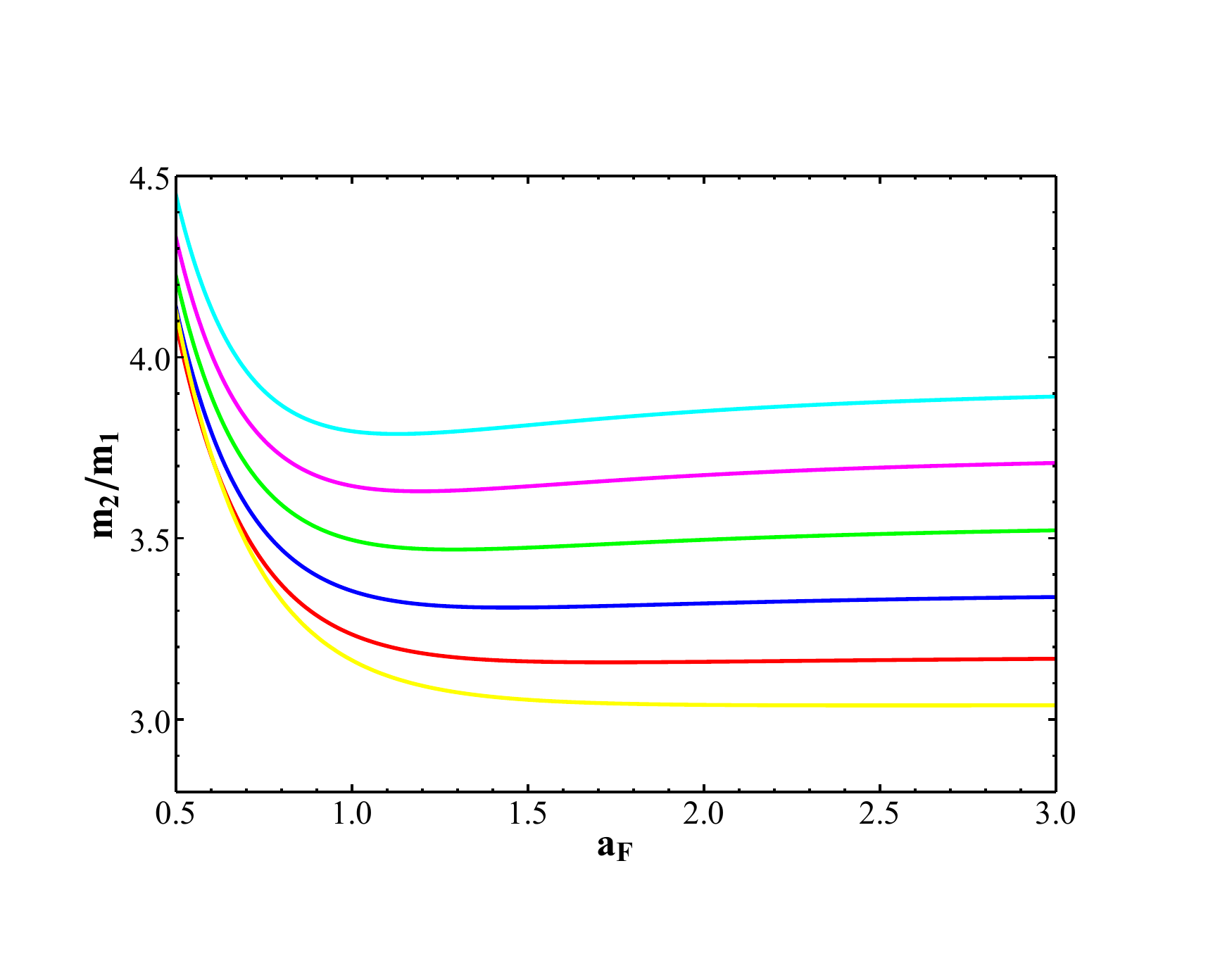}}
\vspace*{-1.0cm}
\caption{(Top) The mass ratio of the lowest two dark photon KK states, $m_{2}/m_{1}=x^F_2/x^F_1$, as a function of $\tau$ for $a_F=$3(cyan), 5/2(magenta), 2(green), 3/2(blue), 1(red), and 1/2(yellow), 
respectively. (Bottom) As in the previous panel, but now as a function of $a_F$ assuming $\tau=$3(cyan), 5/2(magenta), 2(green), 3/2(blue), 1(red), and 1/2(yellow), respectively.}
\label{fig2}
\end{figure}

\begin{figure}[htbp]
\centerline{\includegraphics[width=5.0in,angle=0]{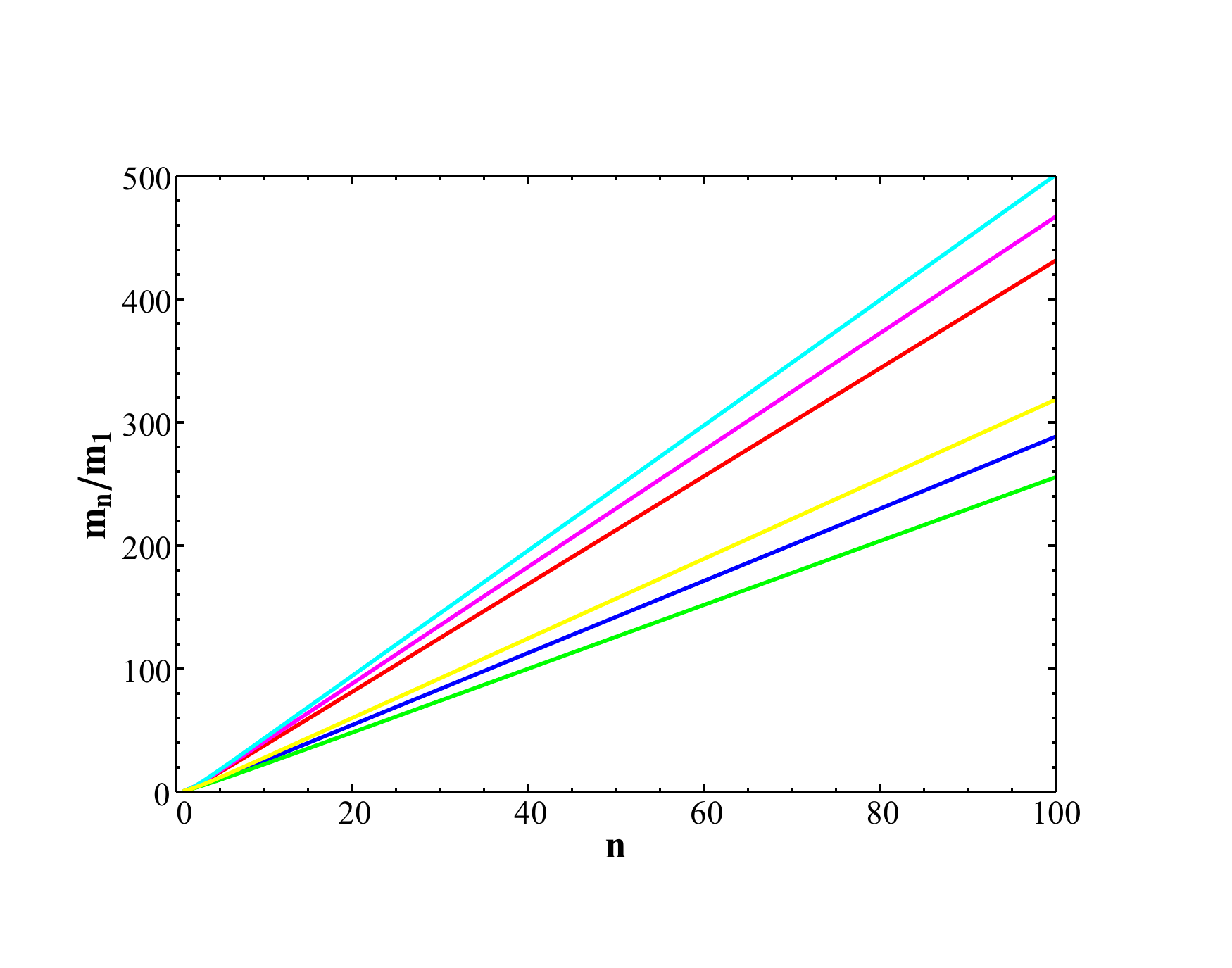}}
\vspace*{-1.30cm}
\caption{Approximate linear growth of the relative dark photon KK mass ratio $m_n/m_1$ as a function of $n$ for various choices of $(\tau ,a_F)$ =(1/2,1/2) [red], (1/2,1) [blue], (1/2,3/2) [green], 
(1,1/2) [magenta], (3/2,1/2) [cyan] and (1,1) [yellow], respectively.}
\label{fig3}
\end{figure}

Beyond the position of the lowest-lying root of Eq.(\ref{FlatEigenvalues}), the particular spectrum of the more massive KK modes are obviously of significant interest. A clear phenomenological signal for the types of models we are considering is the experimental observation of the dark photon KK excitations, perhaps most importantly that of the second dark photon KK excitation. Hence, knowing where the `next' state beyond the lowest lying member of the KK tower may lie is of a great deal of importance, \ie, where do we look for the dark photon KK excitations if the lowest KK state is discovered? In Fig.~\ref{fig2} we display the ratio $m_{2}/m_{1}=x^F_2/x^F_1$ as a functions of $a_F,\tau$ and we see that for a reasonable variation of these parameters this mass ratio lies in the range $3-4$.  Note that for fixed $a_F$ this ratio increases with increasing $\tau$ (mostly since since $x^F_1$ is pushed lower). Meanwhile, for any fixed value of $\tau$, this ratio sharply declines with increasing $a_F$ in the region $a_F \lesssim 1$ (largely because $x^F_1$ itself decreases sharply in this regime), while for $a_F \gsim 1$ the ratio slowly increases with increasing $a_F$.
Non-zero values of $a_F,\tau$ particularly influence the low mass end of the dark photon KK mass spectrum as, \eg, $a_F\neq 0$ provides the mass for the lightest KK mode in the present case. 
However, beyond the first few KK levels the masses of the dark photon KK states, in particular the ratio $m_{n}/m_{1}$ grows roughly linearly with increasing $n$ with a slope that is 
dependent on the values of the parameters $a_F,\tau$ as is shown in Fig.~\ref{fig3}. It is actually straightforward to see the eventual linear trend of the lines in Fig.~\ref{fig3} analytically, using the root equation Eq.(\ref{FlatEigenvalues}). In particular, note that as $x^F_n \rightarrow \infty$, Eq.(\ref{FlatEigenvalues}) approaches
\begin{align}
    \textrm{tanc}(\pi x^F_n) = -\frac{\tau}{\pi (1+a^2_F \tau)},
\end{align}
where $\textrm{tanc}(z) \equiv \textrm{tan}(z)/z$. It is well known that the difference between consecutive solutions of $\textrm{tanc}(z) = C$, for some constant $C$, approaches $\pi$ for very large $z$. So, we can see that for high-mass KK modes, the difference between consecutive solutions of Eq.(\ref{FlatEigenvalues}) will approach 1. Hence, the slope of the lines in Fig.~\ref{fig3} can be easily approximated as $\sim (x^F_1)^{-1}$, and will therefore exhibit the inverse of the dependence of $x^F_1$ on the parameters $\tau$ and $a_F$, which we have already observed in Fig.~\ref{fig1}. In addition, we can note that without taking the ratio of $x^F_n$ to $x^F_1$, \emph{any} large-$n$ solution of Eq.(\ref{FlatEigenvalues}) eventually follows the pattern $x^F_n \approx n$.

The next quantities of phenomenological relevance are the relative values of the KM parameters, $\epsilon_n/\epsilon_1$, and the couplings of the dark photon KK tower states to DM, 
$g_{DM}^n/g_D$; note that these latter quantities are found to oscillate in sign.
Before exploring the numerics in detail here, it is useful to note that one can get a feel for the behavior of these ratios by purely analytical methods. In particular, by invoking Eqs.(\ref{gepsilonDefs}),(\ref{Flatvn}), and (\ref{FlatEigenvalues}), it is possible to derive the expressions
\begin{align}\label{Flatgepsilon}
    \bigg( \frac{\epsilon_n}{\epsilon_1} \bigg)^2 &= \bigg(\frac{a_F^4+(x^F_n)^2}{a_F^4+(x^F_1)^2}\bigg)\lambda_F, \\
    \bigg( \frac{g^n_{DM}}{g_D} \bigg)^2 &= \bigg( \frac{(1+(x^F_n)^2 \tau^2)(x^F_n)^2}{(1+(x^F_1)^2 \tau^2)(x^F_1)^2} \bigg) \lambda_F, \nonumber \\
    \lambda_F &\equiv \frac{\pi (1+(x^F_1)^2 \tau^2)(a_F^4+(x^F_1)^2)+(1+a_F^2 \tau)(a_F^2 + (x^F_1)^2 \tau)}{\pi (1+(x^F_n)^2 \tau^2)(a_F^4+(x^F_n)^2)+(1+a_F^2 \tau)(a_F^2 + (x^F_n)^2 \tau)}. \nonumber
\end{align}
From Eq.(\ref{Flatgepsilon}), we can readily take the limits of $(\epsilon_n/\epsilon_1)^2$ and $(g^n_{DM}/g_D)^2$ at large $n$ (and hence large $x^F_n \approx n$). We arrive at the result that as $n\rightarrow \infty$
\begin{align}\label{FlatgepsilonAsymptotic}
    \bigg( \frac{\epsilon_n}{\epsilon_1} \bigg)^2 &\rightarrow \frac{\pi (1+(x^F_1)^2 \tau^2)(a_F^4+(x^F_1)^2)+(1+a_F^2 \tau)(a_F^2+(x^F_1)^2 \tau)}{\pi \tau^2(a_F^4+(x^F_1)^2)}\frac{1}{n^2}, \\
    \bigg( \frac{g^n_{DM}}{g_D} \bigg)^2 &\rightarrow \frac{\pi (1+(x^F_1)^2 \tau^2)(a_F^4+(x^F_1)^2)+(1+a_F^2 \tau)(a_F^2+(x^F_1)^2\tau)}{\pi (1+(x^F_1)^2 \tau^2)(x^F_1)^2}. \nonumber
\end{align}
From the first expression in Eq.(\ref{FlatgepsilonAsymptotic}), we see that the ratio $(\epsilon_n/\epsilon_1)$ falls roughly as $1/n$ for large $n$; this result is readily borne out numerically in the top panel of Fig.~\ref{fig5}, where we also see that even for small $n$, $\epsilon_n$ never significantly exceeds the value of $\epsilon_1$, offering encouraging evidence that the small-kinetic mixing limit we took in Section \ref{Setup} was valid. More rigorously demonstrating this validity, however, will require the use of sum identities we shall derive later in this section.

\begin{figure}[htbp] 
\centerline{\includegraphics[width=5.0in,angle=0]{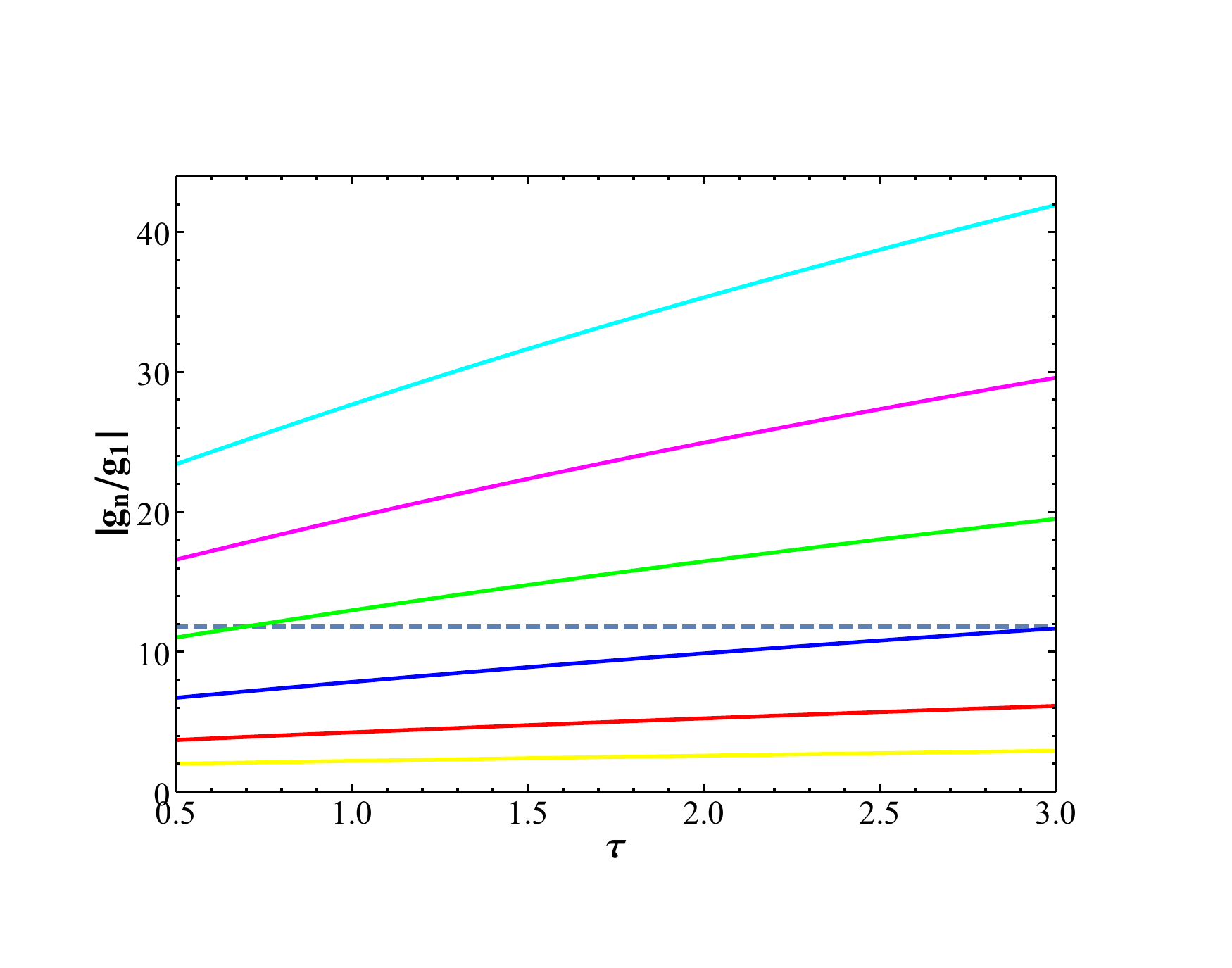}}
\vspace*{-2.0cm}
\centerline{\includegraphics[width=5.0in,angle=0]{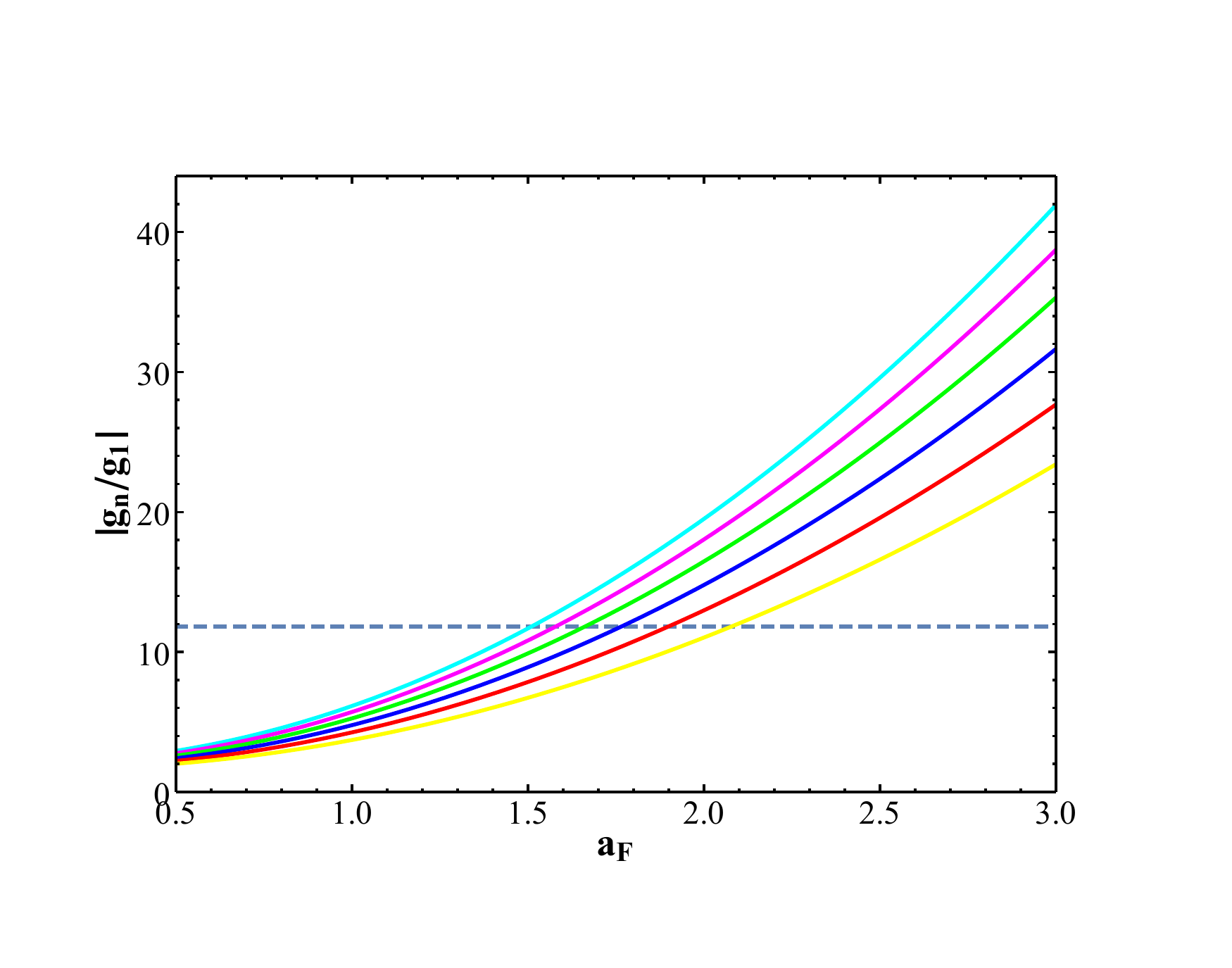}}
\vspace*{-1.0cm}
\caption{(Top) The limit of the ratio $|g^n_{DM}/g_D|$ as $n\rightarrow \infty$, as a function of $\tau$ for $a_F=$3(cyan), 5/2(magenta), 2(green), 3/2(blue), 1(red), and 1/2(yellow), 
respectively. The dashed line denotes the largest possible ratio such that the couplings of the DM particle to the gauge boson KK modes remain perturbative for all KK modes in the theory, assuming $g_D=0.3$ (Bottom) As in the previous panel, but now as a function of $a_F$ assuming $\tau=$3(cyan), 5/2(magenta), 2(green), 3/2(blue), 1(red), and 1/2(yellow), respectively.}
\label{fig4}
\end{figure}

\begin{figure}[htbp] 
\centerline{\includegraphics[width=5.0in,angle=0]{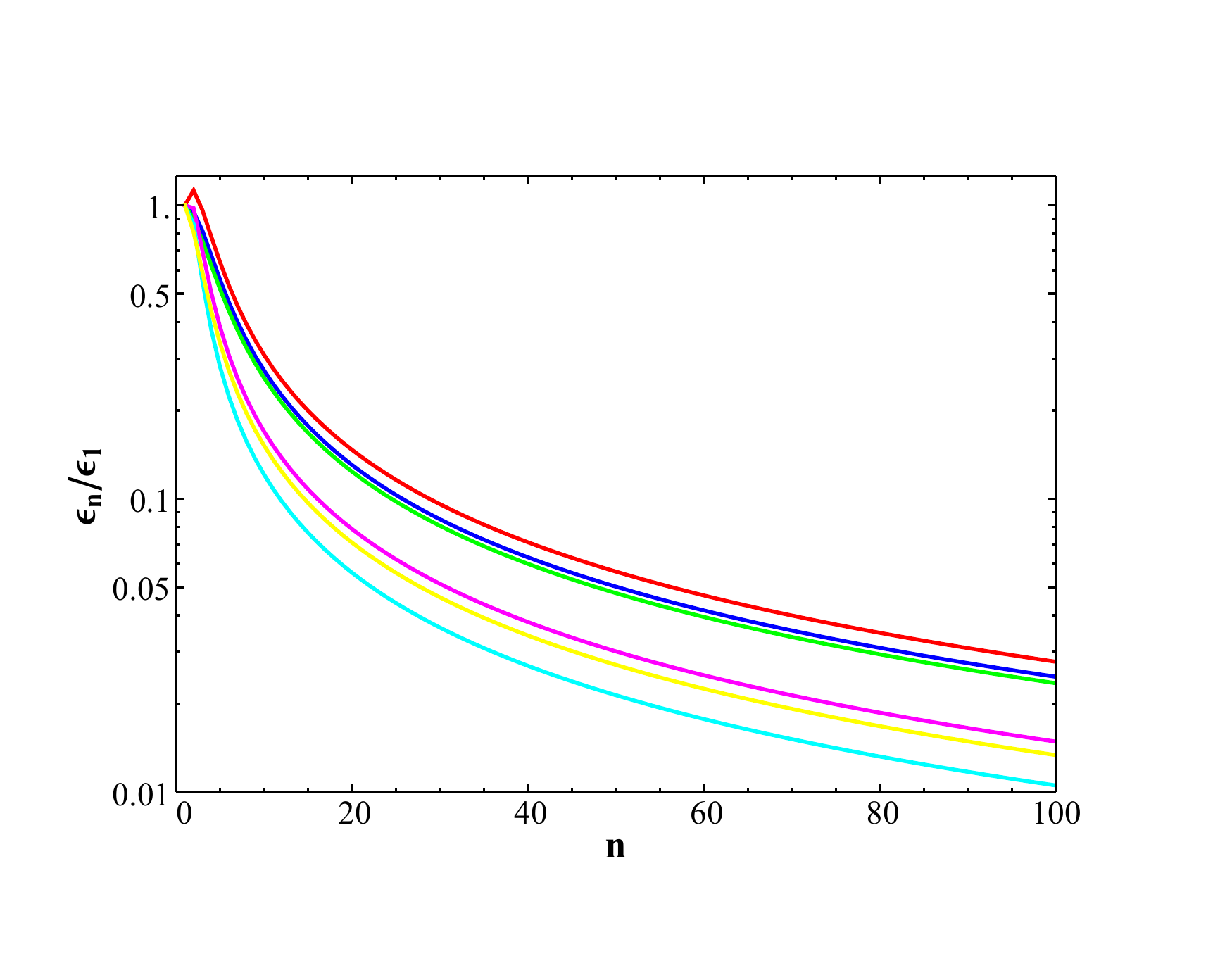}}
\vspace*{-2.0cm}
\centerline{\includegraphics[width=5.0in,angle=0]{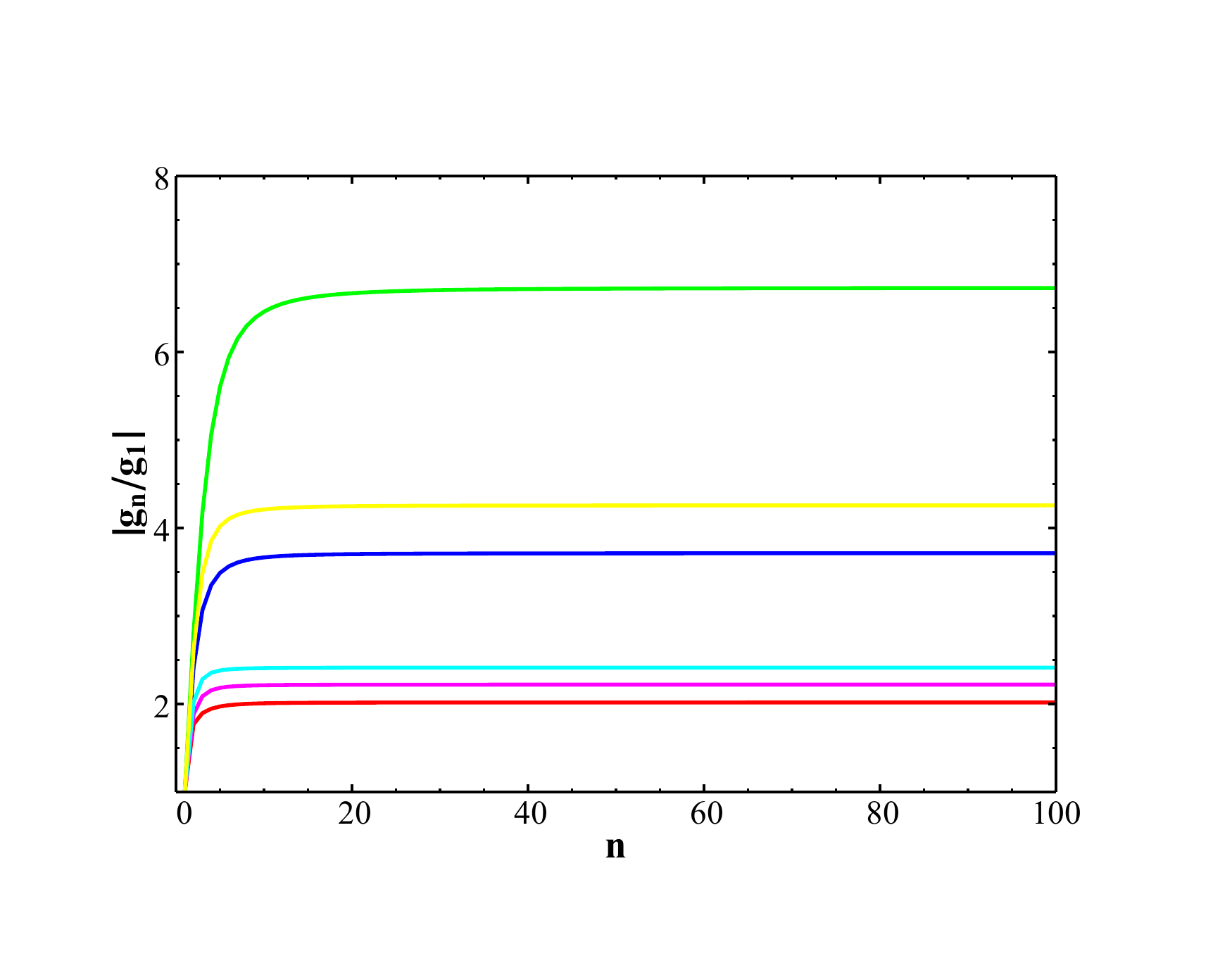}}
\vspace*{-1.0cm}
\caption{(Top) The ratio $\epsilon_n/\epsilon_1$ as a function of $n$ for various choices of $(\tau,a_F)$=(1/2,1/2)[red], (1/2,1)[blue], (1/2,3/2)[green], (1,1/2)[magenta], (3/2,1/2)[cyan], and (1,1)[yellow], respectively.
(Bottom) Same as the top panel but now for the absolute value of the strength of the $n^{\rm th}$ KK coupling of the dark photon to DM in units of $g_D$. Note that this quantity alternates in sign.}
\label{fig5}
\end{figure}

In contrast to the behavior of the effective kinetic mixing terms $\epsilon_n/\epsilon_1$, the ratio $|g^n_{DM}/g_D|$ approaches a constant non-zero value as $n \rightarrow \infty$. The precise value of this asymptotic limit of the ratio $|g^n_{DM}/g_D|$ is naturally of quite significant phenomenological interest: If $|g_{DM}^n/g_{DM}|$ is large, one might be concerned that even for a reasonable value of $g_D \lesssim 1$, the DM particle may experience some non-perturbative couplings to the various KK modes.\footnote{We also note that a large $|g_{DM}^n/g_{DM}|$ may raise concerns about non-convergence of various sums over all KK modes, such as those that appear in Eqs.(\ref{sigmae}) and (\ref{sigmavrel}), however, as we shall see later in this section, these sums remain well-defined.} In Fig.~\ref{fig4}, we explore the $\tau$ and $a_F$ dependence of this asymptotic coupling limit numerically. Notably, we find that the coupling ratio increases sharply as $a_F$ increases. For comparison's sake, in both panels of Fig.~\ref{fig4}, we have depicted as a dashed line the \emph{maximum} $|g^n_{DM}/g_D|$ that would be allowed such that all couplings would remain perturbative (that is, have a structure constant $(g^n_{DM})^2/(4 \pi) < 1$) given a choice of $g_D =0.3$, that is, assuming that the coupling of DM to the first KK mode of the dark photon field has approximately the same coupling constant as the electroweak force. In the figure then, we see that such a choice of $g_D$ is only permitted when $a_F \lsim 3/2$; much larger and the DM interactions with large-$n$ KK modes become strongly coupled. In both Figs.~\ref{fig4} and \ref{fig5}, however, we see that limiting our choice of $a_F$ to $a_F \lsim 3/2$ leads to substantially more modest asymptotic values of $|g^n_{DM}/g_D|$, of $\lsim 10$. Because $\lvert g^n_{DM}/g_D \rvert$ rises quadratically (or more accurately, the square of this ratio rises quartically) with increasing $a_F$, these conditions would be 
only slightly less restrictive if a somewhat smaller value of $g_D$, \eg, $g_D=0.1$ were chosen.

To continue our discussion of the phenomenology of our construction, we must now also find the sum $F(y,y',s)$, which we remind the reader is defined in Eq.(\ref{FDefinition}), for the flat space case, which we can accomplish by inserting $f(y)=1$ into Eq.(\ref{FDiffEq}), yielding
\begin{align}\label{FDiffEqFlat}
    \partial_y^2 F(y,y',s) &= R \delta(y-y')- s F(y,y',s), \nonumber\\
    \partial_y F(y,y',s)|_{y=0} &= -s \tau R F(0,y',s), \\
    \partial_y F(y,y',s)|_{y=\pi R} &= - m_V^2 R F(\pi R, y',s), \nonumber
\end{align}
from which the solution
\begin{align}\label{FlatFSolution}
    F(y,y',s) &= R^2 \frac{[\cos(\sqrt{s}y_<)-R \sqrt{s} \sin(\sqrt{s} y_<)][\sqrt{s} R \cos(\sqrt{s}(y_>-\pi R))-a_F^2 \sin(\sqrt{s}(y_> - \pi R))]}{R \sqrt{s}(-a_F^2 +s R^2 \tau)\cos(\pi R \sqrt{s})+s R^2 (1+a_F^2 \tau)\sin(\pi R \sqrt{s})}, \\
    y_> &\equiv \textrm{max}(y,y'), \;\;\; y_< \equiv \textrm{min}(y,y') \nonumber
\end{align}
can be straightforwardly derived. We see that, as expected, the sum $F(y,y',s)$ has poles whenever $s=m_n^2$, as can be seen from the mass eigenvalue condition Eq.(\ref{FlatEigenvalues}); in other words, our sum of propagators possesses poles exactly where the individual propagators have poles. Additionally, equipped with this sum, it is possible to derive in closed form the sum $\sum_n \epsilon_n^2/\epsilon_1^2$, which we recall from I and Section \ref{Setup} must be $\lsim 10$ in order for our assumption of small kinetic mixing (KM) to be valid. Taking the limit of $F(y,y',s)$ as $s\rightarrow 0$, we arrive at the result
\begin{align}
    -F(y,y',0) = \sum_n \frac{v_n(y)v_n(y')}{m_n^2} = R^2 \bigg( \frac{1}{a_F^2}+\pi \bigg) - \theta(y-y') R y-\theta(y'-y) R y'.
\end{align}
Differentiating this sum with respect to $y$ at $y=0$ and applying the SM-brane boundary condition given in Eq.(\ref{FlatEOM}), we rapidly arrive at
\begin{align}\label{FlatepsilonSum}
    \sum_n v_n (0)^2 &= \frac{1}{\tau} \\
    \rightarrow \sum_n \frac{\epsilon_n^2}{\epsilon_1^2} &= \sum_n \frac{v_n (0)^2}{v_1(0)^2} = \frac{1}{\tau v_1(0)^2}. \nonumber
\end{align}
The form of the sum in the second line of Eq.(\ref{FlatepsilonSum}) already then confirms what has previously been observed in I, namely, that a nontrivial positive BLKT is necessary for the consistency of our KM analysis. The sum sharply increases to infinity as $\tau\rightarrow 0$, indicating that an insufficiently large $\tau$ will result in the sum being unacceptably large, namely $\gsim O(10)$. Furthermore, a \emph{negative} $\tau$ would suggest a still more worrying scenario, indicating the need for at least one KK state to be ghost-like (have a negative norm squared).
\begin{figure}[htbp] 
\centerline{\includegraphics[width=5.0in,angle=0]{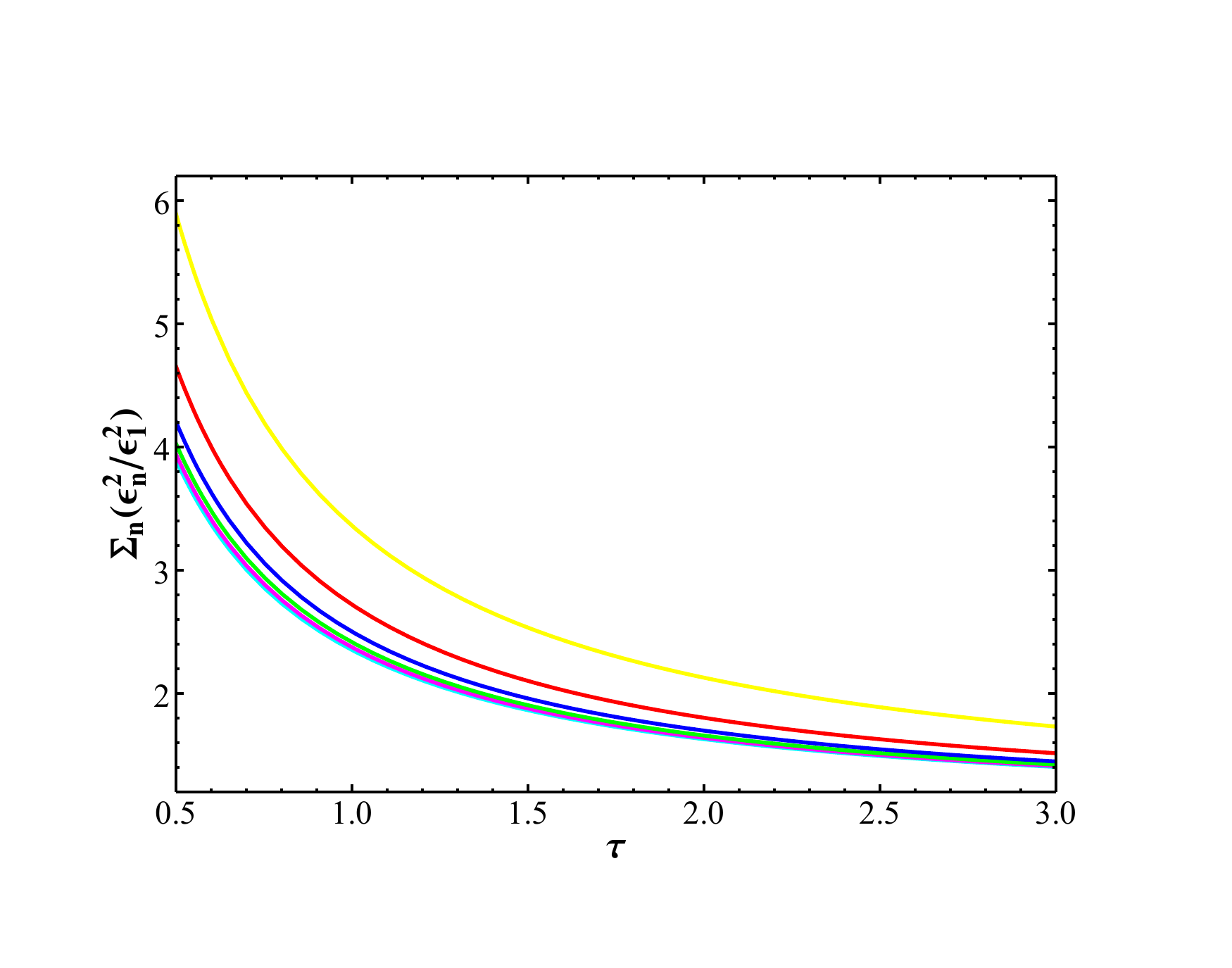}}
\vspace*{-2.0cm}
\centerline{\includegraphics[width=5.0in,angle=0]{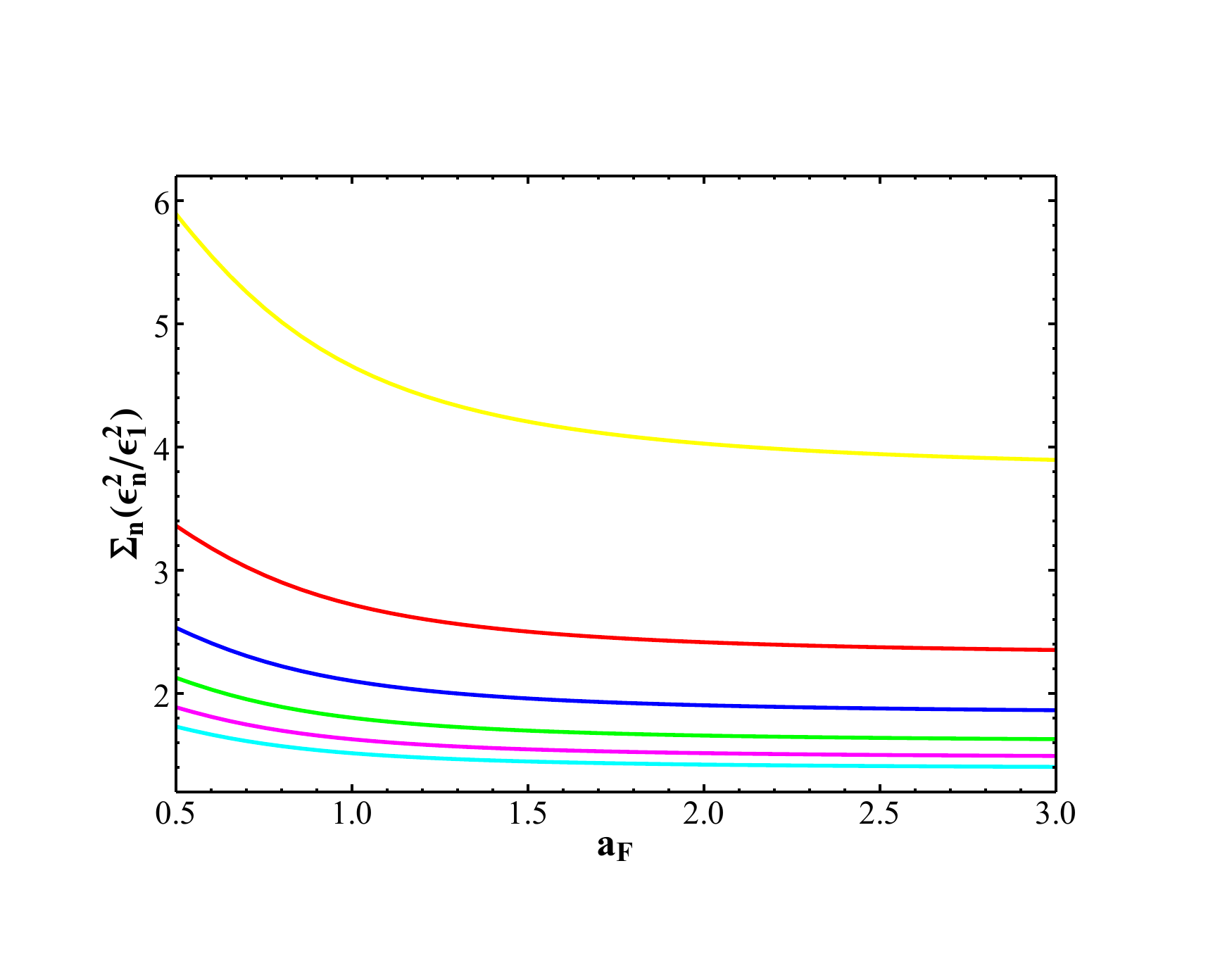}}
\vspace*{-1.0cm}
\caption{(Top) The sum $\sum_n \epsilon_n^2/\epsilon_1^2$ over all $n$, as a function of $\tau$ for $a_F=$3(cyan), 5/2(magenta), 2(green), 3/2(blue), 1(red), and 1/2(yellow), 
respectively. (Bottom) As in the previous panel, but now as a function of $a_F$ assuming $\tau=$3(cyan), 5/2(magenta), 2(green), 3/2(blue), 1(red), and 1/2(yellow), respectively.}
\label{figFlatepsilonSum}
\end{figure}
To determine if our kinetic mixing treatment is valid for the full parameter space we consider, we depict the sum $\sum_n \epsilon_n^2/\epsilon_1^2$ in Fig.~\ref{figFlatepsilonSum}. Our results here explicitly confirm those observed in I, namely, that for selections of $(\tau, a_F)$ such that $\tau \gsim 1/2$, the summation $\sum_n \epsilon_n^2/\epsilon_1^2$ remains small enough not to vitiate our treatment of kinetic mixing: The sum remains $\lsim O(10)$.

\begin{figure}[htbp]
\centerline{\includegraphics[width=5.0in,angle=0]{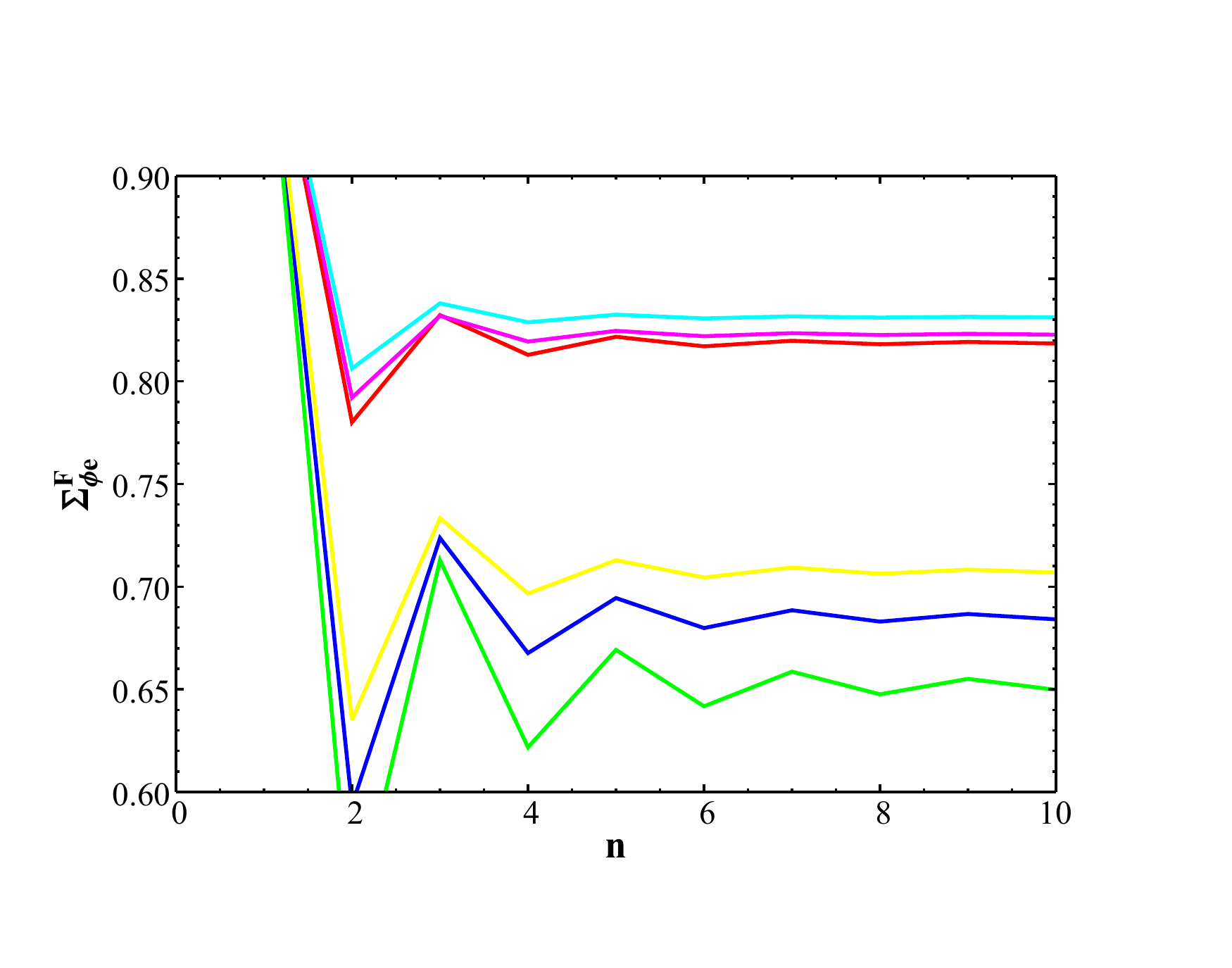}}
\vspace*{-1.0cm}
\caption{The explicit KK sum form of $\Sigma^F_{\phi e}$ defined in Eq.(\ref{FlatsigmaeNum}), which encapsulates the dependence of the DM-electron scattering cross section on parameters of the model of the extra dimension, where only terms coming from the first $n$ KK modes are included in the sum, for the choices of $(\tau ,a_F)$ =(1/2,1/2) [red], (1/2,1) [blue], (1/2,3/2) [green], (1,1/2) [magenta], (3/2,1/2) [cyan] and (1,1) [yellow], respectively.}
\label{fig6}
\end{figure}

Next, we apply the results of Eqs.(\ref{Flatvn}), (\ref{FlatEigenvalues}), and (\ref{FlatFSolution}) to find the DM-$e^-$ scattering cross section, to explore the possibility of direct detection of the DM. Inserting Eq.(\ref{FlatFSolution}) into Eq.(\ref{sigmae}) yields
\begin{align}\label{Flatsigmae}
    \sigma_{\phi e} &= \frac{4 \alpha_{em} m_e^2 (g_D \epsilon_1)^2}{v_1 (0)^2 v_1 (\pi R)^2}\frac{R^4}{a_F^4}\\
    &=\frac{4 \alpha_{em} m_e^2 (g_D \epsilon_1)^2}{v_1 (0)^2 v_1 (\pi R)^2}\frac{(x^F_1)^4}{m_1^4 a_F^4}, \nonumber
\end{align}
where in the second line we have substituted the parameter $m_1$, the mass of the lowest-lying KK mode of the dark photon field, for the compactification radius $R$. We can now suggestively rewrite this expression as
\begin{align}\label{FlatsigmaeNum}
    \sigma_{\phi e} &= (2.97 \times 10^{-40} \; \textrm{cm}^2)\bigg( \frac{100 \; \textrm{MeV}}{m_1}\bigg)^4 \bigg( \frac{g_D \epsilon_1}{10^{-4}} \bigg)^2 \Sigma^F_{\phi e}, \\
    \Sigma^F_{\phi e} &\equiv \frac{(x^F_1)^4}{v_1(0)^2 v_1(\pi R)^2 a_F^4} = \bigg\lvert \sum_{n=0}^\infty \frac{(x^F_1)^2 v_n(0) v_n(\pi R)}{(x^F_n)^2 v_1(0) v_1(\pi R)} \bigg\rvert^2. \nonumber
\end{align}
Note here that the quantity $\Sigma^F_{\phi e}$ depends \emph{only} on the model parameters $(\tau, a_F)$, while the rest of the expression above is independent of them. While the closed form of $\Sigma^F_{\phi e}$ is convenient for calculation, we have also included an explicit expression for this quantity in terms of an infinite sum over KK modes -- notably, because the quantity $g^n_{DM} \epsilon_n$ (or alternatively, $v_n(\pi R) v_n(0)$) alternates in sign and decreases sharply with increasing $n$, we can see in Fig.~\ref{fig6} that the sum rapidly converges, coming within $O(10^{-2})$ corrections to the value of the closed form of $\Sigma^F_{\phi e}$ even when the sum is truncated at $n=10$. 

Looking at the numerical coefficient of $\Sigma^F_{\phi e}$ in Eq.(\ref{FlatsigmaeNum}), meanwhile, we see that for $m_1 \sim O(100 \; \textrm{MeV})$ and $g_D \epsilon_1 \sim 10^{-4}$, the DM-$e^-$ scattering cross section easily avoids current direct detection constraints as long as the quantity $\Sigma^F_{\phi e} \leq 1$ \cite{Essig:2017kqs,Essig:2015cda,TTYu,Aprile:2019xxb,Agnes:2018oej}, although it does lie within the possible reach of future experiments such as SuperCDMS \cite{Essig:2015cda}. Anticipating that $m_{DM} \approx m_1/2$ (which we shall shortly see is necessary in order to enjoy the resonant enhancement of the annihilation cross section we require to recreate the relic density), we note that if we assume $g_D \epsilon_1 =10^{-4}$, the quantity $\sigma_{\phi e}/\Sigma^F_{\phi e}$ (that is, the direct detection cross section divided by the variable which parameterizes the parameters related to the geometry of the extra dimension) is at least an order of magnitude below the most stringent boundaries of \cite{Essig:2017kqs,Essig:2015cda,TTYu,Aprile:2019xxb,Agnes:2018oej} for any $m_1 \gsim O(\textrm{a few}) \; \textrm{MeV}$. So, our sole remaining task to demonstrate that this model escapes direct detection bounds is to demonstrate that $\Sigma^F_{\phi e} \leq O(1)$.

\begin{figure}[htbp]
\centerline{\includegraphics[width=5.0in,angle=0]{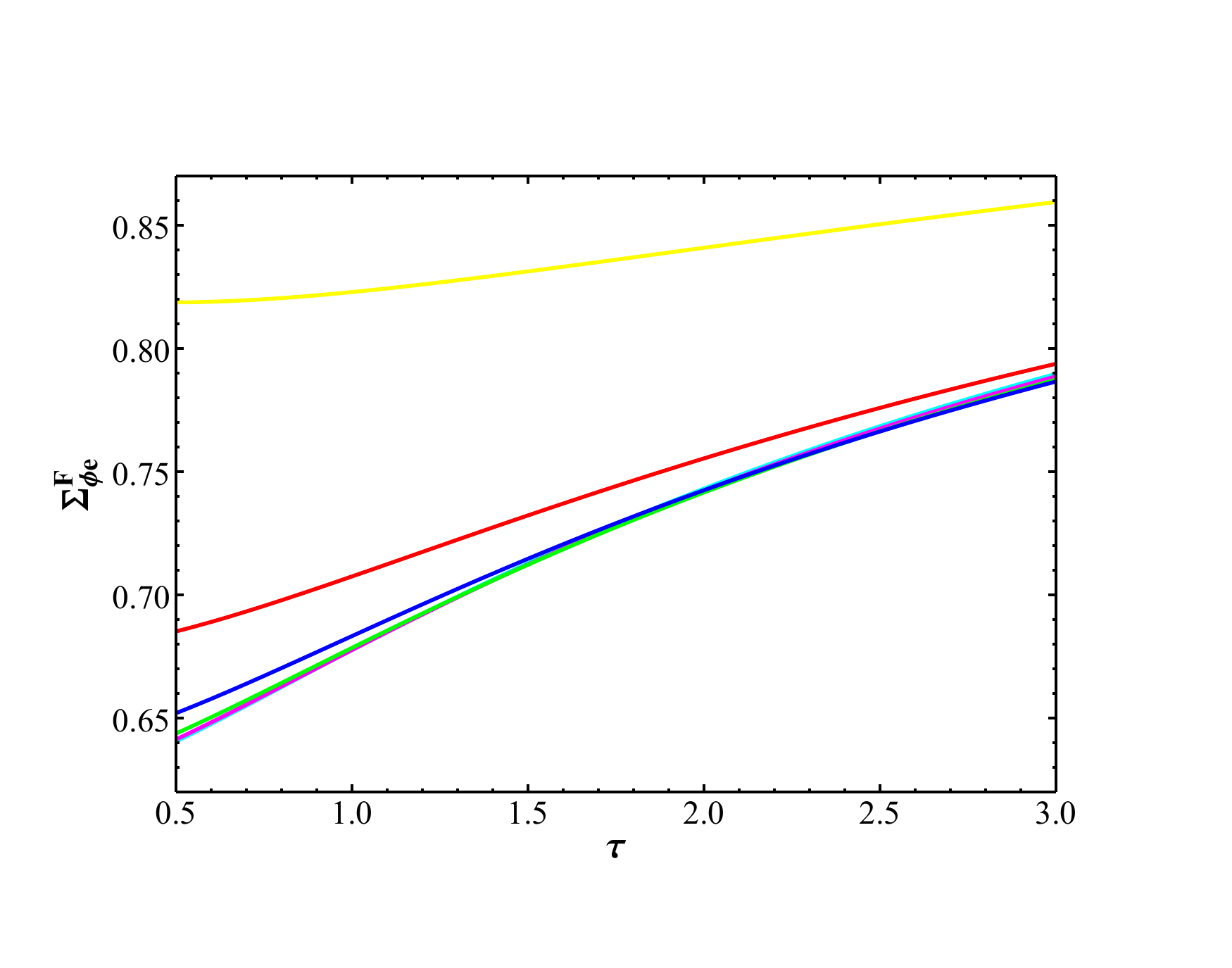}}
\vspace*{-2.0cm}
\centerline{\includegraphics[width=5.0in,angle=0]{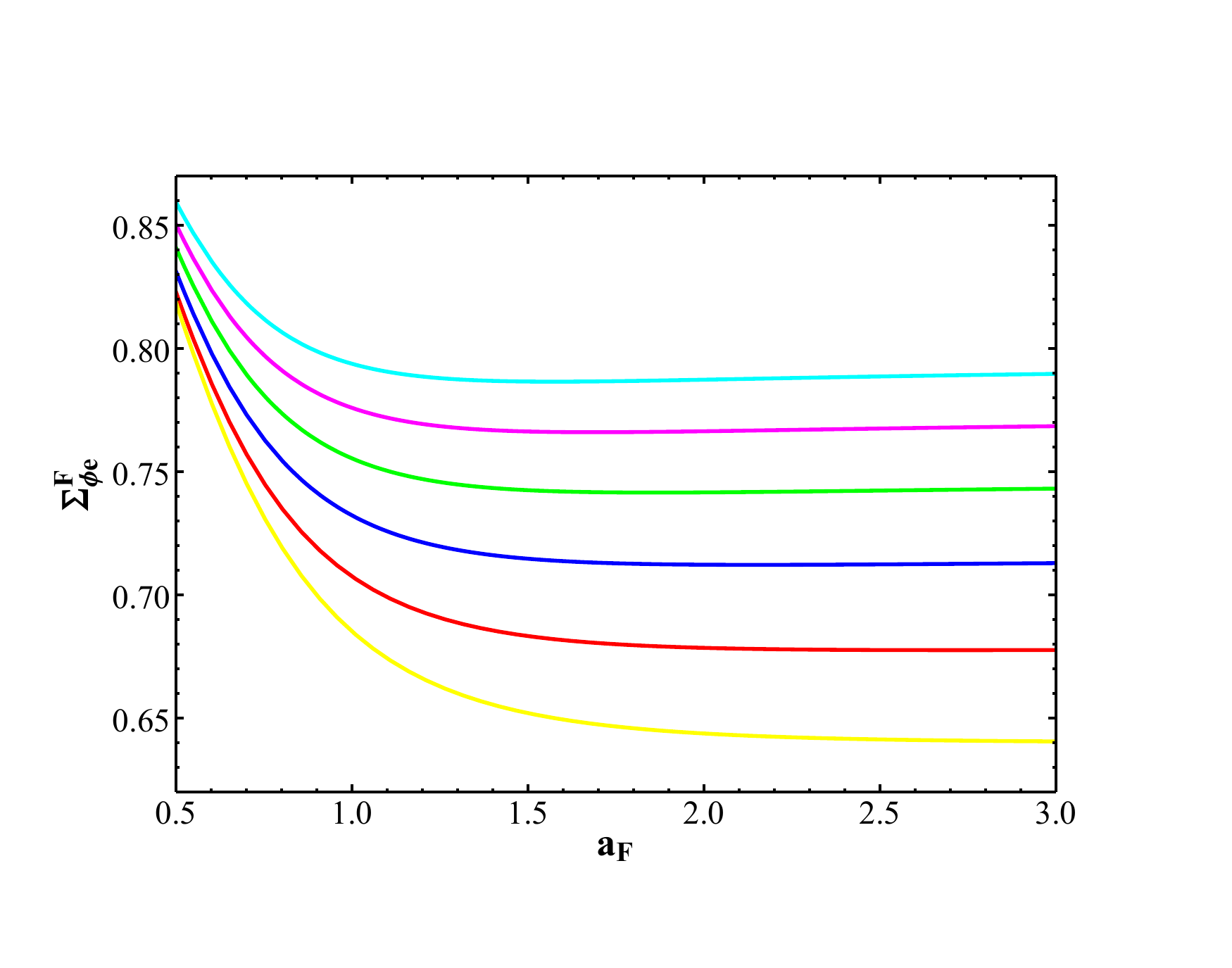}}
\vspace*{-1.0cm}
\caption{(Top) The sum $\Sigma^F_{\phi e}$ defined in Eq.(\ref{FlatsigmaeNum}), which encapsulates the dependence of the DM-electron scattering cross section on parameters of the model of the extra dimension, as a function of $\tau$ for $a_F=$3(cyan), 5/2(magenta), 2(green), 3/2(blue), 1(red), and 1/2(yellow), 
respectively. (Bottom) As in the previous panel, but now as a function of $a_F$ assuming $\tau=$3(cyan), 5/2(magenta), 2(green), 3/2(blue), 1(red), and 1/2(yellow), respectively.}
\label{fig7}
\end{figure}

We can see that $\Sigma^F_{\phi e}$ does in fact stay below $O(1)$ for a broad range of parameters in Fig.~\ref{fig7}; for every choice of $(\tau,a_F)$ that we are considering here, $\Sigma^F_{\phi e}$ lies between 0.6 and 0.9 implying that the 
KK states lying above the lightest one do not make critical contributions to this cross section. 
Hence, this model can easily evade present DM direct detection constraints for reasonable choices of $m_1 \sim 100 \; \textrm{MeV}$ and $g_D \epsilon_1 \sim 10^{-4}$.

\begin{figure}[htbp]
\centerline{\includegraphics[width=5.0in,angle=0]{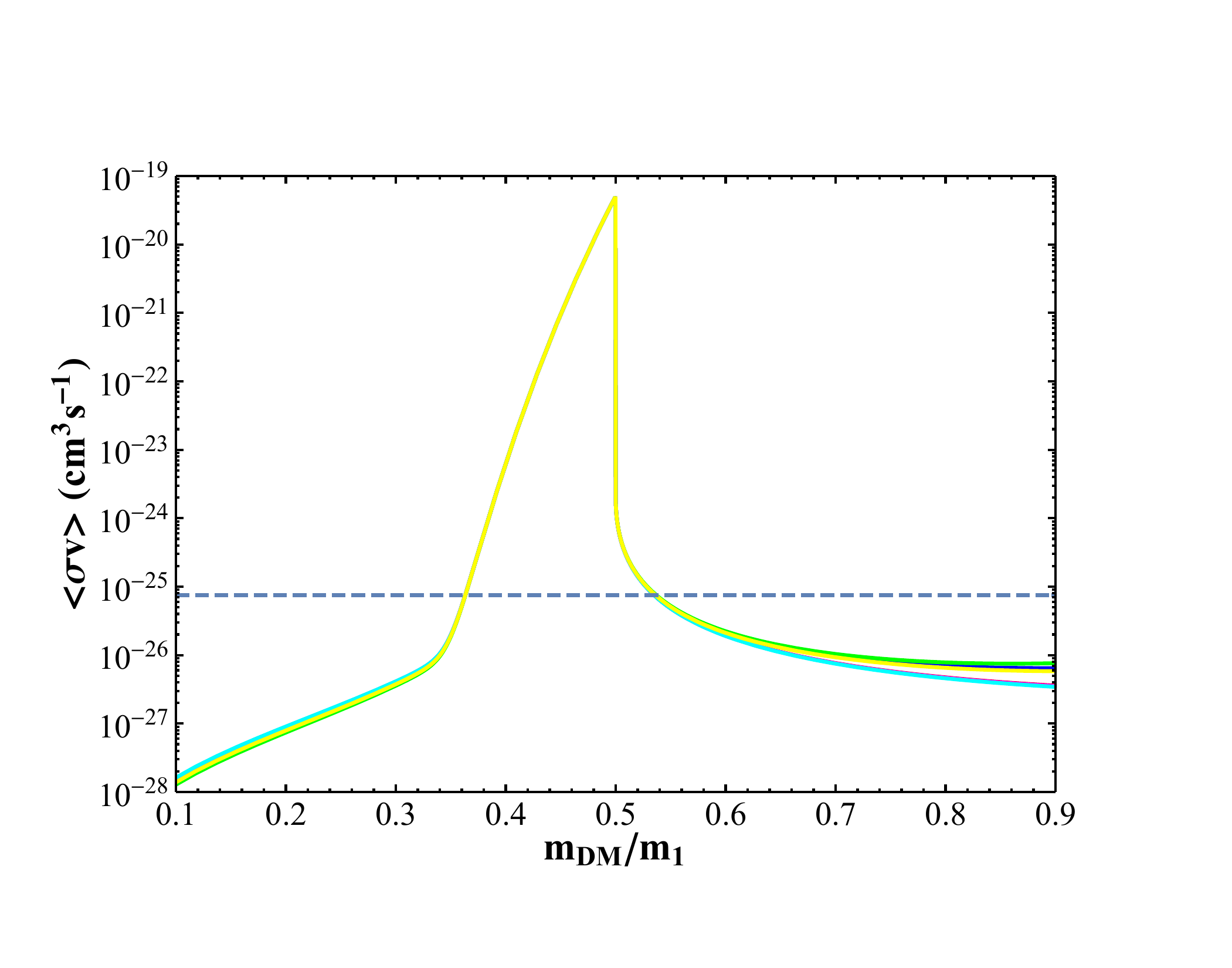}}
\vspace*{-1.0cm}
\caption{The thermally averaged, velocity-weighted cross section in $\textrm{cm}^3/\textrm{s}$ for the annihilation process $\phi^\dagger \phi \rightarrow f \bar{f}$, where the final-state fermions $f$ are electrons, for the choices of $(\tau ,a_F)$ =(1/2,1/2) [red], (1/2,1) [blue], (1/2,3/2) [green], (1,1/2) [magenta], (3/2,1/2) [cyan] and (1,1) [yellow], respectively. The dashed line denotes the value for this cross section necessary to produce the observed relic abundance of DM after freeze out.}
\label{fig8}
\end{figure}

Our brief phenomenological survey of the flat space scenario now concludes with a discussion of the thermally averaged annihilation cross section at freeze-out, that is, demonstrating that this construction is capable of producing the correct relic density of DM in the universe. To begin, we insert Eq.(\ref{FlatFSolution}) into the expression for the $\phi^\dagger \phi \rightarrow f \bar{f}$ (where $f$ is some fermion species) velocity-weighted annihilation cross section of Eq.(\ref{sigmavrel}). This yields the result
\begin{align}\label{Flatsigmavrel}
    \sigma v_{lab} &= \frac{1}{3}\frac{g_D^2 \epsilon_1^2 \alpha_{\textrm{em}}Q_f^2}{v_1(\pi R)^2 v_1 (0)^2 }\frac{(s+2m_f^2)(s-4 m_{DM}^2)\sqrt{s(s-4 m_f^2)}R^4}{s(s-2 m_{DM}^2)}\\
    &\times\Bigg\lvert \frac{1}{R^2}F(0,\pi R, s) - \frac{v_1(\pi R) v_1 (0)}{sR^2-(x^F_1)^2}+\frac{v_1(\pi R) v_1 (0)}{sR^2-(x^F_1)^2+i (x^F_1)R \Gamma_1}\Bigg\rvert^2, \nonumber \\
    \frac{1}{R^2}F(0,\pi R, s) &= \frac{1}{(-a_F^2 + s R^2 \tau)\cos(\pi R \sqrt{s})+R \sqrt{s}(1+a_F^2 \tau)\sin(\pi R \sqrt{s})}. \nonumber
\end{align}
We can then use this expression in the single integral formula for a thermally averaged annihilation cross section given in Eq.(\ref{singleIntegralAvg}), and compare the results to the approximate necessary cross section to reproduce the (complex) DM relic density with a $p-$wave annihilation process, namely $\simeq 7.5 \times 10^{-26} \; \textrm{cm}^3/\textrm{s}$ \cite{Saikawa:2020swg}.\footnote{Note that due to the sub-GeV mass of the DM, the familiar required annihilation cross section of $\sim 3 \times 10^{-26} \; \textrm{cm}^3/\textrm{s}$ is inaccurate, as discussed in \cite{Saikawa:2020swg}.} We note that this quantity is the \emph{only} one in our analysis which has any direct dependence on the mass of the DM, $m_{DM}$, (assuming, as we do, that the DM particle's mass is substantially greater than that of the electron). In fact, because we must rely on resonant enhancement in order to achieve the correct relic density, we see that with all the other parameters fixed our results for the thermally averaged cross section are extremely sensitive to $m_{DM}$ and largely agnostic to differing choices of $(\tau, a_F)$. In Fig.~\ref{fig8}, we depict the thermally averaged velocity-weighted cross section as a function of the DM mass $m_{DM}$, requiring, as we have argued must be the case in Section \ref{Setup}, that $m_{DM} < m_1$. For demonstration purposes, we have selected that $m_1 = 100 \; \textrm{MeV}$, $x_F = m_{DM}/T = 20$, $g_D = 0.3$, $(g_D \epsilon_1) = 10^{-4}$ (where our choices of $m_1$ and $\epsilon_1$ have been informed by the constraints on direct detection), and have included only the possibility of the DM particles annihilating into an $e^+ e^-$ final state.

Notably, the cross sections depicted are largely independent of the choices of ($\tau, a_F$) near values of $m_{DM}/m_1$ that produce the correct relic abundance (that is, relatively near the $m_1$ resonance of the cross section). In fact, for \emph{all} parameter space points we depict here, it is possible to produce the correct cross section when $m_{DM} \sim 0.36 m_1$ or $m_{DM} \sim 0.54 m_1$; however, other values would be required if we also varied $m_1$ or $g_D\epsilon_1$  By leveraging the resonance, therefore, our model is clearly able to reproduce the observed relic abundance for a wide variety of reasonable points in parameter space. We also note that the annihilation cross section here displays an extremely sharp decline when very close to the resonance peak. This is a consequence of the total decay width of the first KK excitation of the dark photon field becoming progressively smaller, as the width of the decay to a pair of DM particles becomes suppressed by a shrinking phase space factor, eventually approaching 0 when $m_{DM}=m_1/2$. In the absence of a kinematically allowed decay to the DM pairs, the decay into an electron-positron pair, which has a width of $\simeq \alpha_{em} \epsilon_1^2 m_1/3$, or $O(10^{-10})m_1$ if $\epsilon_1 \sim 10^{-4}$, becomes the dominant decay channel for the lightest KK mode of the dark photon field; this state is thus extremely narrow under these circumstances.

\section{Warped Space Model Analysis}\label{WarpedAnalysis}

We now consider the possibility that the extra dimension is not flat, but rather has a Randall-Sundrum-like geometry with a curvature scale $k$. In this case, $f(y)$ in the metric of Eq.(\ref{genericMetric}) shall be $e^{-k y}$, but our analysis closely follows that of the flat space scenario. The warped geometry does, however, necessitate additional care in certain aspects of model construction, which we should address before moving forward with our discussion.

First, in the warped space scenario, because $f(y)$ is non-trivial, we need two parameters to describe the metric rather than the single parameter, $R$, that we used in the flat-space analysis. We shall find the most convenient parameters with which to describe our metric are $kR$, the product of the curvature scale and the compactification radius, and the so-called 
``KK mass", $M_{KK} \equiv k \, \textrm{exp}(-k R \pi)$. Second, unlike the flat-space case, our choice to place the SM on the $y=0$ brane and the DM on the $y=\pi R$ brane is no longer arbitrary. Specifically, we note that naturalness suggests that $\sim M_{KK}$ is a natural scale for mass terms localized on the $y= \pi R$ brane, and that the lowest-mass Kaluza-Klein (KK) modes of any bulk fields should also in general be $O(M_{KK})$, while the natural scale for mass terms localized on the $y=0$ brane should be $\sim M_{KK}\textrm{exp}(k R \pi)$, which is exponentially larger \cite{Randall:1999ee,Csaki:2004ay}. In our construction, then, naturalness suggests that we localize the higher-scale physics (the SM, with a scale of roughly $O(250 \; \textrm{GeV})$) on the $y=0$ brane, and the lower-scale DM sector with a scale of $O(0.1-1 \; {\rm GeV})$ localized on the $y=\pi R$ brane. Furthermore, the hierarchy between the two scales roughly sets the value of the product $kR$, namely, we must require that $e^{-k R \pi}\sim O(0.1-1 \; {\rm GeV})/O(250 \; {\rm GeV})$. Thus we will require that $kR \approx 1.5-2$. We note in passing, therefore, that in contrast to the flat space model, the warped space construction offers the aesthetically appealing characteristic of explaining the mild hierarchy between the brane-localized vev of the SM Higgs and the brane-localized mass parameters of the DM and dark photon fields appearing on the opposite brane.

With these concerns addressed, we can now move on to determining the bulk profiles and sums of KK modes required for our analysis. First, we note that the equations of motion for the bulk profile $v_n(y)$ become{\footnote {Here we use the label ``W'' to denote the values relevant for the warped scenario.}} 
\begin{align}
    \partial_y [e^{-2 k y} ~\partial_y v_n(y)] = -m_n^2 v_n(y), \\
    (\partial_y+m_n^2 \tau R)v_n(y)|_{y=0}=0, \nonumber\\
    (e^{-2 k R \pi}~ \partial_y +m_V^2 R)v_n(\pi R)|_{y=\pi R}=0. \nonumber
1\end{align}
The solution to these equations can be written,
\begin{align}\label{Warpedvn}
    &v_n(y) = A_n z^W_n ~\zeta^{(n)}_1 (z^W_n), \\
    &z^W_n \equiv x^W_n e^{k(y-\pi R)}, \;\; x^W_n \equiv \frac{m_n}{M_{KK}}, \;\; \varepsilon^W_n \equiv x^W_n e^{-k R \pi} \nonumber
\end{align}
where $A_n$ is a normalization factor, and the function $\zeta^{(n)}_\nu(z)$ is given by 
\begin{align}
    &\zeta^{(n)}_\nu (z) \equiv \alpha_n J_\nu(z)-\beta_n Y_\nu(z), \\
    &\alpha_n \equiv [(Y_0\big( \varepsilon^W_n\big)+\big( \varepsilon^W_n\big) k R \tau Y_1 \big( \varepsilon^W_n \big) ], \nonumber \\
    &\beta_n \equiv [(J_0\big( \varepsilon^W_n\big)+\big( \varepsilon^W_n\big) k R \tau J_1 \big( \varepsilon^W_n \big) ], \nonumber
\end{align}
with $J_\nu$, $Y_\nu$ denoting order-$\nu$ Bessel functions of the first and second kind, respectively. Notice that $v_n(y)$ then automatically satisfies its boundary condition at the brane $y=0$, while the allowed values of $x^W_n$ (and hence the masses of the KK tower modes $m_n$) are then found with the boundary condition at $y= \pi R$, which can be simplified to
\begin{align}\label{WarpedEigenvalues}
    x^W_n \zeta^{(n)}_0 (x^W_n) &= -a_W^2 \zeta^{(n)}_1(x^W_n), \\
    a_W &\equiv \sqrt{kR} \frac{m_n}{M_{KK}} \nonumber.
\end{align}
The normalization constant $A_n$ can be found using the orthonormality relation of Eq.(\ref{orthoRelation}), yielding
\begin{align}
    A_n = \frac{\sqrt{2kR}}{\Bigg[ (z^W_n)^2 [\zeta^{(n)}_1(z^W_n)^2-\zeta^{(n)}_0(z^W_n)\zeta^{(n)}_2 (z^W_n)]\rvert^{z^W_n = x^W_n}_{z^W_n = \varepsilon^W_n}+2 \tau k R (\varepsilon^W_n)^2 \zeta^{(n)}_1(\varepsilon^W_n)^2\Bigg]^{1/2}}.
\end{align}

\begin{figure}[htbp]
\centerline{\includegraphics[width=3.5in]{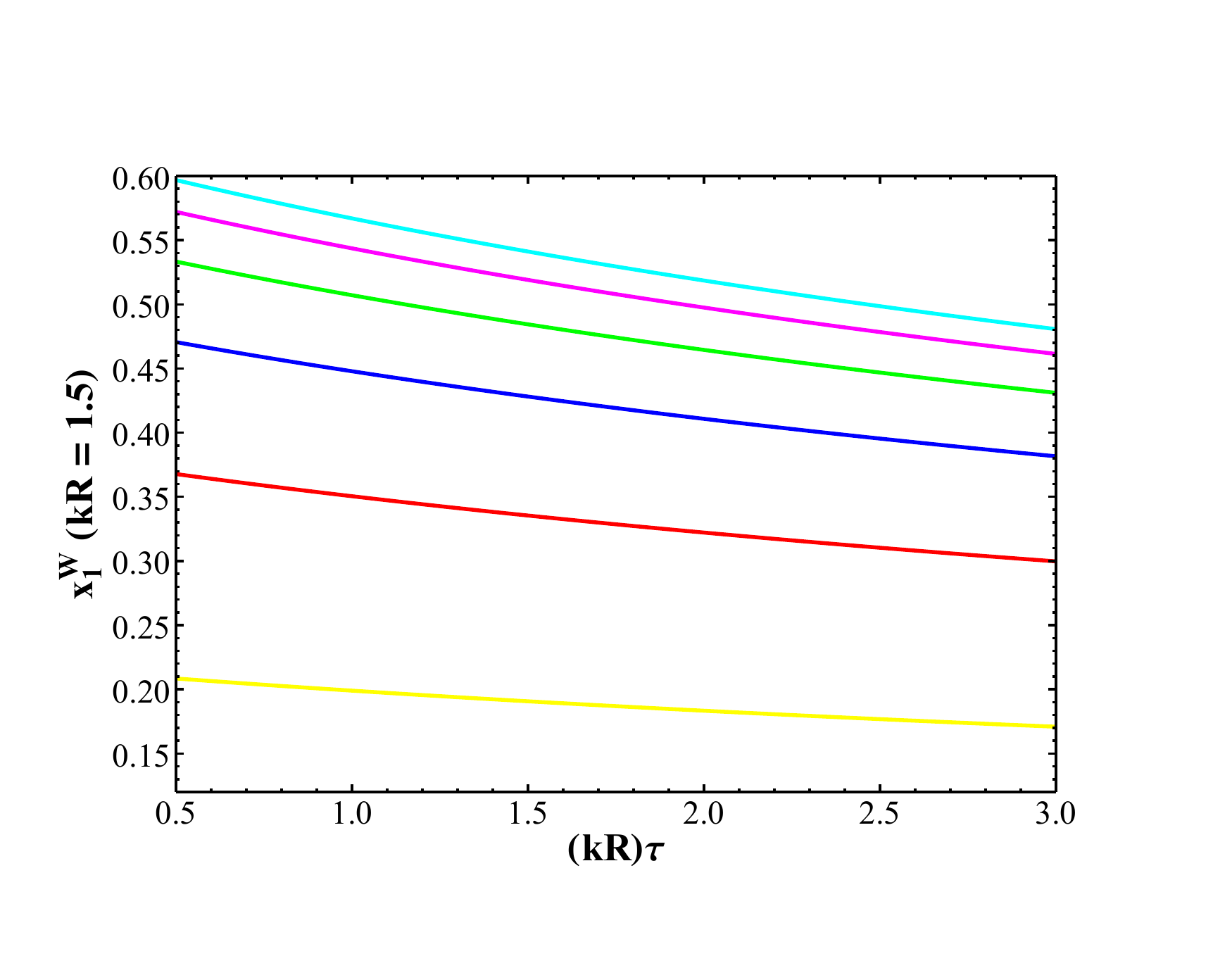}
\hspace{-0.75cm}
\includegraphics[width=3.5in]{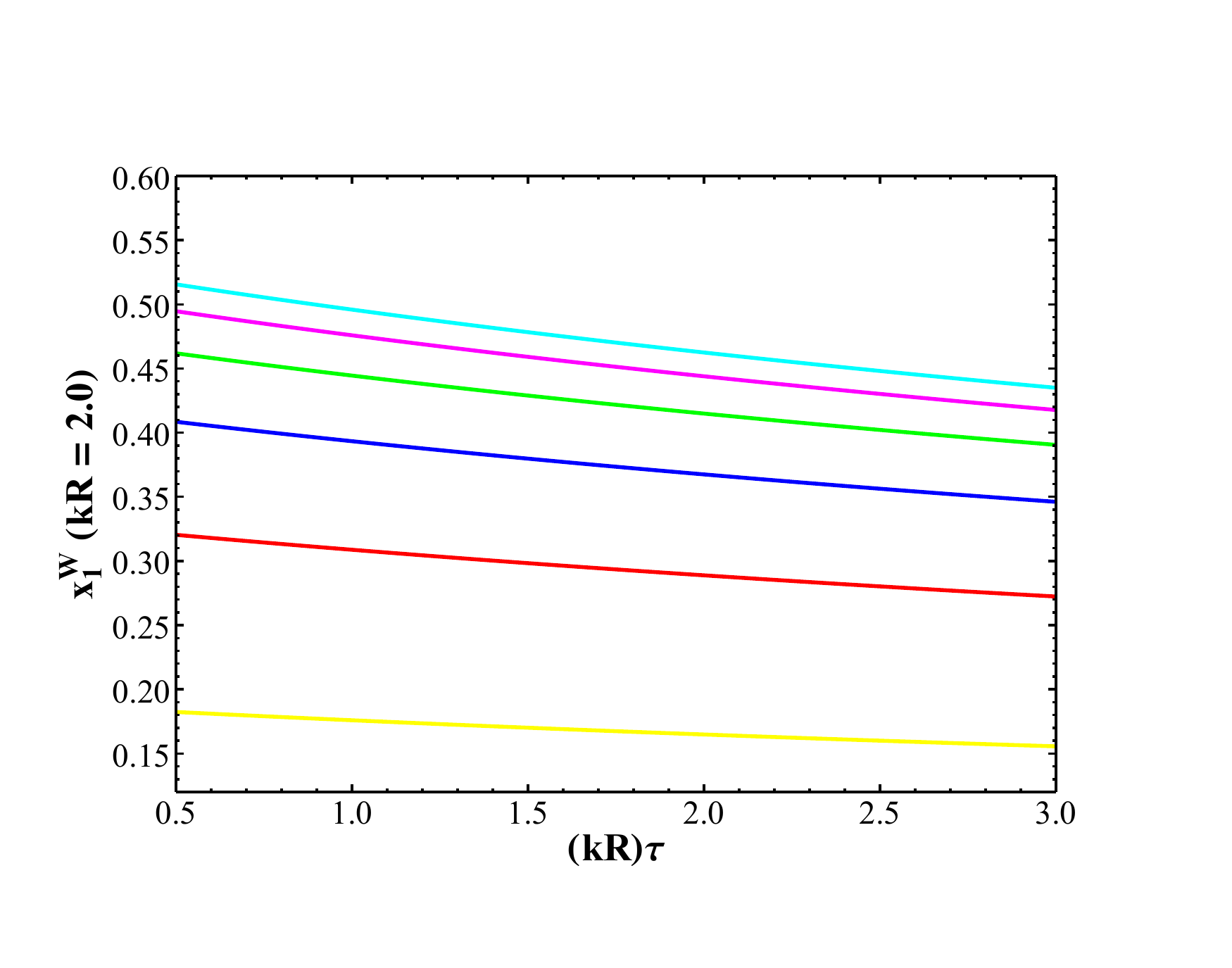}}
\vspace*{-0.25cm}
\centerline{\includegraphics[width=3.5in]{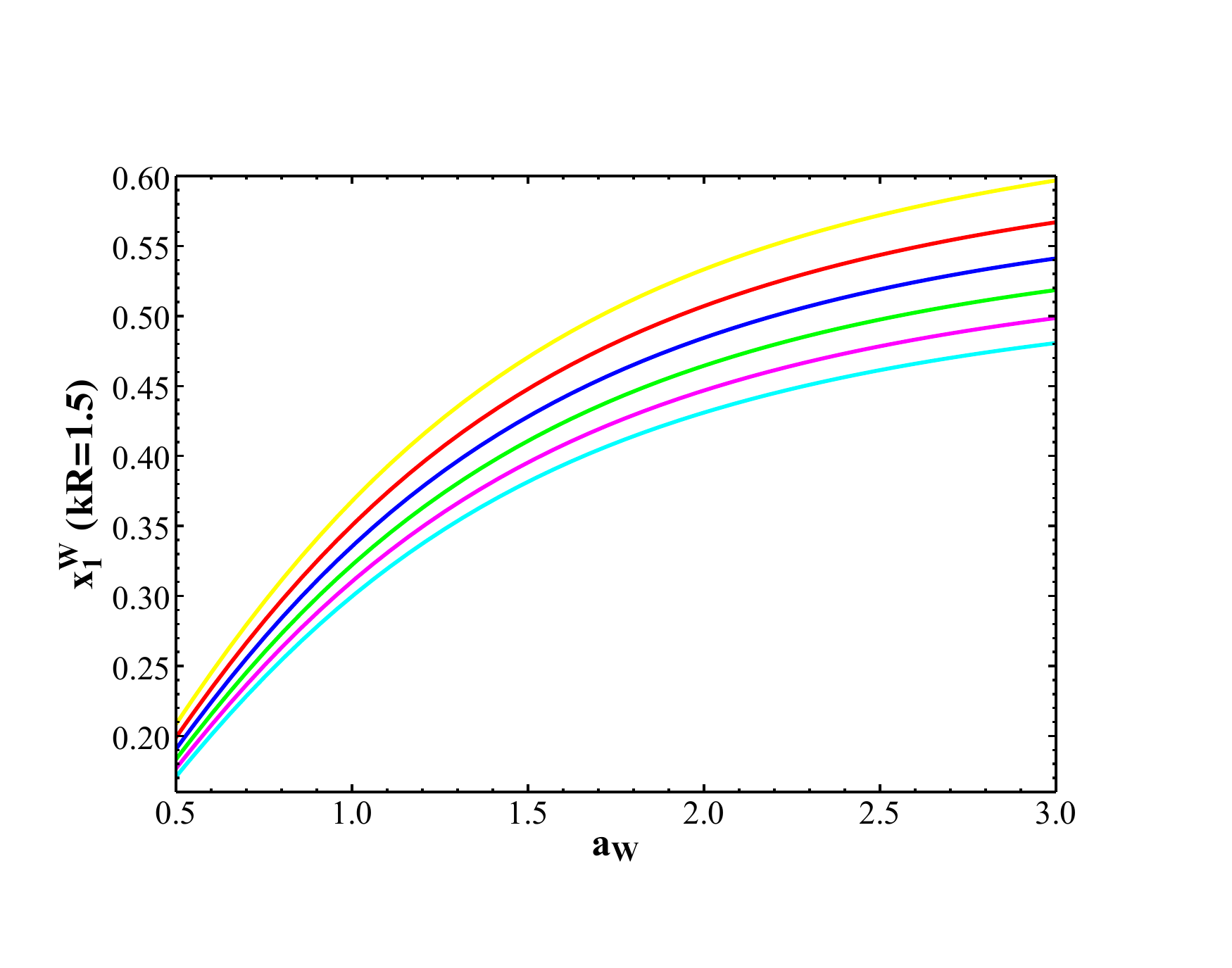}
\hspace{-0.75cm}
\includegraphics[width=3.5in]{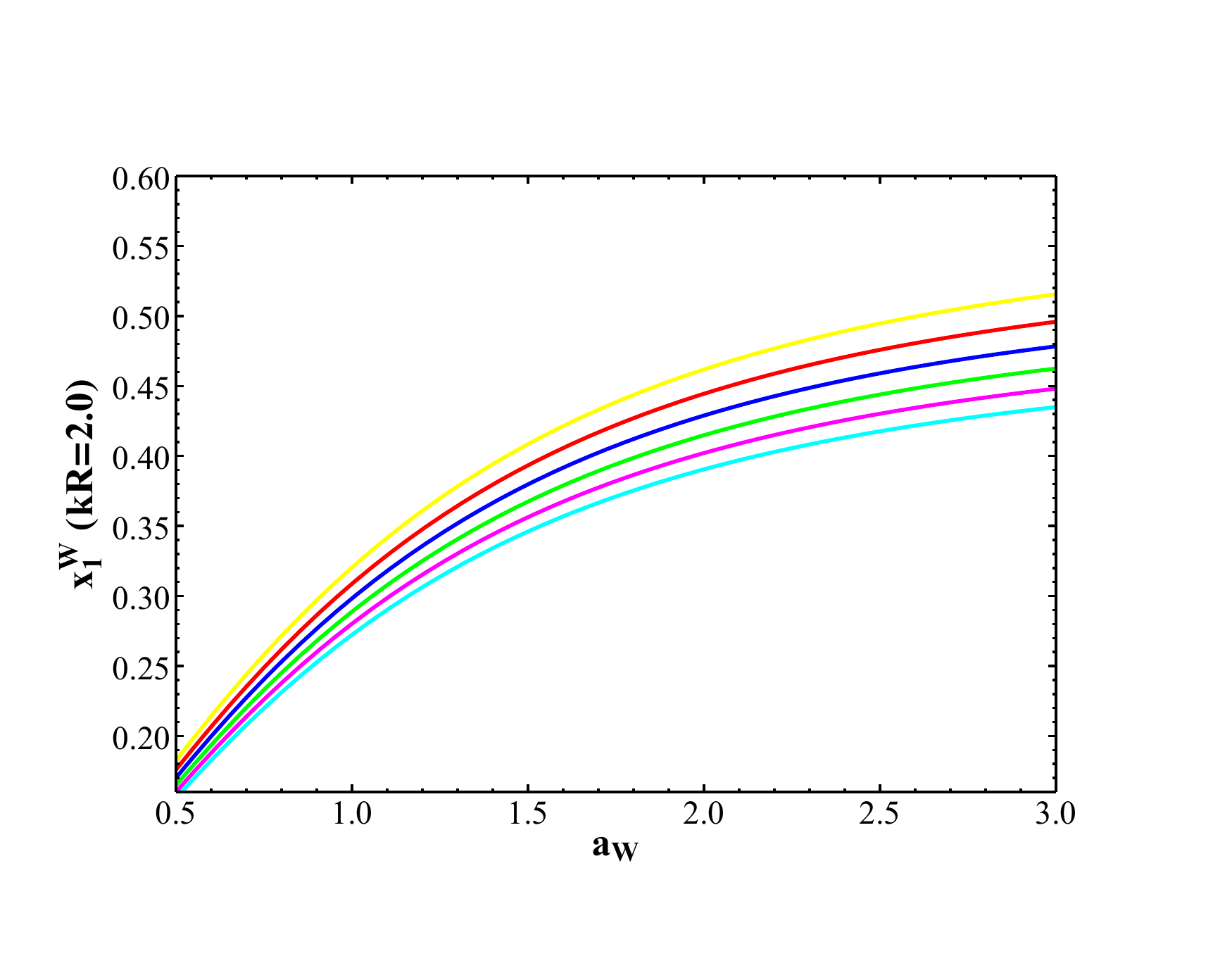}}
\caption{(Top Left) Value of the root $x^W_1$ assuming $kR = 1.5$ as a function of $\tau$ for various choices of $a_W$, from bottom to top, $a_W=$1/2, 1, 3/2, 2, 5/2 and 3, respectively.
(Top Right) The same as the top left, but assuming $kR = 2.0$
(Bottom Left) Value of the root $x^W_1$ assuming as a function of $a_W$ for various choices of $\tau$, from top to bottom, $(kR)\tau=$1/2, 1, 3/2, 2, 5/2 and 3, respectively.
(Bottom Right) The same as the bottom left, but assuming $kR =2.0$}
\label{fig9}
\end{figure}

Using Eqs.(\ref{Warpedvn}) and (\ref{WarpedEigenvalues}), we can now continue on to an exploration of the phenomenology of various KK modes, much as we have done in Section \ref{FlatAnalysis} for the scenario with a flat extra dimension. We begin, as in the case of flat space, by determining the dependencies of the lowest-lying root of Eq.(\ref{WarpedEigenvalues}), $x^W_1$, as a function of the parameters $(\tau,a_W)$, depicted in Fig.~\ref{fig9}. Note that in Fig.~\ref{fig9} and subsequent calculations, we have elected to specify the parameter $(k R) \tau$ (that is $\tau$ scaled by the quantity $kR$) rather than $\tau$. This is because in practice, expressions featuring the brane term $\tau$ in this setup will always do so through the quantity $(kR) \tau$; we therefore find, as has been the case in other work with Randall-Sundrum brane terms \cite{blkts}, that $(kR) \tau$ is the more natural parameter to use.

Qualitatively, we observe largely similar behavior for the root $x^W_1$ in Fig.~\ref{fig9} as we observed in $x^F_1$ in Fig.~\ref{fig1}, namely that $x^W_1 \lsim 1$ for the range of $(\tau, a_W)$ parameters we probe, and that $x^W_1$ increases with increasing $a_W$ and decreases with increasing $\tau$. It is interesting to note that the specific values of $x^W_1$ are somewhat sensitive to the specific value of $kR$: In particular, when $kR = 2.0$, the values of $x^F_1$ for a given choice of $(kR) \tau$ and $a_W$ is approximately 15\% lower than these values in a scenario where $kR = 1.5$.

\begin{figure}[htbp]
\centerline{\includegraphics[width=3.5in]{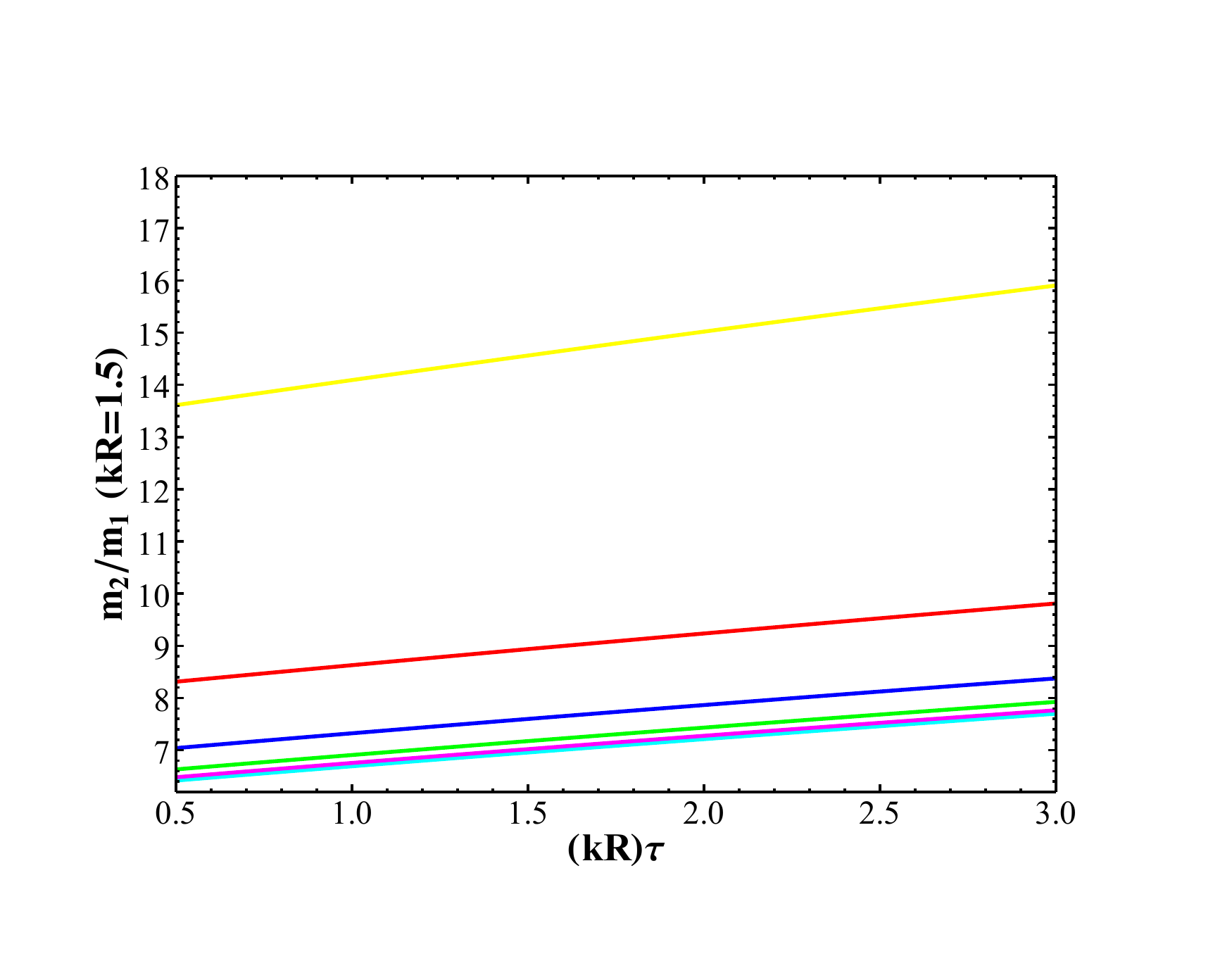}
\hspace{-0.75cm}
\includegraphics[width=3.5in]{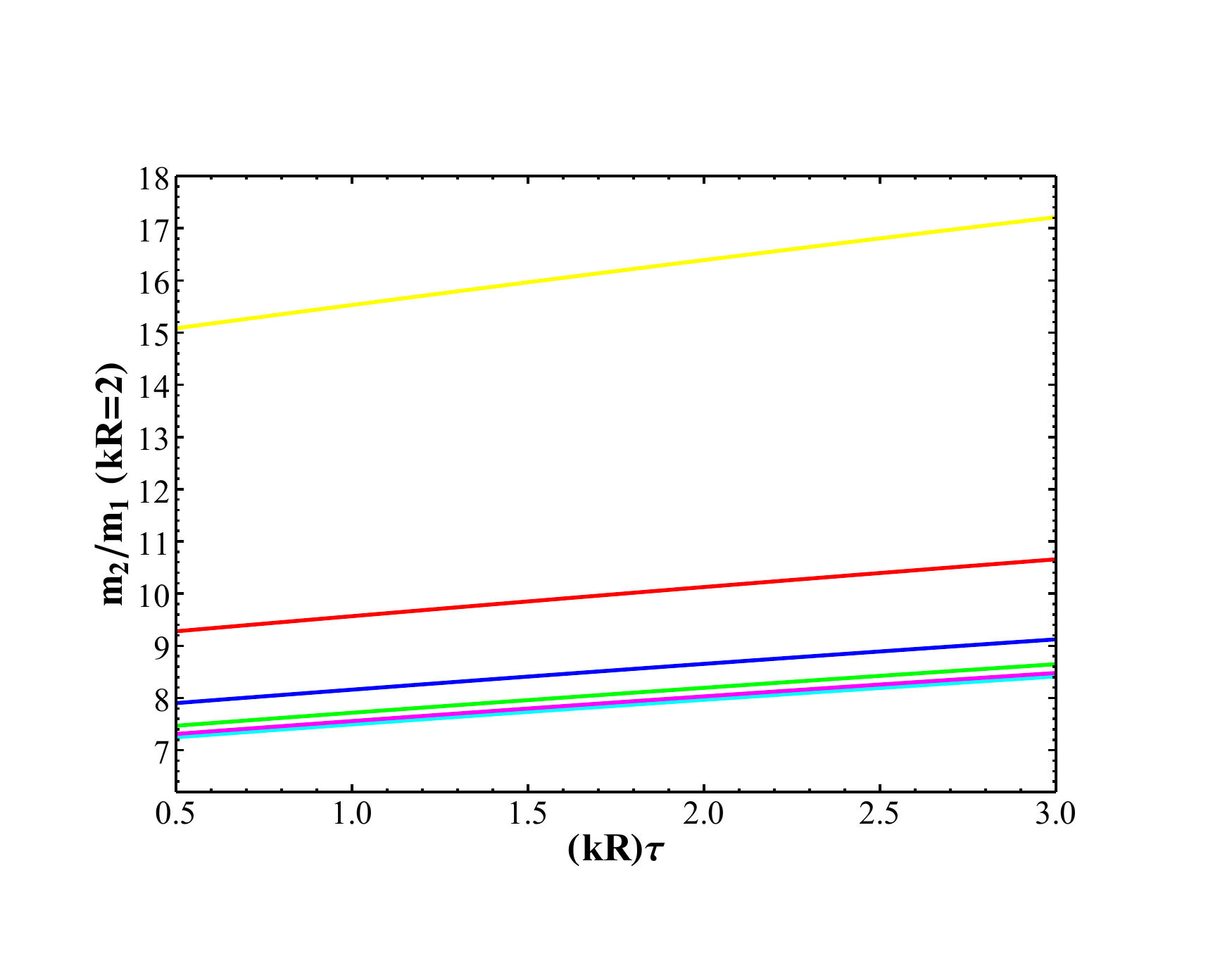}}
\vspace*{-0.25cm}
\centerline{\includegraphics[width=3.5in]{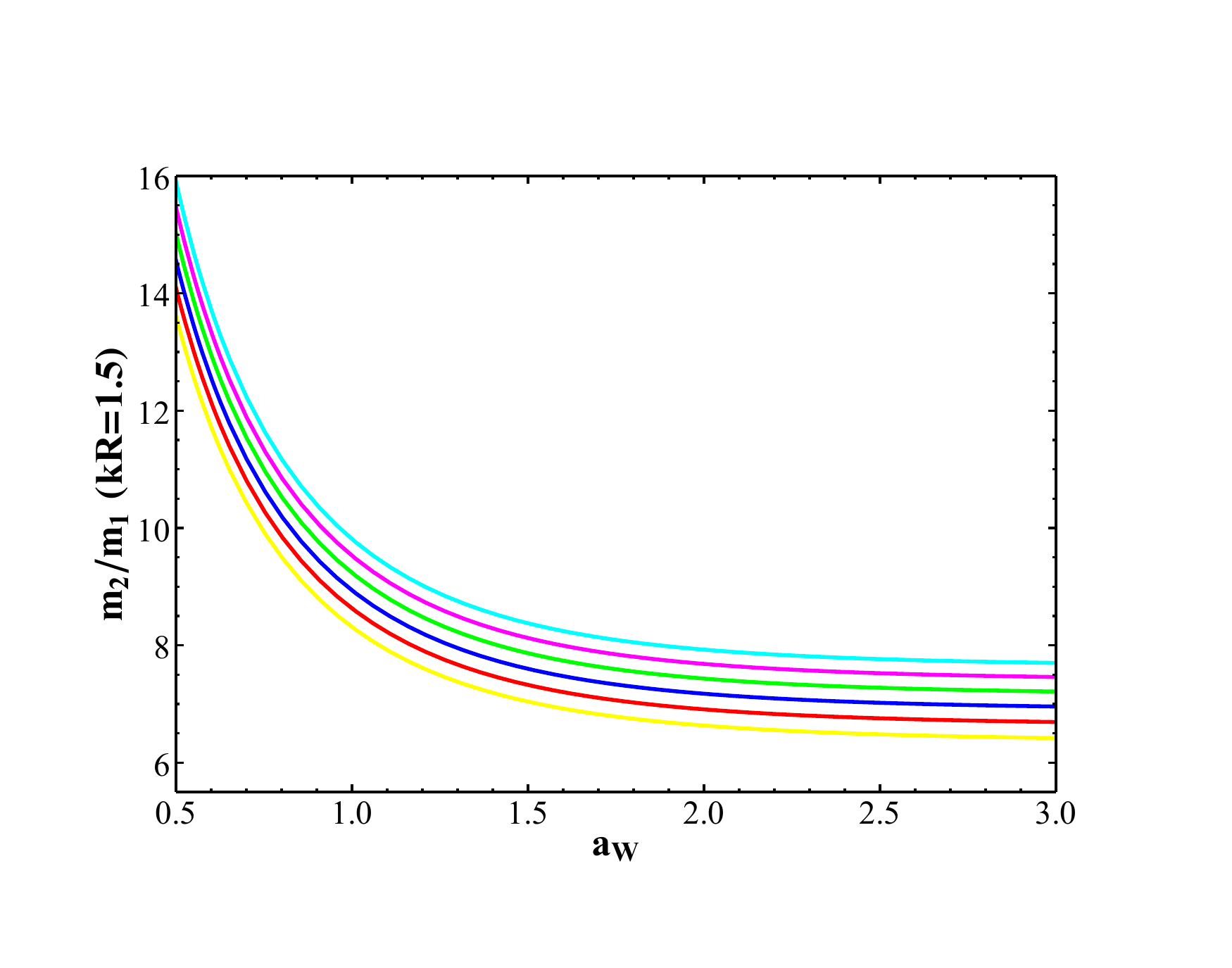}
\hspace{-0.75cm}
\includegraphics[width=3.5in]{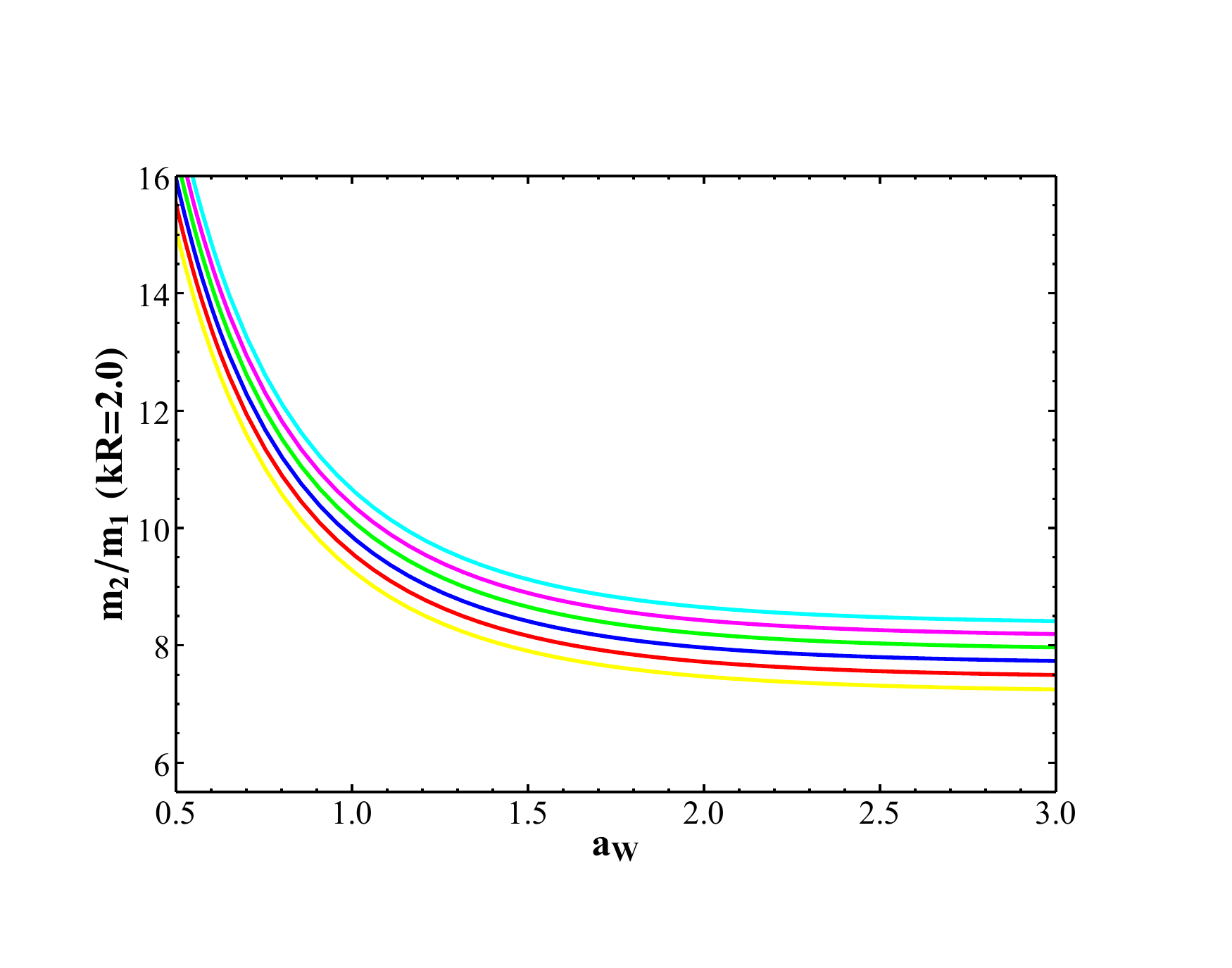}}
\caption{(Top Left) The mass ratio of the lowest two dark photon KK states, $m_{2}/m_{1}=x^W_2/x^W_1$ assuming $kR=1.5$, as a function of $(kR) \tau$ for $a_W=$3(cyan), 5/2(magenta), 2(green), 3/2(blue), 1(red), and 1/2(yellow), 
respectively.
(Top Right) As in the top left, but now assuming $kR = 2.0$.
(Bottom Left) As in the top left, but now as a function of $a_W$ assuming $(kR)\tau=$3(cyan), 5/2(magenta), 2(green), 3/2(blue), 1(red), and 1/2(yellow), respectively.
(Bottom Right) As in the bottom left, but assuming $kR = 2.0$.}
\label{fig10}
\end{figure}

Next, we discuss the quantity $m_2/m_1$, the ratio of the mass of the second KK mode of the dark photon field to that of the first KK mode; as in our discussion of this ratio in the flat space scenario, this quantity continues to possess substantial phenomenological importance due to the potential of the second KK mode to be an experimental signal for the existence of extra dimensions. In Fig.~\ref{fig10}, we depict this mass ratio's dependence on the quantities $\tau$ and $a_W$. The most salient difference between the results here and those for the flat space case discussed in Section \ref{FlatAnalysis} lies in the typical magnitude of the ratio itself: With a flat extra dimension, we found that reasonable selections for $\tau$ and $a_F$ resulted in ratios $m_2/m_1 \sim 3-4$. In the warped setup, we find that the same ratio now typically lies within the range of $m_2/m_1 \sim 6-16$. This represents one of the primary distinctions between the warped and flat constructions, namely, that for a given mass of the lightest KK mode of the dark photon, $m_1$, \emph{the mass of the second KK mode} $m_2$ \emph{is significantly greater in the case of a warped extra dimension than it is in the case of a flat one}. Beyond this observation, we also note that changing $kR$ in our computations below has an effect roughly in line with what we might expect from the results depicted in Fig.~\ref{fig1}, namely, that a larger value of $kR$ slightly increases the ratio $m_2/m_1$, likely because the value of the root $x^W_1$ is somewhat  reduced.

\begin{figure}[htbp]
\centerline{\includegraphics[width=5.0in,angle=0]{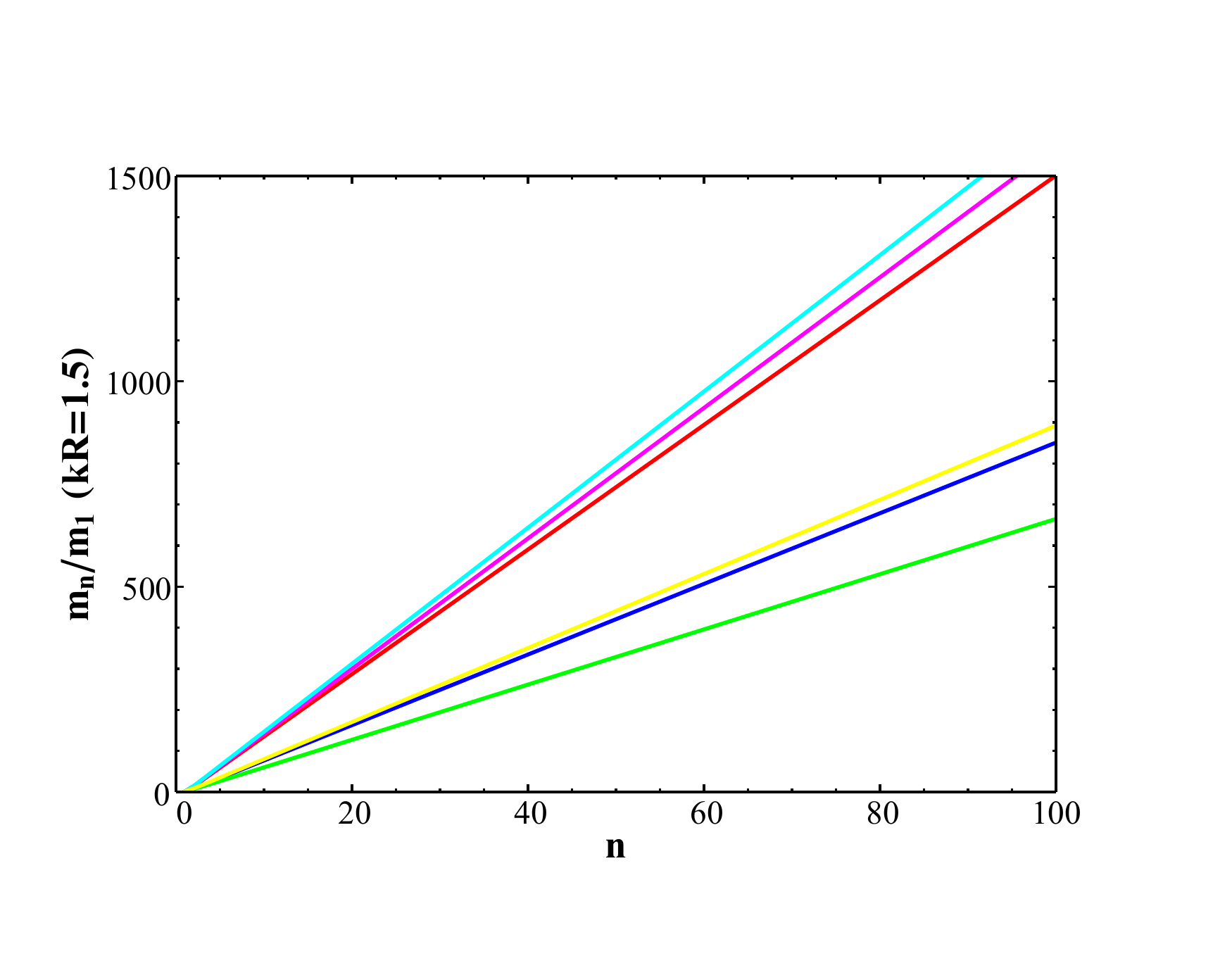}}
\vspace*{-2.0cm}
\centerline{\includegraphics[width=5.0in,angle=0]{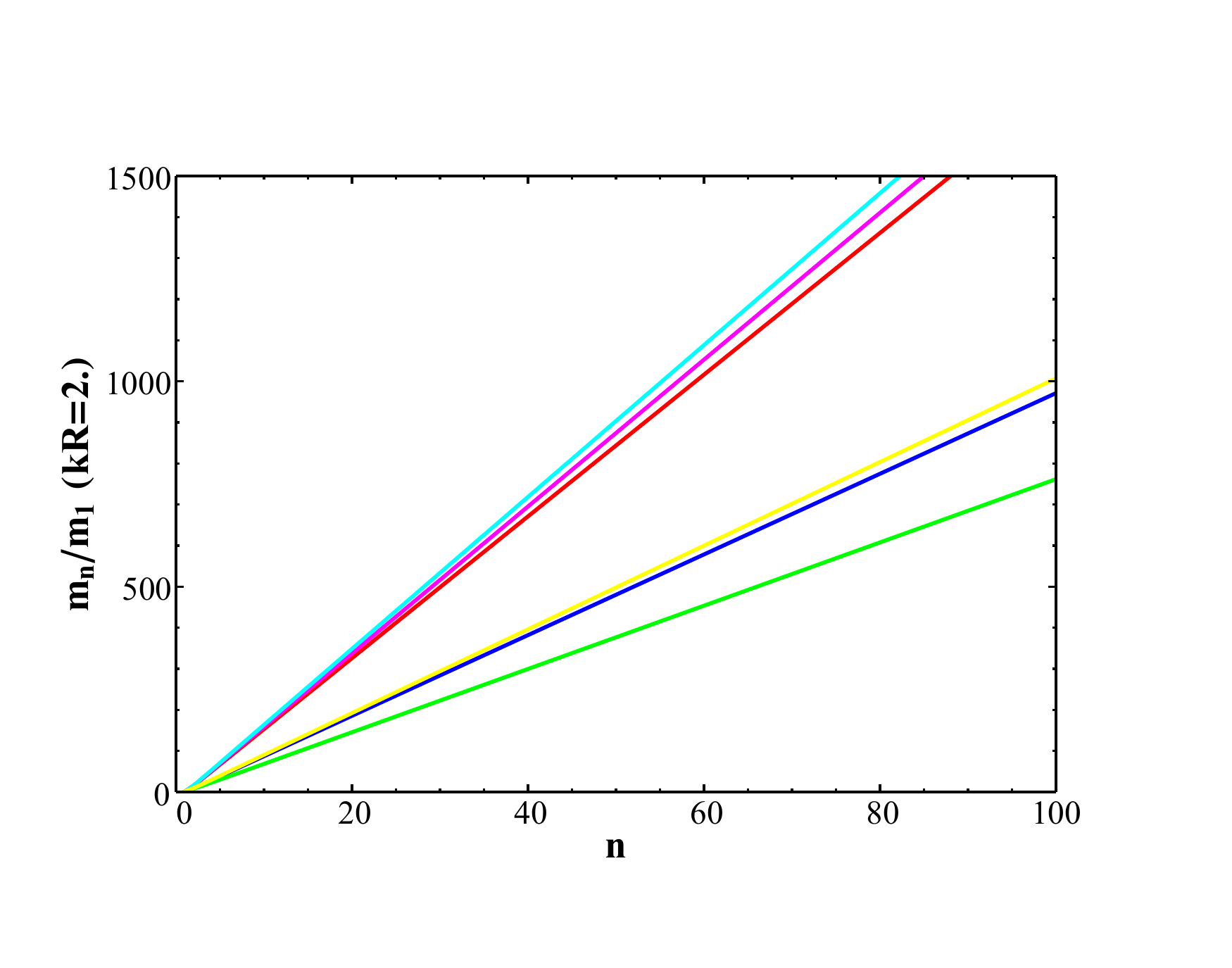}}
\vspace*{-1.0cm}
\caption{(Top) Approximate linear growth of the relative dark photon KK mass ratio $m_n/m_1$ as a function of $n$ assuming $kR = 1.5$ for various choices of $((kR)\tau ,a_W)$ =(1/2,1/2) [red], (1/2,1) [blue], (1/2,3/2) [green], 
(1,1/2) [magenta], (3/2,1/2) [cyan] and (1,1) [yellow], respectively.
(Bottom) As in the previous panel, but assuming $kR=2.0$}
\label{fig11}
\end{figure}

We complete our exploration of the relative masses of various KK modes just as we have in the flat space scenario, namely, by exploring the growth of $m_n$ as $n$ increases. We depict the results in Fig.~\ref{fig11} for both $kR=1.5$ and $kR=2.0$, for various selections of $(kR)\tau$ and $a_W$. The most salient contrast between these results and those in the flat space analysis again lies in the magnitude of the mass ratio: In the warped setup, $m_n/m_1$ increases significantly more sharply with $n$ than it does in the flat space, such that at large $n$, typical values of $m_{n}/m_1$ are approximately three times larger for a warped extra dimension than they are for a flat one. The dominant share of this discrepancy is determinable from the mass eigenvalue equation Eq.(\ref{WarpedEigenvalues})-- numerically, it can be readily seen that the difference between successive roots of this equation approaches $\pi$ as $n$ becomes large, so the eventual slope of the line depicted in Fig.~\ref{fig11} should be roughly $\pi (x^W_1)^{-1}$. This is compared to the analogous slope in the flat space scenario, which, as discussed in Section \ref{FlatAnalysis}, should be approximated by $(x^F_1)^{-1}$. Because the typical values of $x^F_1$ and $x^W_1$ are roughly comparable, this in turn suggests that the slope of the lines in Fig.~\ref{fig11} should be steeper by roughly a factor of $O(\pi)$ than their flat space counterparts in Fig.~\ref{fig3}. Before moving on, we also note that the same behavior with increasing $kR$ that we observed in the ratio $m_2/m_1$ appears again as we consider more massive KK modes, namely, that increasing $kR$ will increase the value of the ratios of heavier KK mode masses to that of the lightest mode.

\begin{figure}[htbp]
\centerline{\includegraphics[width=3.5in]{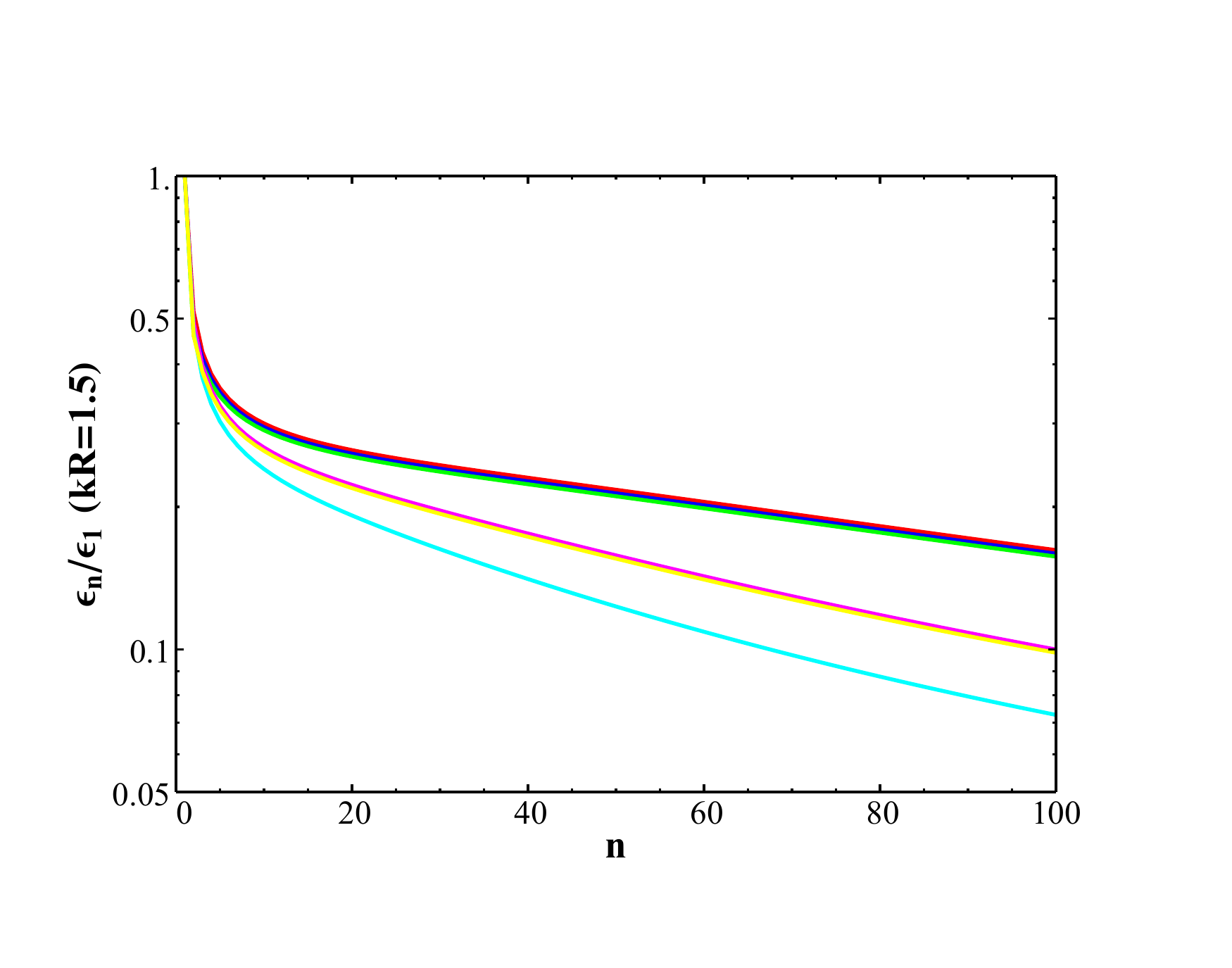}
\hspace{-0.75cm}
\includegraphics[width=3.5in]{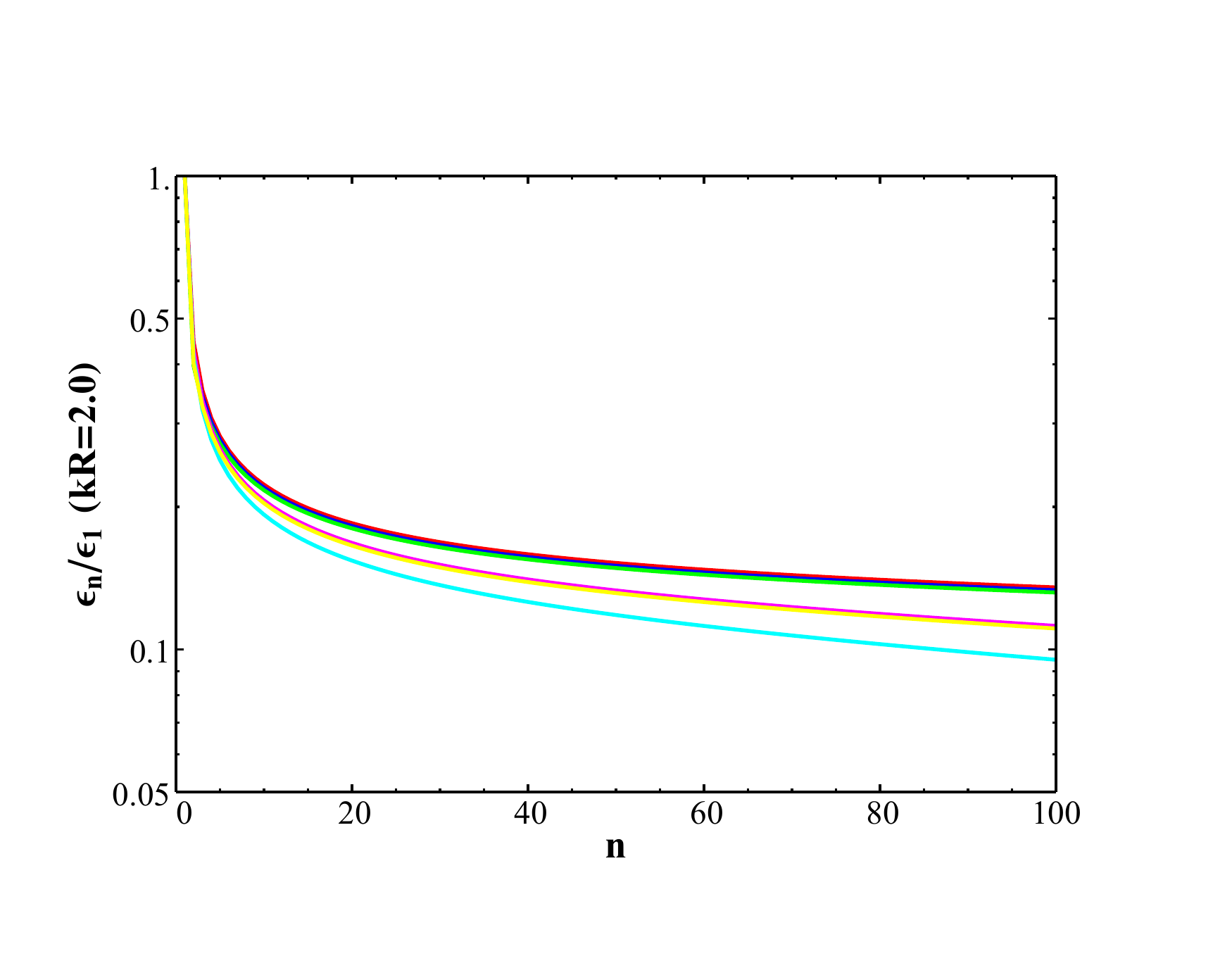}}
\vspace*{-0.25cm}
\centerline{\includegraphics[width=3.5in]{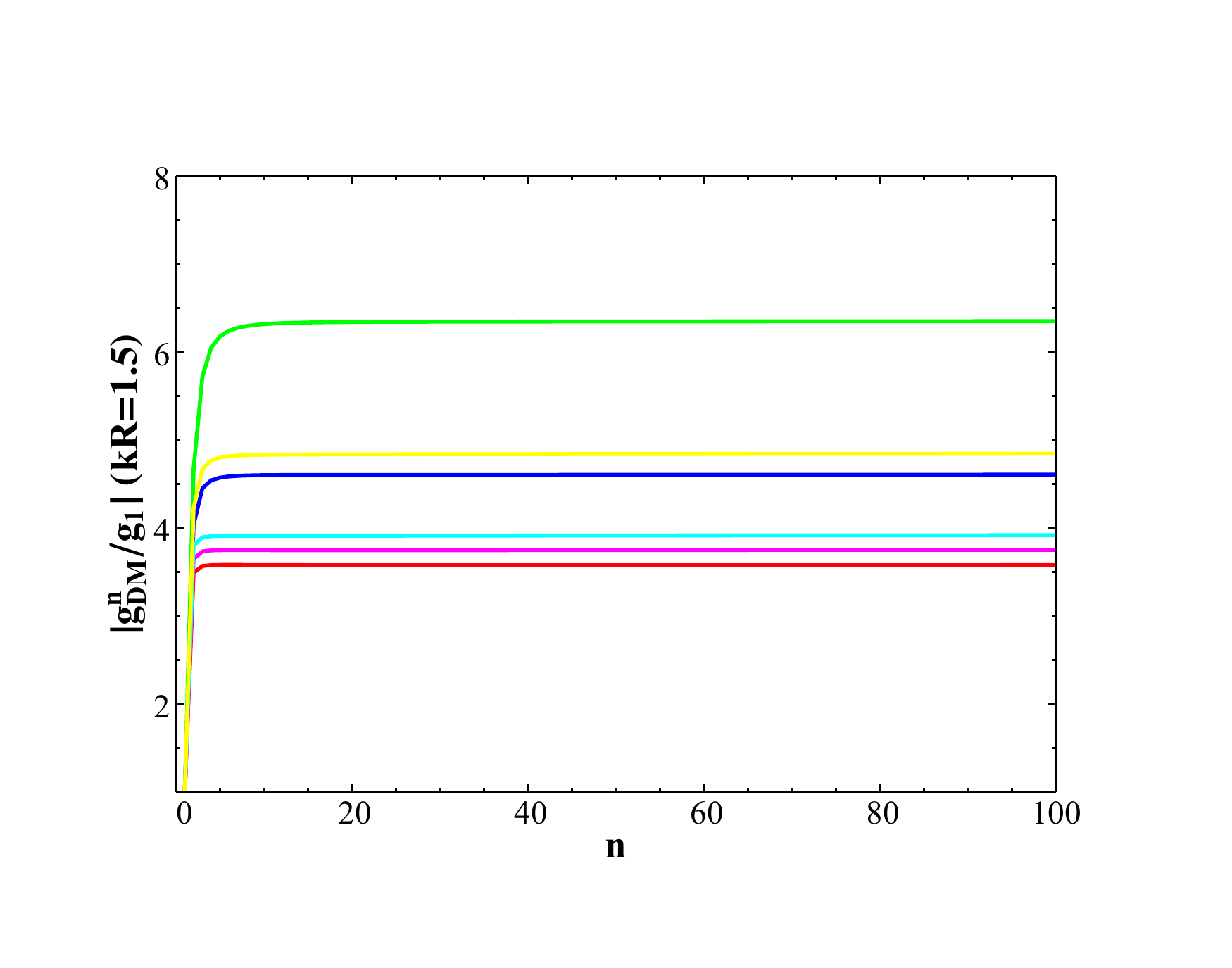}
\hspace{-0.75cm}
\includegraphics[width=3.5in]{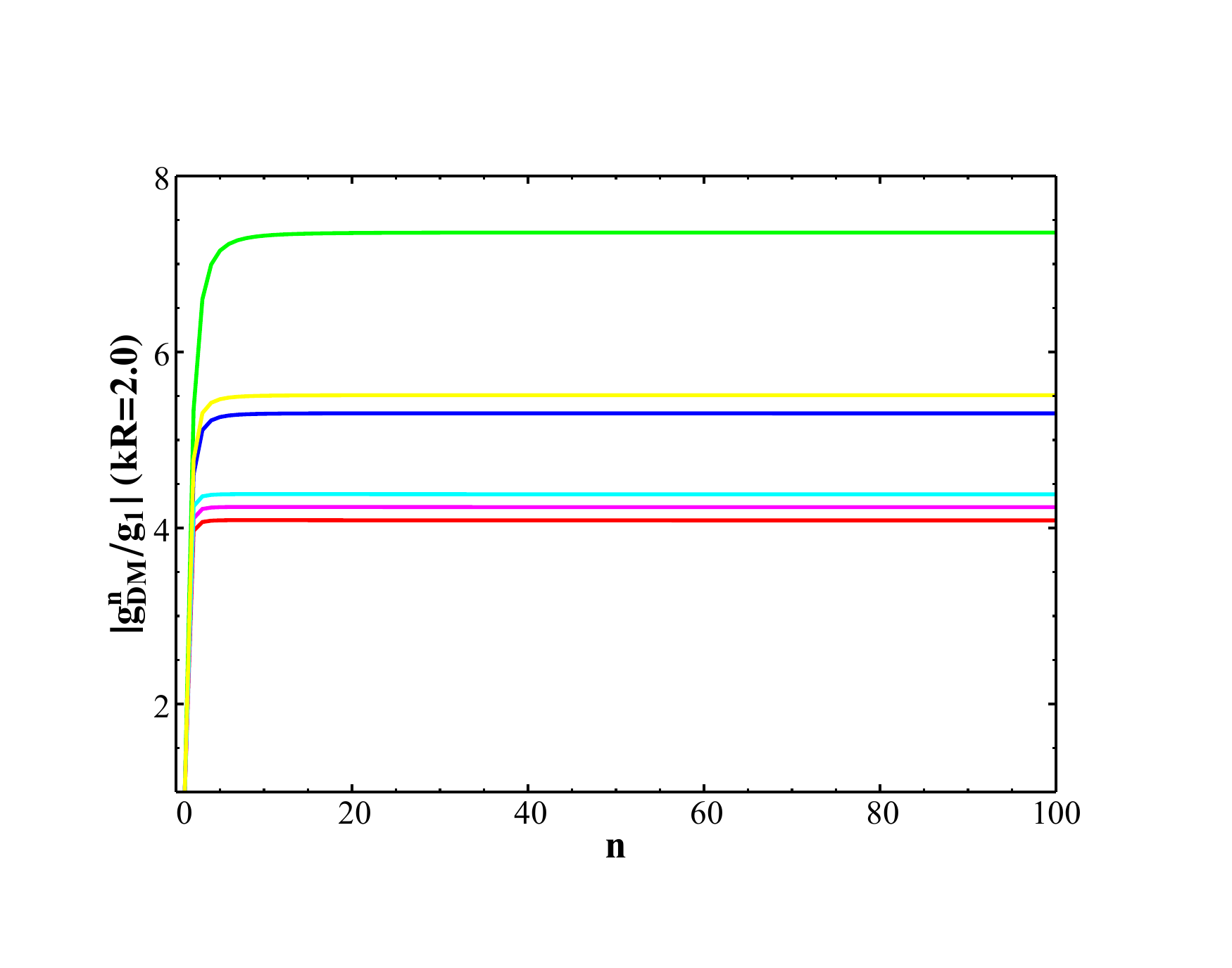}}
\caption{(Top Left) The ratio $\epsilon_n/\epsilon_1$, assuming $kR=1.5$, as a function of $n$ for various choices of $((kR)\tau ,a_W)$ =(1/2,1/2) [red], (1/2,1) [blue], (1/2,3/2) [green], 
(1,1/2) [magenta], (3/2,1/2) [cyan] and (1,1) [yellow], respectively.
(Top Right) The same as the top left, but assuming $kR = 2.0$
(Bottom Left) The ratio $g^n_{DM}/g_D$, assuming $kR=1.5$, as a function of $n$ for various choices of $((kR)\tau ,a_W)$ =(1/2,1/2) [red], (1/2,1) [blue], (1/2,3/2) [green], 
(1,1/2) [magenta], (3/2,1/2) [cyan] and (1,1) [yellow], respectively.
(Bottom Right) The same as the bottom left, but assuming $kR =2.0$}
\label{fig12}
\end{figure}

\begin{figure}[htbp]
\centerline{\includegraphics[width=3.5in]{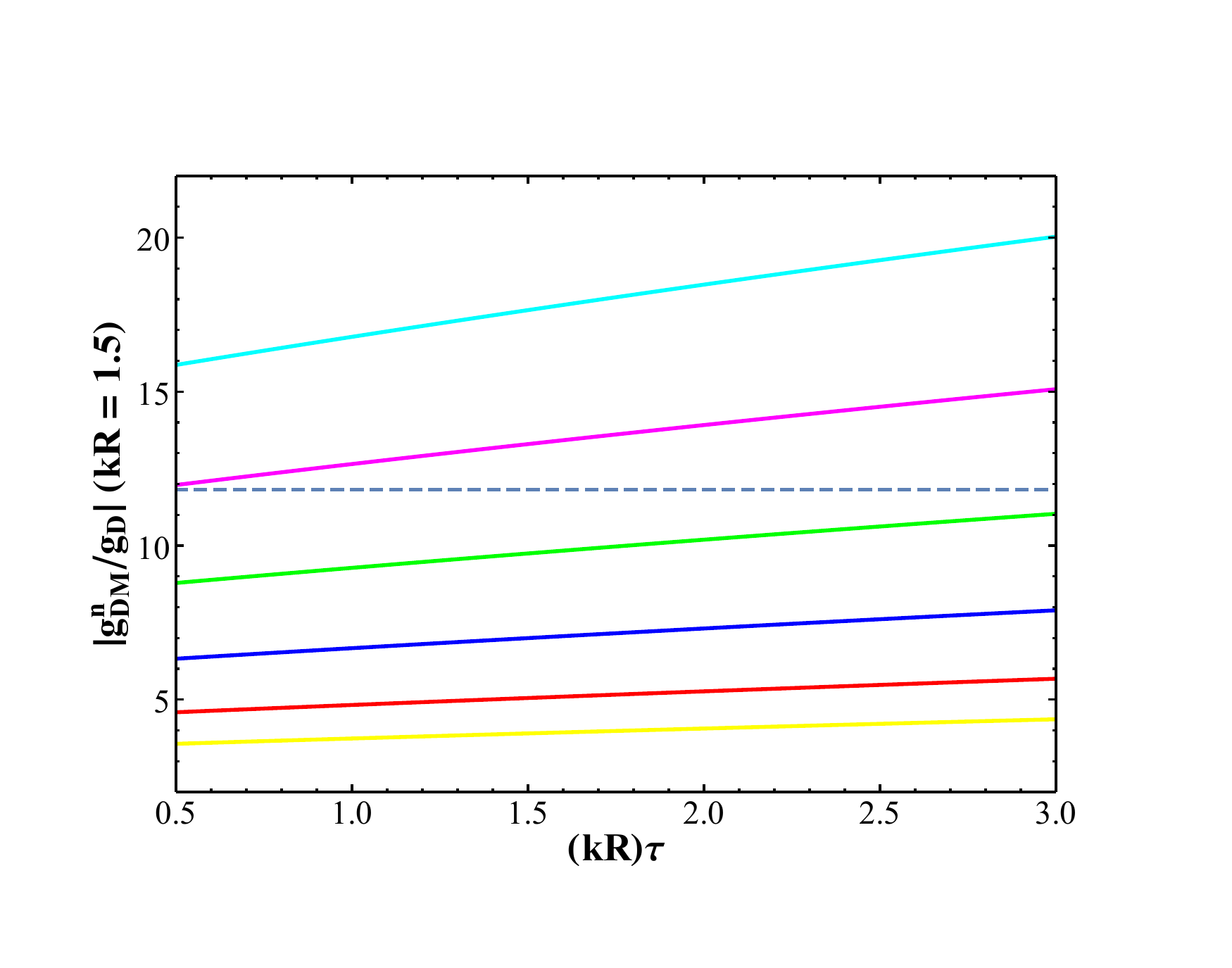}
\hspace{-0.75cm}
\includegraphics[width=3.5in]{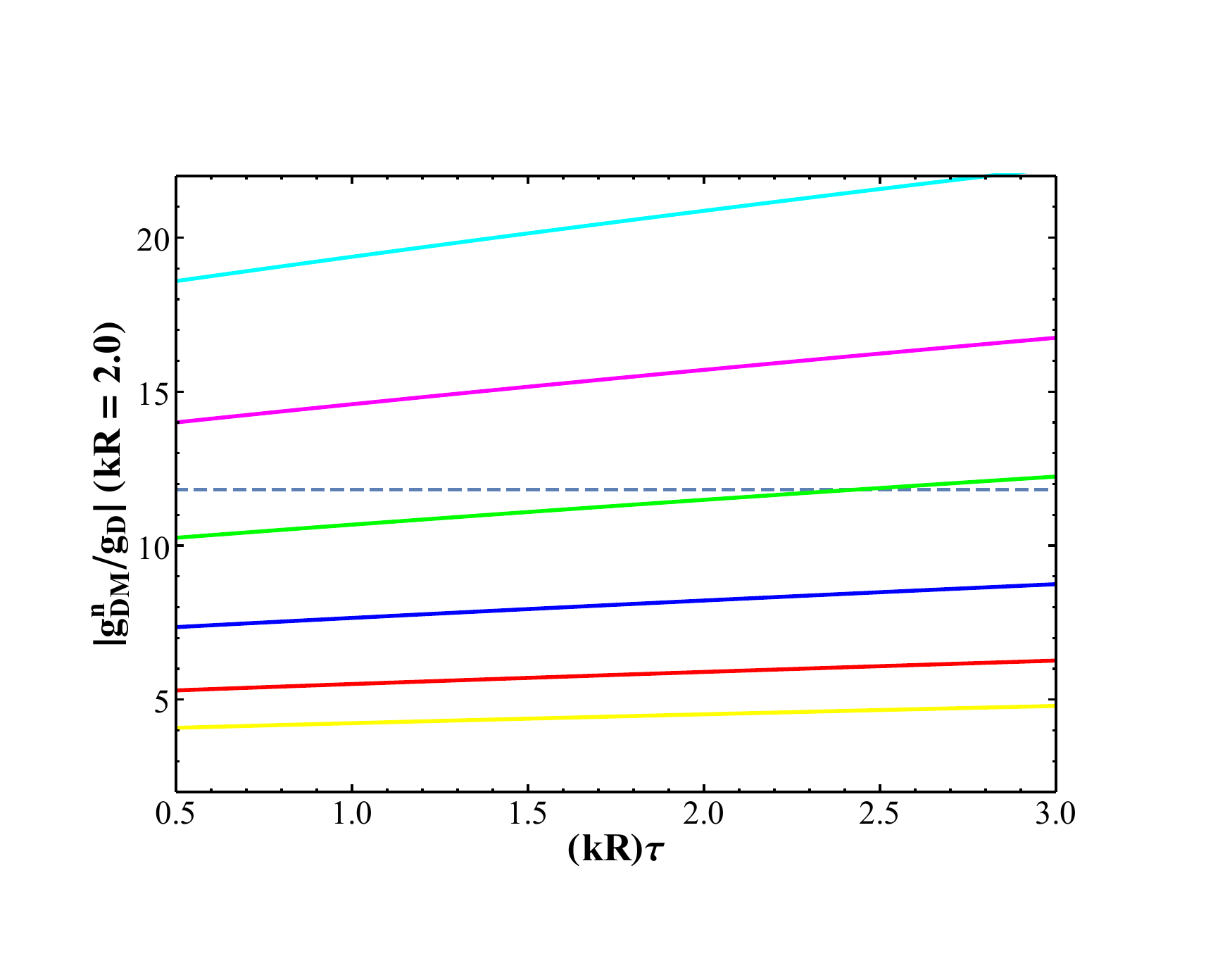}}
\vspace*{-0.25cm}
\centerline{\includegraphics[width=3.5in]{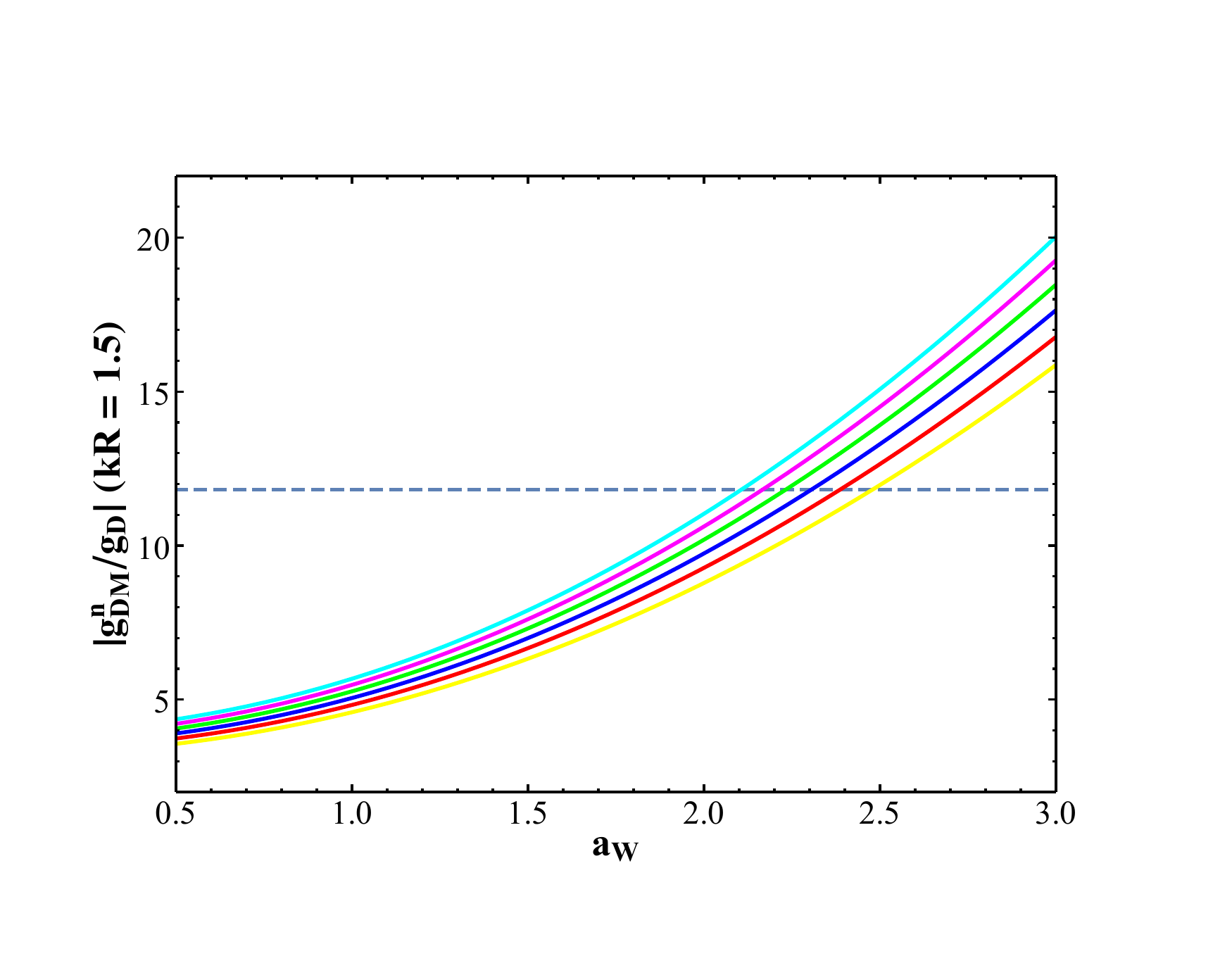}
\hspace{-0.75cm}
\includegraphics[width=3.5in]{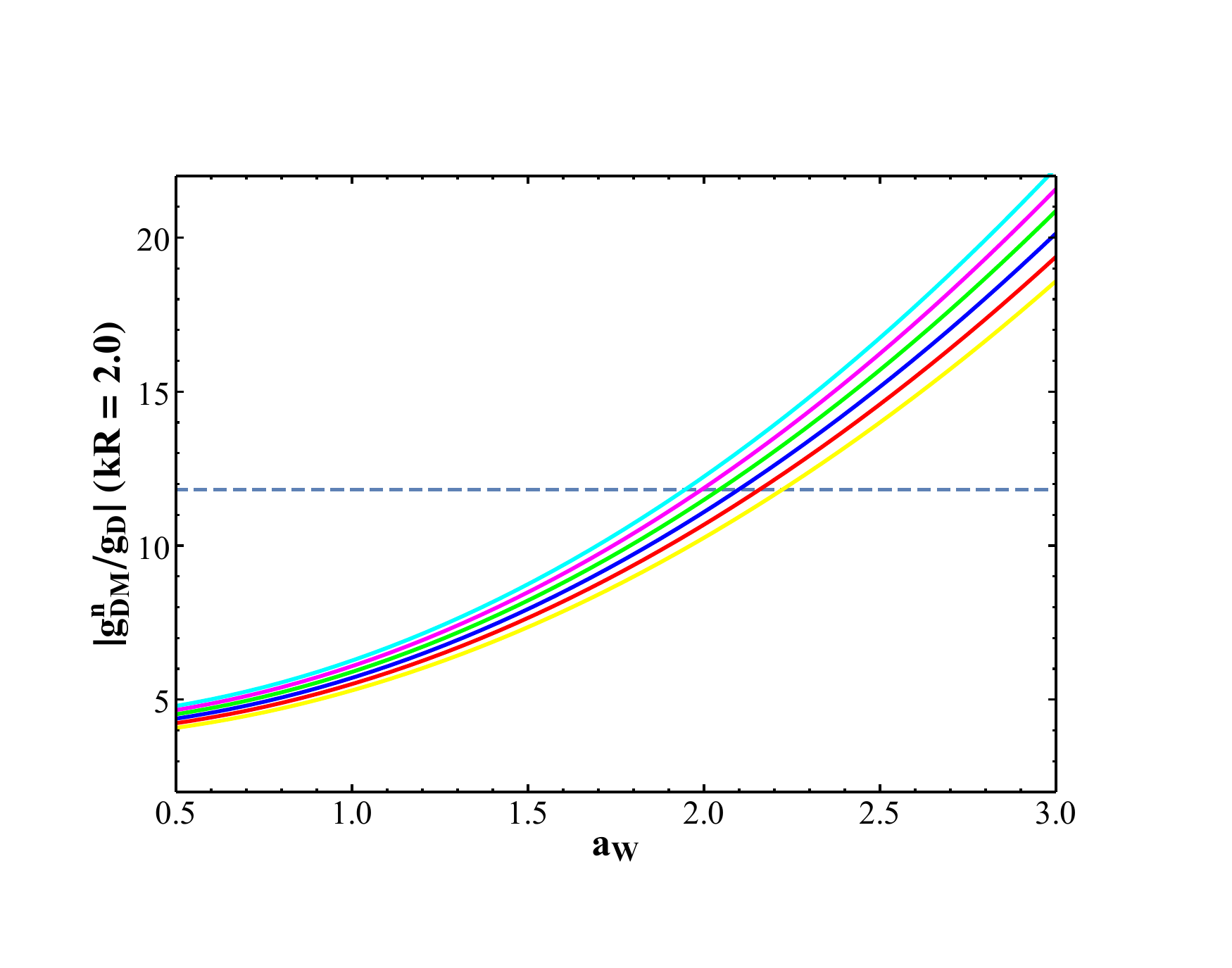}}
\caption{(Top Left) The approximate asymptotic value of $|g^n_{DM}/g_D|$ given by Eq.(\ref{WarpedApproxgn}) for large $n$, assuming $kR = 1.5$, as a function of $(kR)\tau$ for various choices of $a_W=$3(cyan), 5/2(magenta), 2(green), 3/2(blue), 1(red), 1/2(yellow).
(Top Right) The same as the top left, but assuming $kR = 2.0$
(Bottom Left) The approximate asymptotic value of $|g^n_{DM}/g_D|$ given by Eq.(\ref{WarpedApproxgn}) for large $n$, assuming $kR=2.0$, as a function of $a_W$ for various choices of $(kR)\tau=$3(cyan), 5/2(magenta), 2(green), 3/2(blue), 1(red), 1/2(yellow). The dashed line represents the maximum value that this ratio can obtain and still have all KK couplings remain perturbative for $g_D=0.3$.
(Bottom Right) The same as the bottom left, but assuming $kR =2.0$}
\label{fig13}
\end{figure}

Having addressed the masses of the various dark photon KK modes, we now move on to discuss the effective kinetic mixing and DM coupling terms that arise in this construction. In Fig.~\ref{fig12}, we depict the behavior of the ratios $\epsilon_n/\epsilon_1$ and $|g^n_{DM}/g_D|$ as a function of the KK mode $n$ (we note that once again, as in the flat space scenario, the values of $g^n_{DM}$ oscillate in sign). The results are qualitatively quite similar to the flat space scenario depicted in Fig.~\ref{fig5}. In particular, we find once again that while $\epsilon_n/\epsilon_1$ consistently decreases for large $n$, $|g^n_{DM}/g_D|$ again approaches a non-zero asymptotic value. This asymptotic value for $|g^n_{DM}/g_D|$, much like its flat space analogue, can be explored further by semi-analytical means. By using Eqs.(\ref{Warpedvn}) and (\ref{WarpedEigenvalues}), as well as the identities,
\begin{align}
    J_1(z) Y_0(z)-J_0(z)Y_1(z) = \frac{2}{\pi z}, \\
    \zeta^{(n)}_2(z) = \frac{2}{z}\zeta^{(n)}_1(z)-\zeta^{(n)}_0(z), \nonumber
\end{align}
it is possible to determine that as $n$ becomes very large, the ratio $|g^n_{DM}/g_D|$ becomes well-approximated by the expression
\begin{align}\label{WarpedApproxgn}
    \bigg\lvert \frac{g^n_{DM}}{g_D} \bigg\rvert &\approx \frac{1}{(x^W_1)}\bigg( (x^W_1)^2 + 2 a_W^2+a_W^4-(1+(kR)^2\tau^2 (x^W_1)^2 e^{-2 kR \pi})\bigg(\mathcal{J}\bigg)^2\bigg)^{\frac{1}{2}},\\
    \mathcal{J} &\equiv \frac{x^W_1 J_0(x^W_1)+a_W^2 J_1(x^W_1)}{J_0(x^W_1 e^{-kR \pi})+(kR)\tau x^W_1 e^{-kR \pi}J_1(x^W_1 e^{-kR \pi})}. \nonumber
\end{align}
In Fig.~\ref{fig13}, we depict the dependence of this approximate asymptotic value on $\tau$ and $a_W$. The behavior of this quantity is quite similar to the analogous results Fig.~\ref{fig4} for the flat space scenario, in particular, we observe a sharp increase in the ratio here as $a_W$ increases, just as the corresponding ratio in the flat space case increases sharply with increasing $a_F$. We note that the typical maximum values that we observe in Fig.~\ref{fig13}, however, are roughly a factor of 2 smaller than those we observed in Fig.~\ref{fig4}, however, as $a_F$ and $a_W$ are not directly comparable quantities, the significance of this diminished range is not obvious. Again, as in Fig.~\ref{fig4}, we have included a dashed line which denotes the maximum value that this ratio can attain such that all $g^n_{DM}$ remain perturbative for the choice $g_D=0.3$; in this case, we see that such a requirement effectively excludes choices of $a_W\gsim 2$.

Just as in our analysis of the flat space setup, we can now move on from discussing individual KK modes' masses and couplings to the basic predictions of phenomenologically important processes. In order to do this, we must first evaluate the sum $F(y,y',s)$ (defined in Eq.(\ref{FDefinition}) for the warped metric, by solving Eq.(\ref{FDiffEq}) with $f(y) = e^{-k y}$ inserted. We arrive at the differential equation
\begin{align}\label{FDiffEqWarped}
    \partial_y [e^{-2 k y} ~\partial_y F(y,y',s)] = R\delta(y-y')- s F(y,y',s), \nonumber\\
    \partial_y F(y,y',s)|_{y=0} = -s \tau R F(0,y',s), \\
    \partial_y F(y,y',s)|_{y=\pi R} = - m_V^2 R e^{- 2 k R \pi} F(\pi R, y',s). \nonumber
\end{align}
By defining the variables $z \equiv (\sqrt{s}/M_{KK})e^{k (y-\pi R)}$ and $z' \equiv (\sqrt{s}/M_{KK}) e^{k (y'-\pi R)}$, we can solve Eq.(\ref{FDiffEqWarped}) in terms of Bessel functions, yielding
\begin{align}\label{WarpedFSolution}
    &F(y,y',s) = -\frac{kR \pi}{2 M_{KK}^2}\frac{e^{k (y+y'-2\pi R)}\xi_1(z_>)\omega_1(z_<)}{z_\pi \omega_0(z_\pi)+a_W^2 \omega_1 (z_\pi)},\\
    &\omega_\nu (z) \equiv [Y_0 (z_0)+\tau k R z_0 Y_1 (z_0)]J_\nu(z)-[J_0 (z_0)+\tau k R z_0 J_1 (z_0)]Y_\nu(z), \nonumber \\
    &\xi_\nu (z) \equiv [z_\pi Y_0 (z_\pi)+a_W^2 Y_1 (z_\pi)]J_\nu (z) - [z_\pi J_0 (z_\pi)+a_W^2 J_1 (z_\pi)]Y_\nu (z), \nonumber \\
    &z_> \equiv \bigg(\frac{\sqrt{s}}{M_{KK}} \bigg) e^{k (\textrm{max}(y,y')-\pi R)}, \;\;\; z_< \equiv \bigg(\frac{\sqrt{s}}{M_{KK}} \bigg) e^{k (\textrm{min}(y,y')-\pi R)}, \nonumber\\
    &z_0 \equiv \bigg(\frac{\sqrt{s}}{M_{KK}} \bigg)e^{-k R \pi}, \;\;\;\;\;\;\;\;\;\;\;\;\;\;\;\;\;\; z_\pi \equiv \bigg(\frac{\sqrt{s}}{M_{KK}} \bigg). \nonumber
\end{align}
We note that in this form, it is readily apparent that $F(y,y',s)$ has poles wherever $\sqrt{s}$ is equal to the mass of a KK mode $m_n$, just as we would expect given the components of its sum and just as we previously observed in the flat-space sum Eq.(\ref{FlatFSolution}).

\begin{figure}[htbp]
\centerline{\includegraphics[width=3.5in]{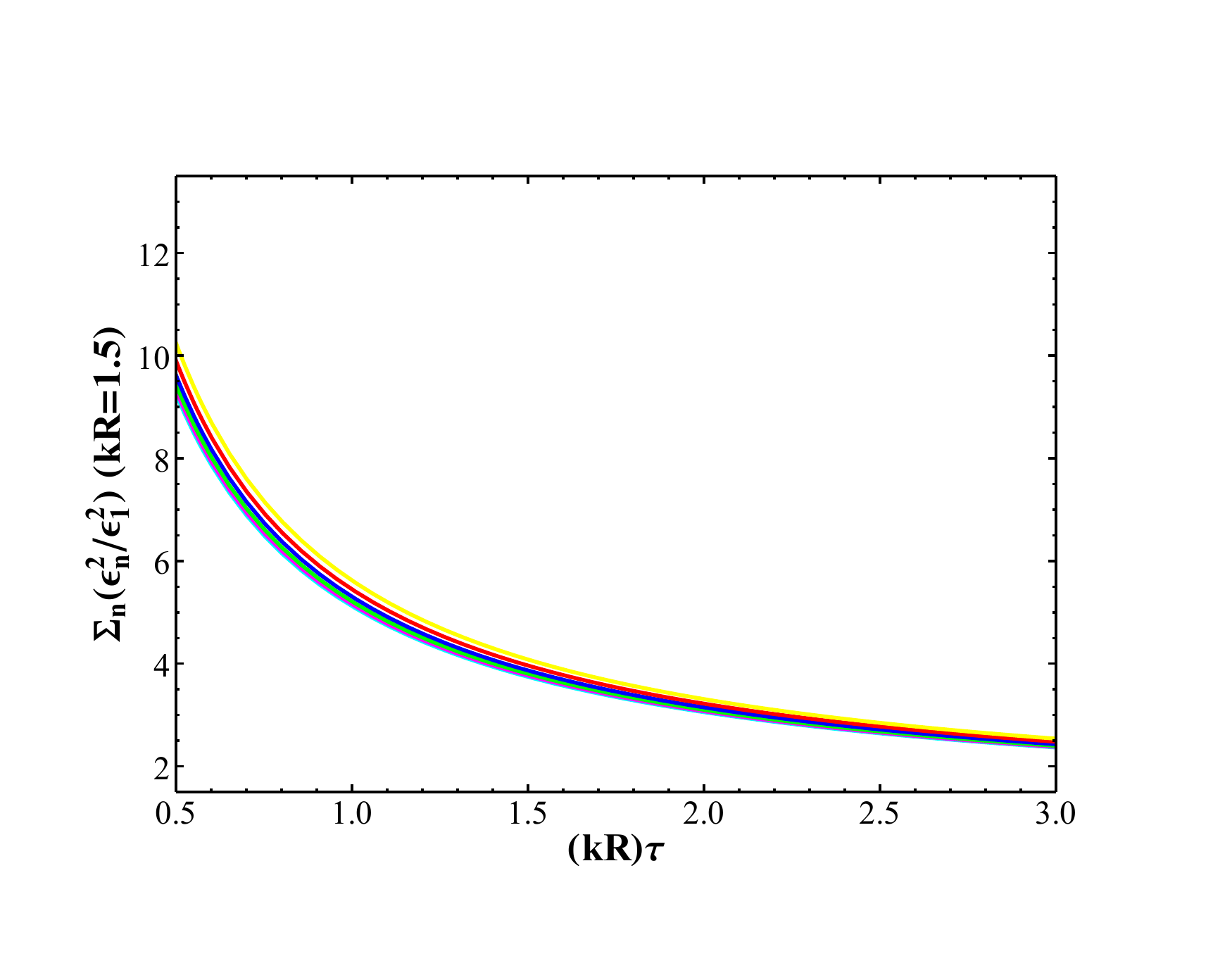}
\hspace{-0.75cm}
\includegraphics[width=3.5in]{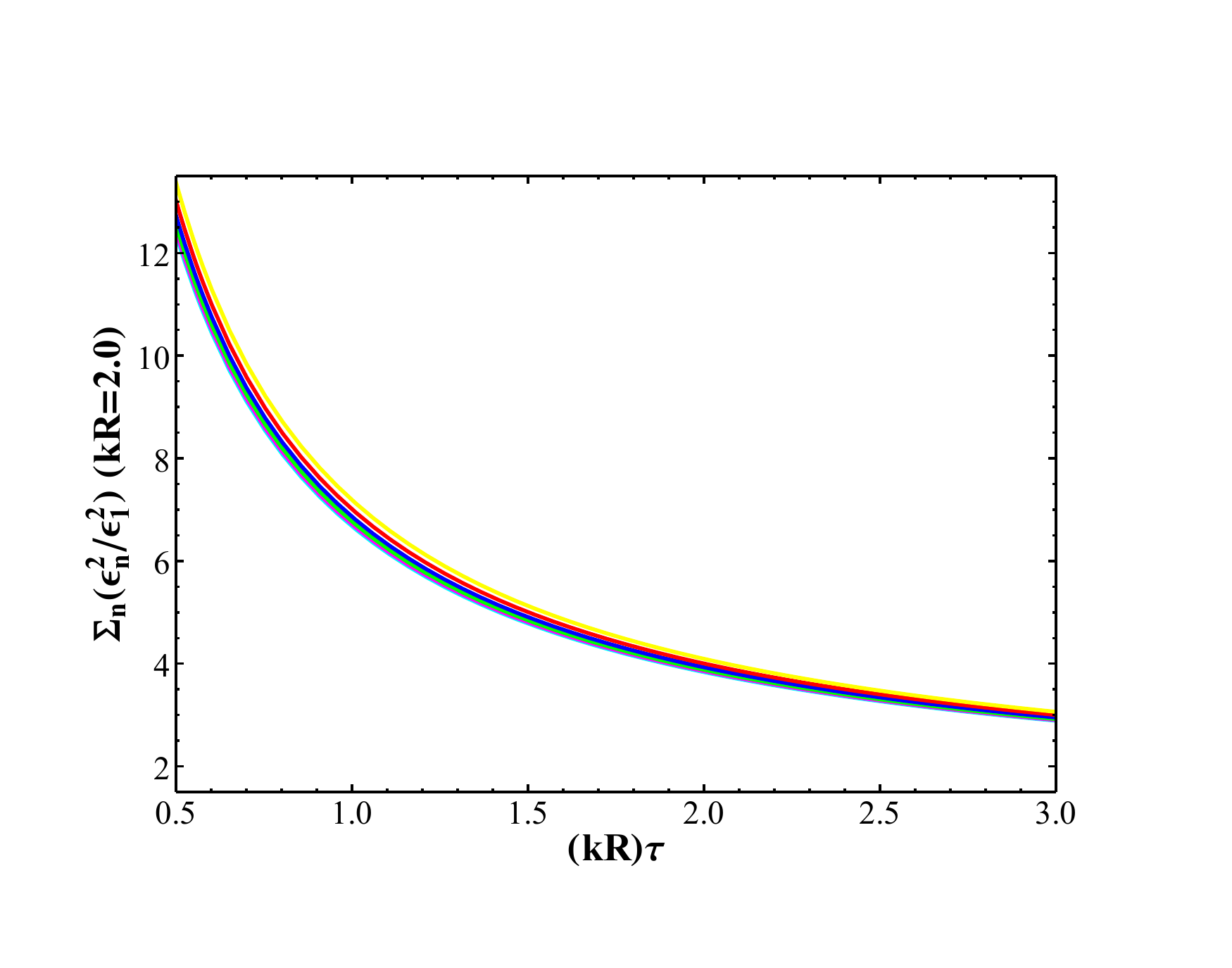}}
\vspace*{-0.25cm}
\centerline{\includegraphics[width=3.5in]{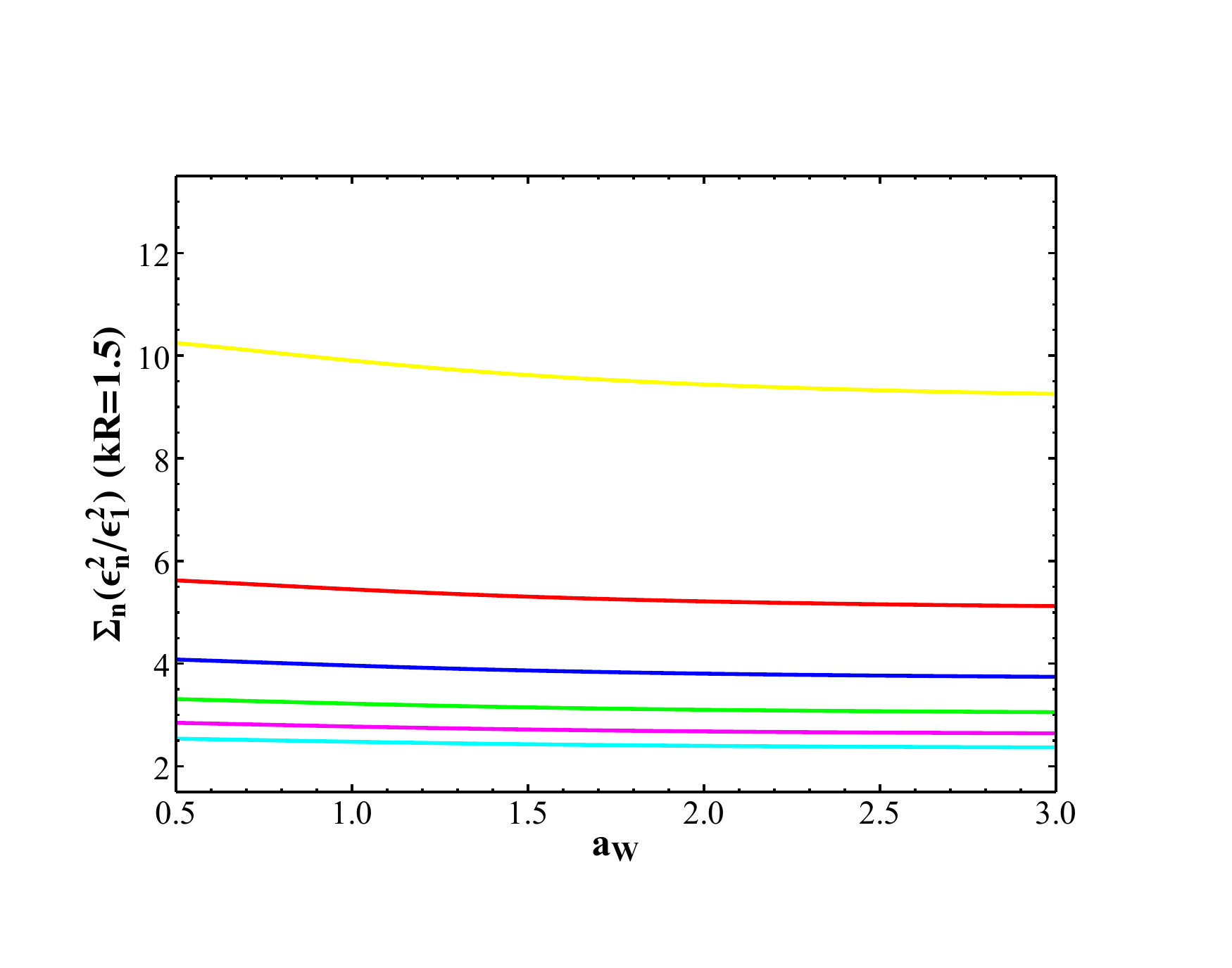}
\hspace{-0.75cm}
\includegraphics[width=3.5in]{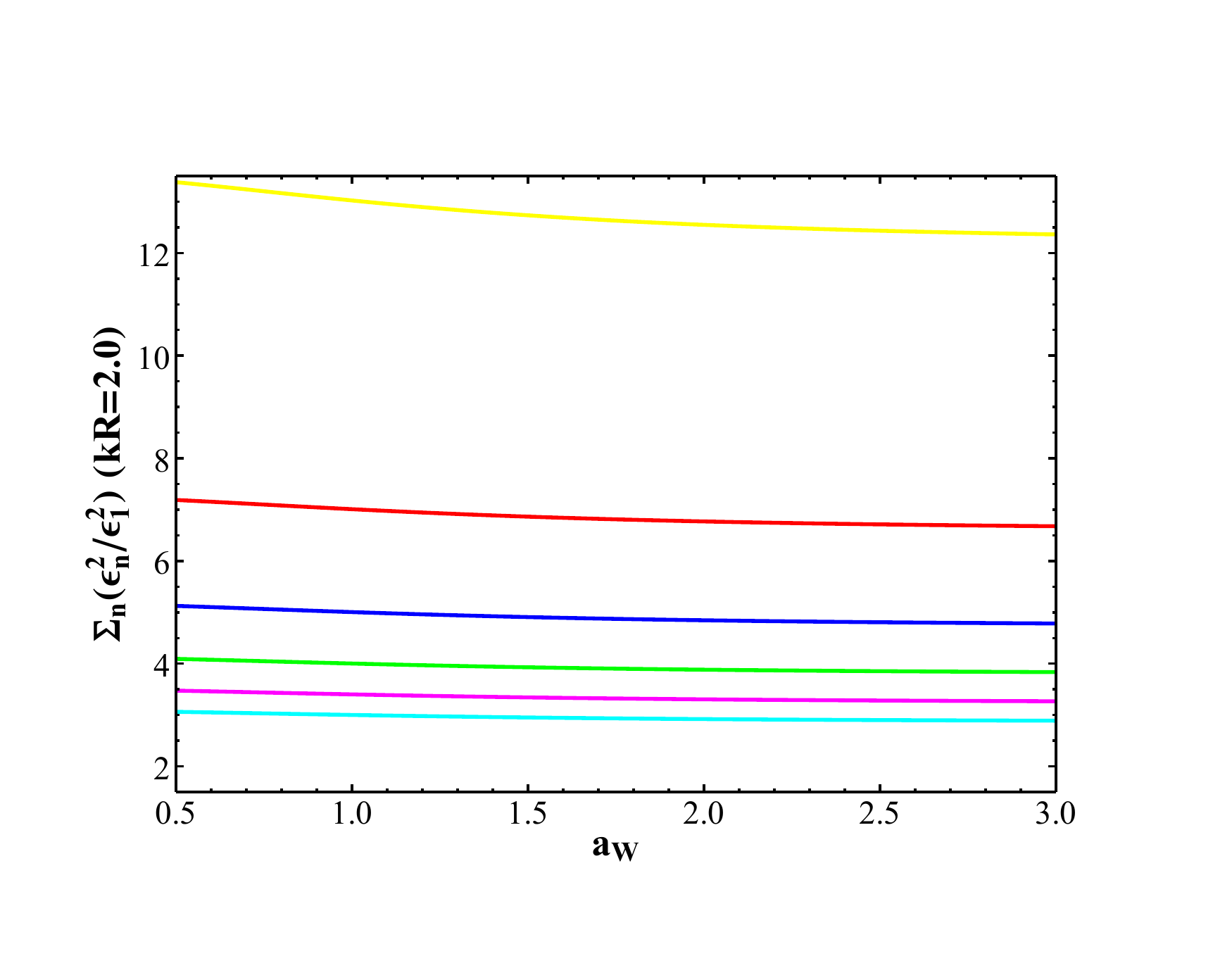}}
\caption{(Top Left) The value of the sum $\sum_n \epsilon_n^2/\epsilon_1^2$ over all $n$ assuming $kR=1.5$, as a function of $(kR) \tau$ for $a_W=$3(cyan), 5/2(magenta), 2(green), 3/2(blue), 1(red), and 1/2(yellow), 
respectively.
(Top Right) As in the top left, but now assuming $kR = 2.0$.
(Bottom Left) As in the top left, but now as a function of $a_W$ assuming $(kR)\tau=$3(cyan), 5/2(magenta), 2(green), 3/2(blue), 1(red), and 1/2(yellow), respectively.
(Bottom Right) As in the bottom left, but assuming $kR = 2.0$.}
\label{figWarpedepsilonSum}
\end{figure}

With a solution for $F(y,y',s)$ in hand, we can then replicate our analysis in Section \ref{FlatAnalysis} to determine whether or not our kinetic mixing treatment is valid in the parameter space we're probing, this time applied to the warped space scenario not considered in I. Through analogous steps to those taken in Section \ref{FlatAnalysis}, we find that the sum $\sum_n \epsilon^2_n/\epsilon^2_1$ in the case of warped spacetime is also given by
\begin{align}
    \sum_n \epsilon^2_n/\epsilon^2_1 = \frac{1}{\tau v_1(0)^2},
\end{align}
the only difference from the flat-space result here being the form of the function $v_1(0)$. The $\tau^{-1}$ dependence of this sum suggests the same requirements as the identical flat space result, then: The brane-localized kinetic term (BLKT) $\tau$ must still be large enough so that its magnitude remains $\lsim 10$ and positive so that the result does not require the existence of ghost states. In Fig.~\ref{figWarpedepsilonSum}, we depict the sum $\sum_n \epsilon_n^2/\epsilon_1^2$ for different values of $\tau$ and $a_W$. Notably, while the sum is generally within reasonable $\lsim 10$ limits, when $(kR)\tau \approx 1/2$, the sum becomes quite close to, and even somewhat exceeds, 10. While the largest values of $\sum_n \epsilon_n^2/\epsilon_1^2$ achieved among the region of parameter space we have explored still aren't quite large enough to render $\epsilon_1^2$ terms in our analysis numerically significant (at least for the $\epsilon_1 \sim 10^{-(3-4)}$ terms we consider here), the sharp rate of increase they enjoy with decreasing $\tau$ near $(kR)\tau=1/2$ suggests that probing significantly below this value is unlikely to yield valid results. On the surface, this may seem to contrast slightly with our results in Section \ref{FlatAnalysis}, in which we found that restricting $\tau$ to values larger than $1/2$ stayed roughly $\lsim 6$. Closer inspection indicates that this discrepancy can largely be attributed to the use of $(kR) \tau$ as the variable we are employing instead of $\tau$: If one compares the maximum value obtained by the warped sum at $(k R) \tau = 3/4$ (for $k R = 1.5$) and $(k R) \tau = 1$ (for $k R = 2.0$), for which the variable $\tau$ itself is simply $1/2$, the results for the sum with both $k R$ values very closely matches that which was observed in the flat space construction of Section \ref{FlatAnalysis}. Hence, in both the flat and warped space cases, our setup's treatment of kinetic mixing easily remains valid for $\tau \gsim 0.5$, although it should be noted that as $kR$ increases, any boundary from these perturbativity concerns on the more natural warped-space parameter $(kR)\tau$, which is often used instead of $\tau$ for warped setups \cite{blkts}, will become increasingly stringent.

\begin{figure}[htbp]
\centerline{\includegraphics[width=3.5in]{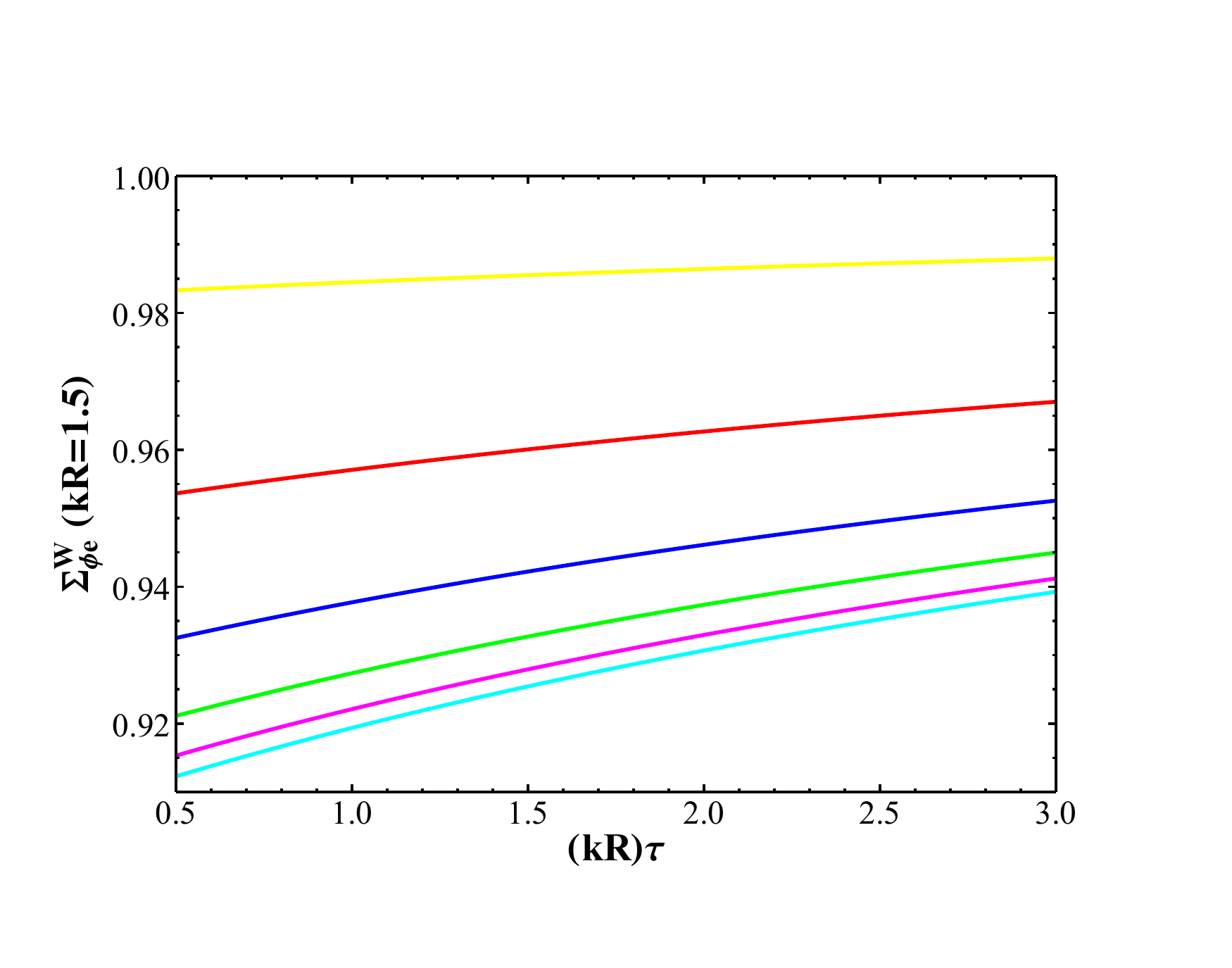}
\hspace{-0.75cm}
\includegraphics[width=3.5in]{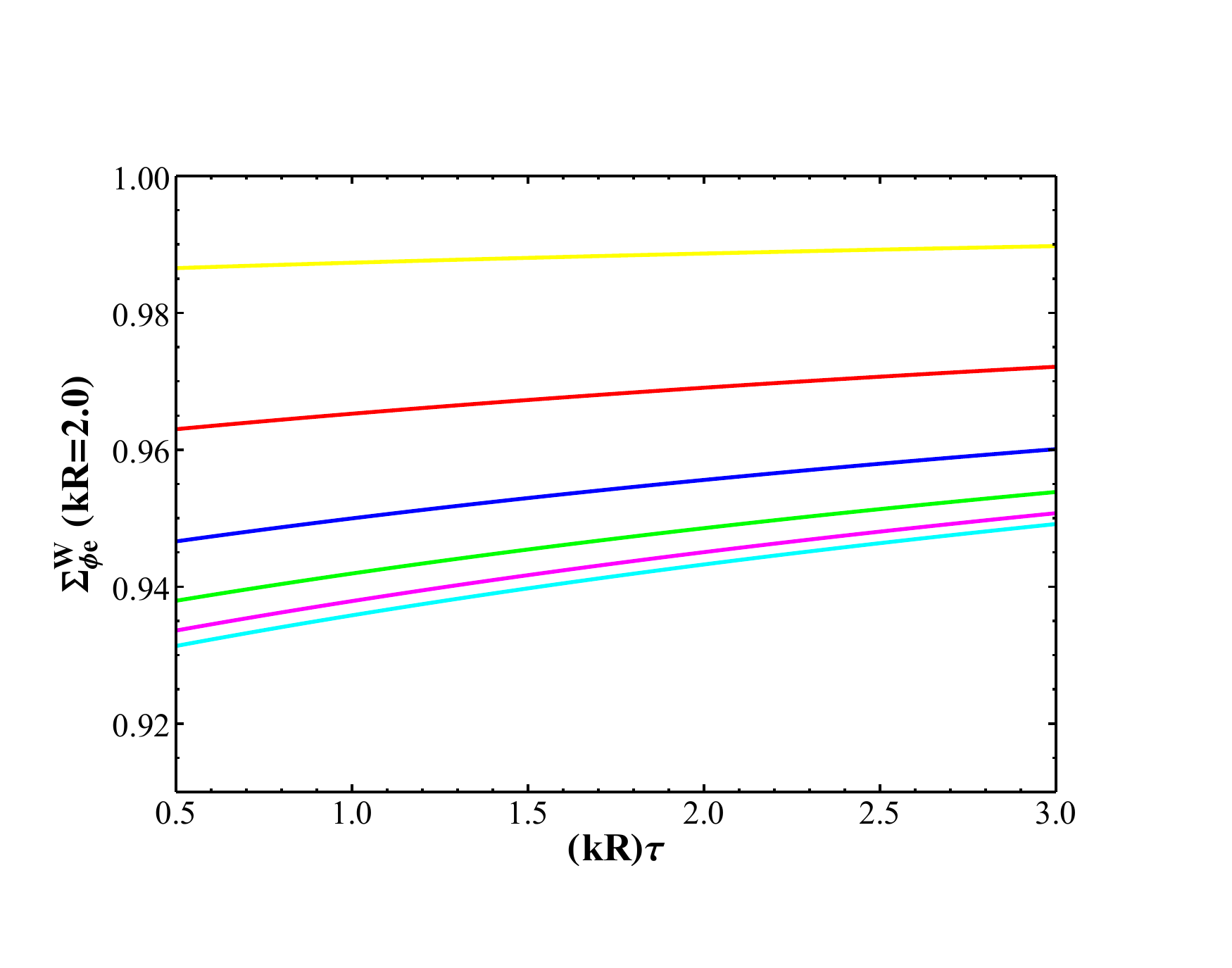}}
\vspace*{-0.25cm}
\centerline{\includegraphics[width=3.5in]{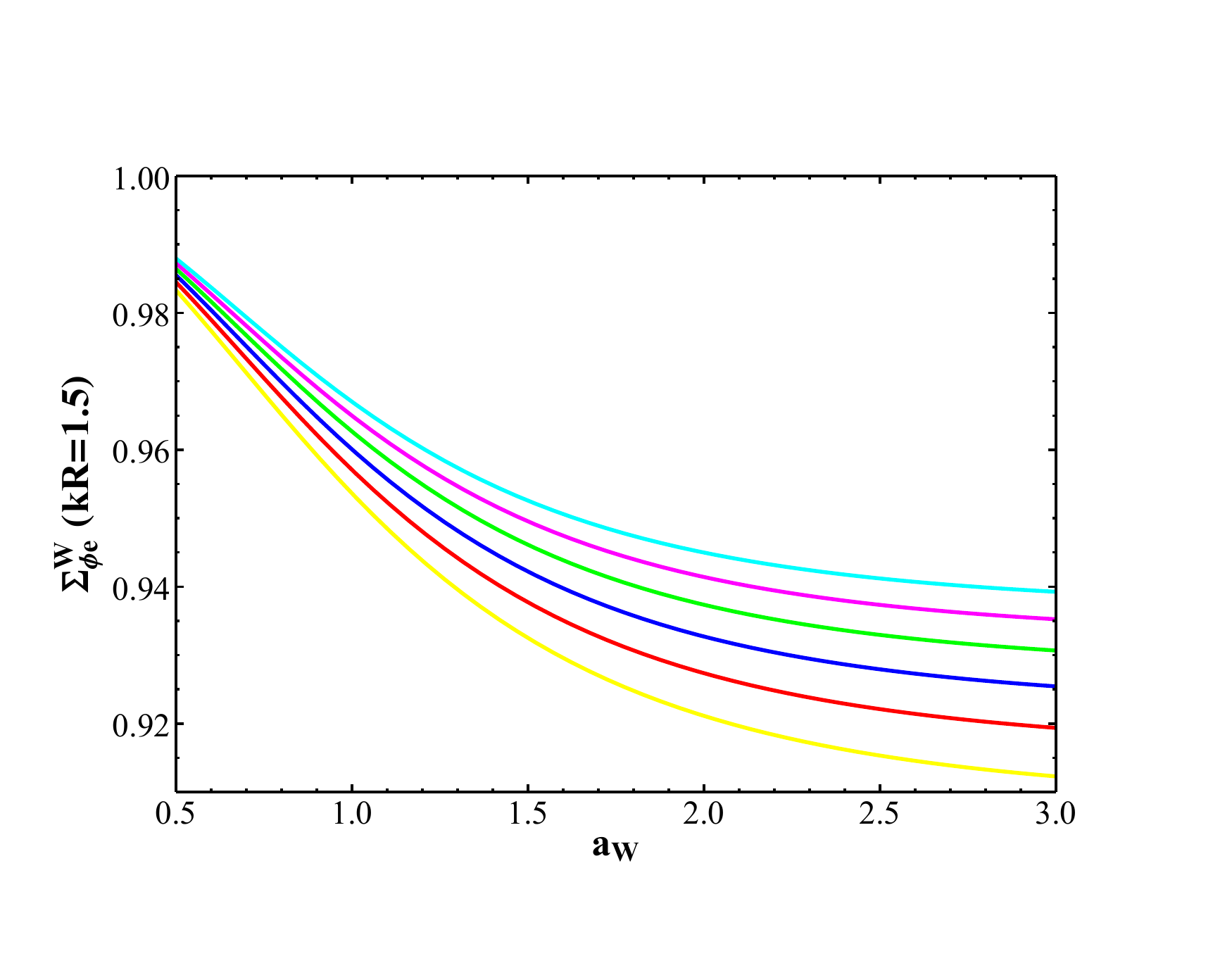}
\hspace{-0.75cm}
\includegraphics[width=3.5in]{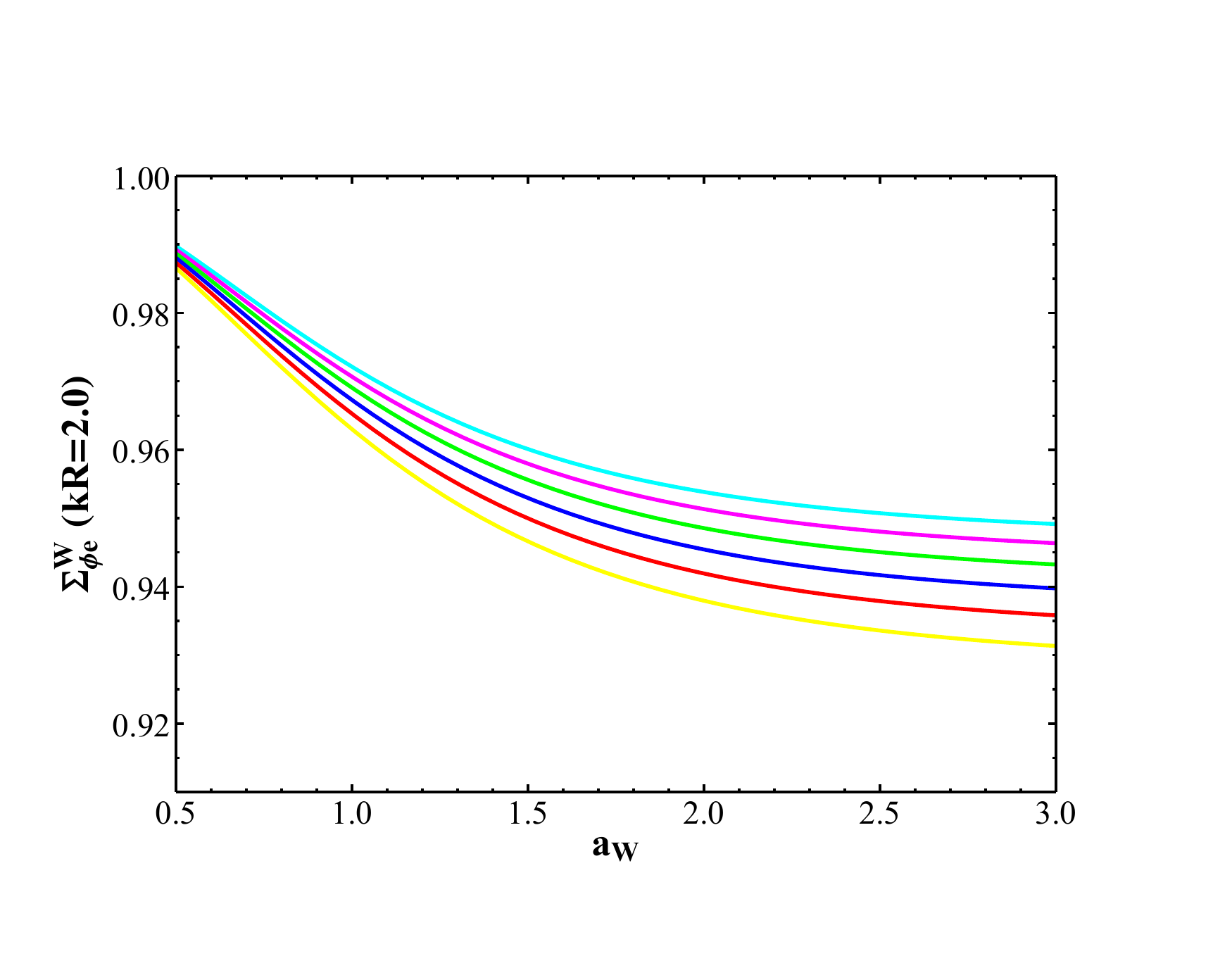}}
\caption{(Top Left) The sum $\Sigma^W_{\phi e}$ defined in Eq.(\ref{Warpedsigmae}), which encapsulates the dependence of the DM-electron scattering cross section on parameters of the model of the extra dimension, assuming $kR=1.5$, as a function of $(kR) \tau$ for $a_W=$3(cyan), 5/2(magenta), 2(green), 3/2(blue), 1(red), and 1/2(yellow), 
respectively.
(Top Right) As in the top left, but now assuming $kR = 2.0$.
(Bottom Left) As in the top left, but now as a function of $a_W$ assuming $(kR)\tau=$3(cyan), 5/2(magenta), 2(green), 3/2(blue), 1(red), and 1/2(yellow), respectively.
(Bottom Right) As in the bottom left, but assuming $kR = 2.0$.}
\label{fig14}
\end{figure}

Moving on, it is then straightforward to find the DM-$e^-$ scattering cross section by inserting our results for $F(y,y',s)$ given in Eq.(\ref{WarpedFSolution}) into Eq.(\ref{sigmae}), arriving at
\begin{align}\label{Warpedsigmae}
    \sigma_{\phi e} &= \frac{4 \alpha_{\textrm{em}} m_e^2 (g_D \epsilon_1)^2}{v_1(\pi R)^2 v_1 (0)^2} \frac{(kR)^2}{a_W^4 M_{KK}^4} = (2.97 \times 10^{-40} \; \textrm{cm}^2)\bigg( \frac{100 \; \textrm{MeV}}{m_1}\bigg)^4 \bigg( \frac{g_D \epsilon_1}{10^{-4}}\bigg)^2 \Sigma^W_{\phi e}, \\
    \Sigma^W_{\phi e} &\equiv \frac{(x^W_1)^4 (kR)^2}{a_W^4 v_1(\pi R)^2 v_1 (0)^2} = \bigg\lvert \sum_{n=0}^\infty \frac{(x^F_1)^2 v_n(0) v_n(\pi R)}{(x^F_n)^2 v_1(0) v_1(\pi R)} \bigg\rvert^2. \nonumber
\end{align}
Notably, this is the same result (up to a normalization convention of the parameter $a_W$ and, of course, different bulk wave functions $v_1(y)$) that we derived for the flat-space case Eq.(\ref{Flatsigmae}). In particular, the sum $F(0,\pi R, 0)$ has \emph{identical} results (again, up to normalization of $a_W$) for the flat- and warped-space scenarios. Just as in the flat space case, the numerical coefficient in front of the quantity $\Sigma^W_{\phi e}$, which now encapsulates all of the cross section's dependence on the parameters $\tau$ and $a_W$, indicates that as long as $\Sigma^W_{\phi e} \lsim O(1)$, the resultant cross section is not constrained by current experimental limits, although we remind the reader that such cross sections may lie within reach of near-term future direct-detection experiments. In Fig.~\ref{fig14}, we depict the dependence of $\Sigma^W_{\phi e}$ on various choices of $\tau$ and $a_W$; we find that just as for the flat space case, this requirement is easily satisfied for every $\tau$ and $a_W$ we consider. 

\begin{figure}[htbp]
\centerline{\includegraphics[width=5.0in,angle=0]{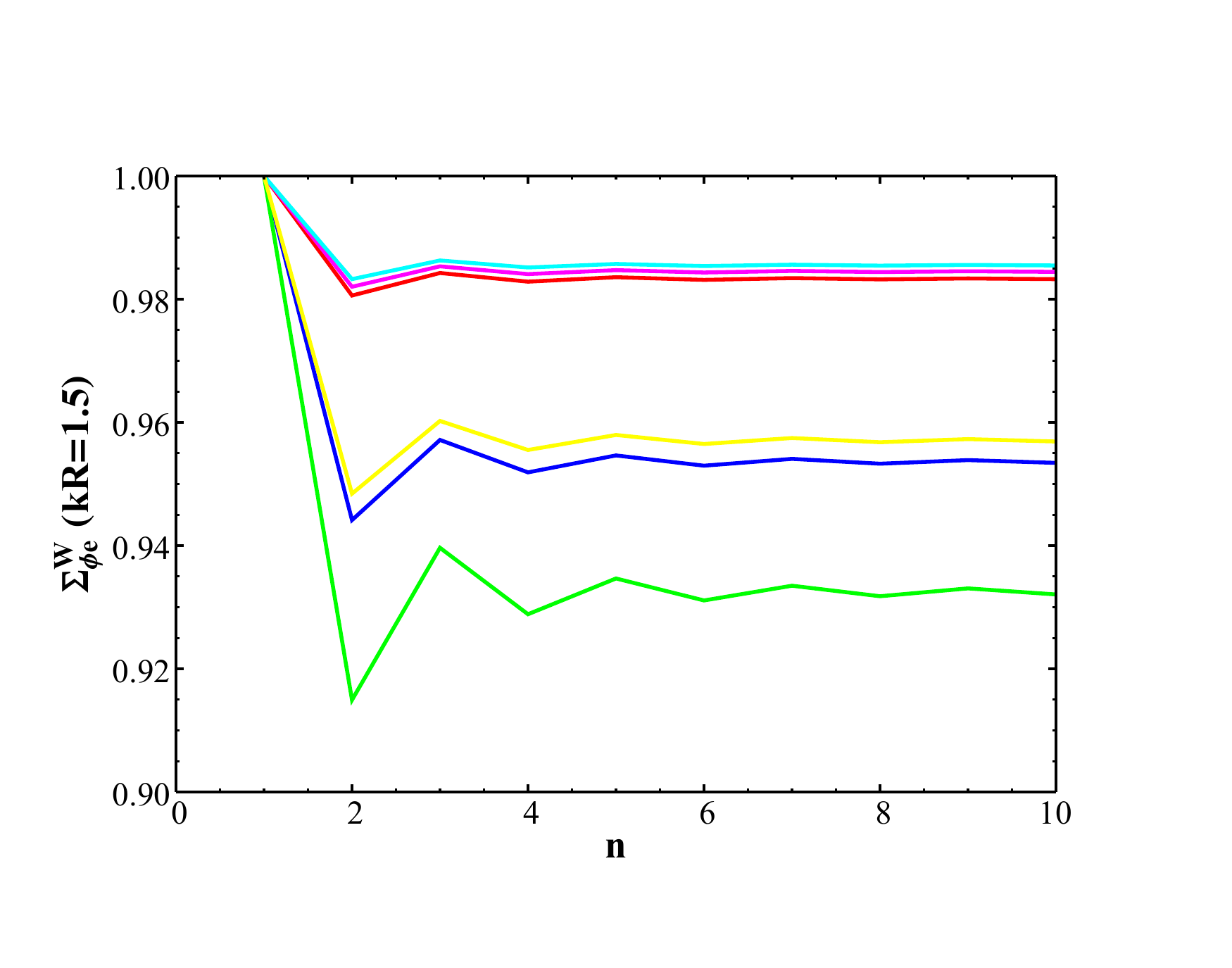}}
\vspace*{-2.0cm}
\centerline{\includegraphics[width=5.0in,angle=0]{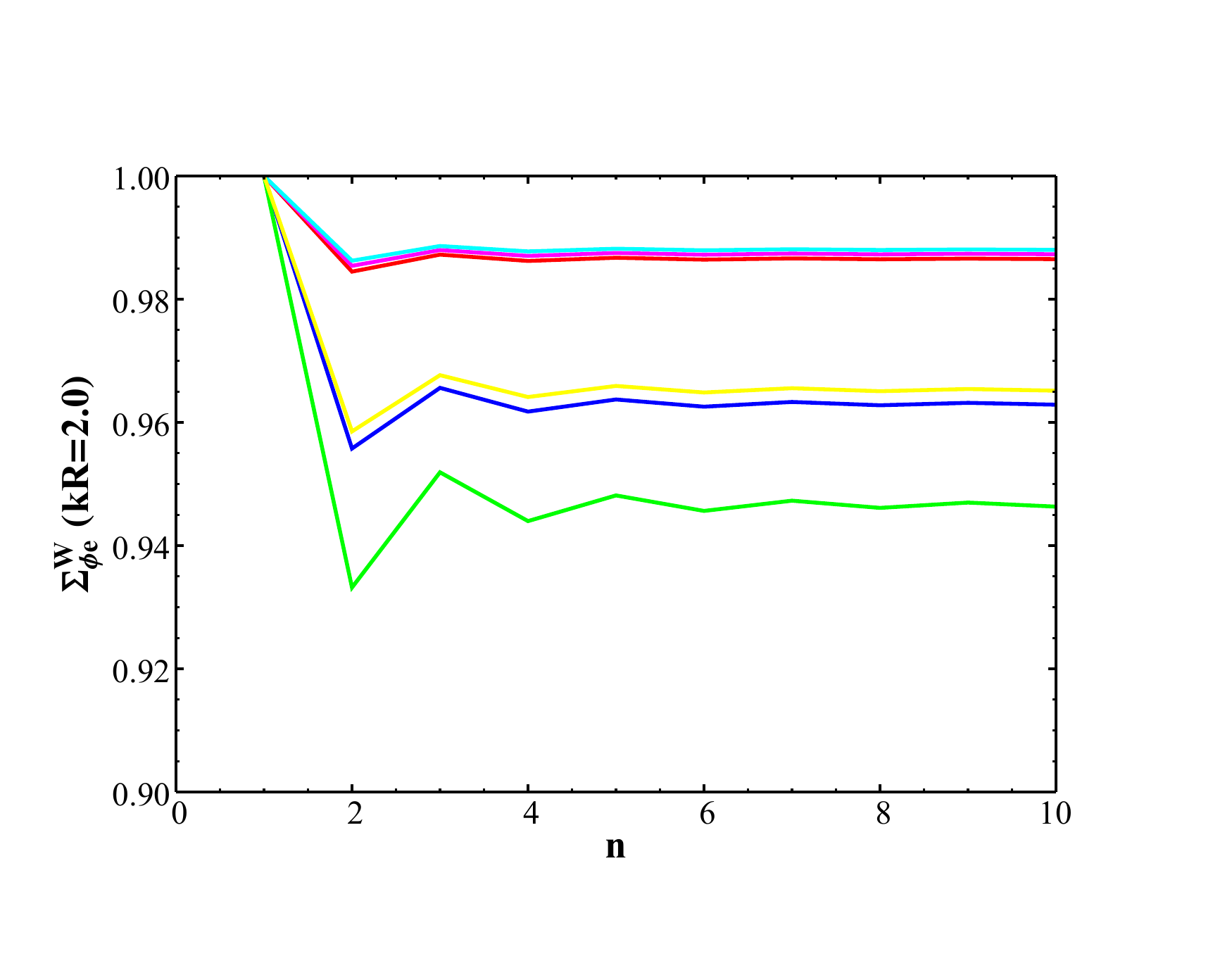}}
\vspace*{-1.30cm}
\caption{(Top) The value of the sum $\Sigma^W_{\phi e}$ defined in Eq.(\ref{Warpedsigmae}), which encapsulates the dependence of the DM-electron scattering cross section on parameters of the model of the extra dimension, truncated at finite $n$, assuming $kR = 1.5$ for various choices of $((kR)\tau ,a_W)$ =(1/2,1/2) [red], (1/2,1) [blue], (1/2,3/2) [green], 
(1,1/2) [magenta], (3/2,1/2) [cyan] and (1,1) [yellow], respectively.
(Bottom) As in the previous panel, but assuming $kR=2.0$}
\label{fig15}
\end{figure}

We also note that the sum over individual KK modes in the computation of $\Sigma^W_{\phi e}$ quickly converges to the closed form expression even when truncated for very low $n$; as depicted in Fig.~\ref{fig15}, $\Sigma^W_{\phi e}$, just like its flat space analogue, converges to within $O(10^{-2})$ corrections to its exact value even when truncated at $n \approx 10$. Hence, just as in the flat space scenario, exchanges of the lightest few dark photon KK modes dominate the direct detection signal.

\begin{figure}[htbp]
\centerline{\includegraphics[width=5.0in,angle=0]{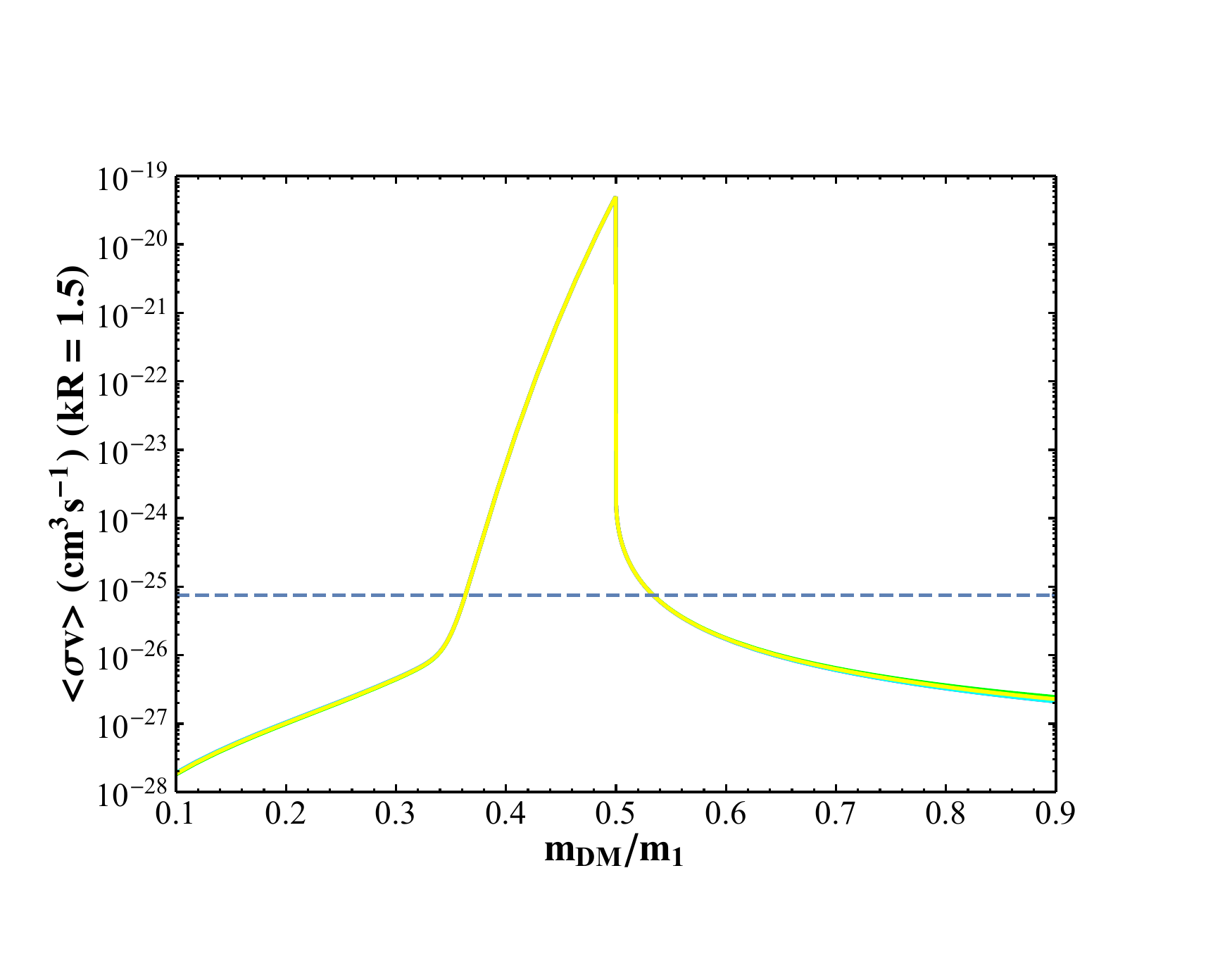}}
\vspace*{-2.0cm}
\centerline{\includegraphics[width=5.0in,angle=0]{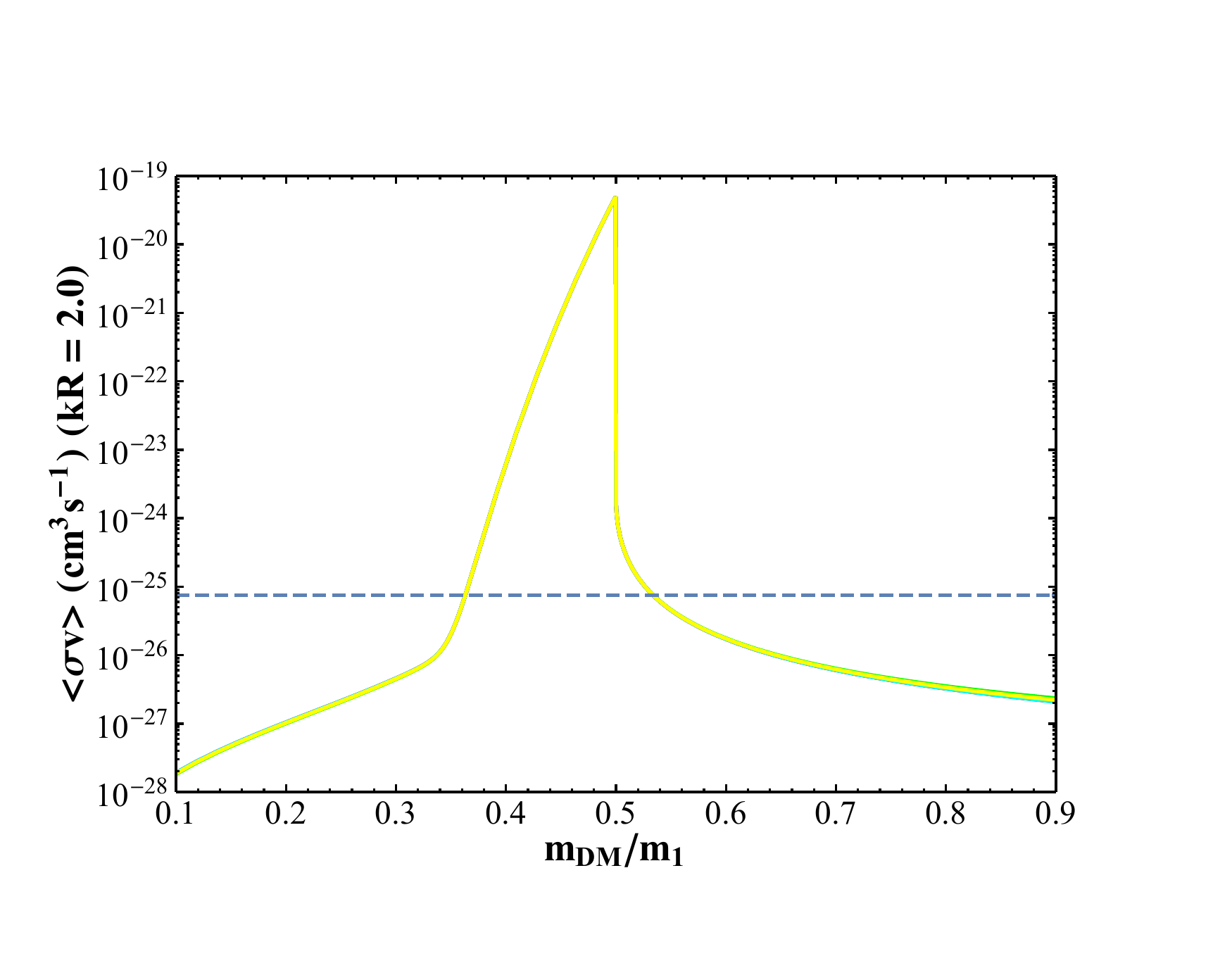}}
\vspace*{-1.0cm}
\caption{(Top) The thermally averaged annihilation cross section in $\textrm{cm}^3/\textrm{s}$, assuming $kR = 1.5$ for various choices of $((kR)\tau ,a_W)$ =(1/2,1/2) [red], (1/2,1) [blue], (1/2,3/2) [green], 
(1,1/2) [magenta], (3/2,1/2) [cyan] and (1,1) [yellow], respectively.
(Bottom) As in the previous panel, but assuming $kR=2.0$}
\label{fig16}
\end{figure}

Finally, we can conclude our discussion of the warped space scenario by considering the thermally averaged annihilation cross section of DM particles into SM fermions. Inserting the relevant value of $F(y,y',s)$ into Eq.(\ref{sigmavrel}) allows us to derive the DM annihilation cross-section, $\sigma v_{lab}$, for the warped space scenario, yielding
\begin{align}\label{Warpedsigmavrel}
    \sigma v_{lab} &= \frac{1}{3}\frac{g_D^2 \epsilon_1^2 \alpha_{\textrm{em}}Q_f^2}{v_1(\pi R)^2 v_1 (0)^2 }\frac{(s+2m_f^2)(s-4 m_{DM}^2)\sqrt{s(s-4 m_f^2)}}{s(s-2 m_{DM}^2)M_{KK}^4}\\
    &\times\Big\lvert \frac{2}{\pi z_\pi} \bigg(\frac{kR}{z_\pi \omega_0(z_\pi)+a_W^2 \omega_1(z_\pi)}\bigg)-\frac{v_1(\pi R) v_1 (0)}{(s/M_{KK}^2)-(x^W_1)^2}+\frac{v_1(\pi R) v_1 (0)}{(s/M_{KK}^2)-(x^W_1)^2+i x^W_1 \Gamma_1/M_{KK}}\Big\rvert^2, \nonumber
\end{align}
where we remind the reader that the functions $\omega_{0,1}(z)$ are defined in Eq.(\ref{WarpedFSolution}). Inserting this result into Eq.(\ref{singleIntegralAvg}), we can straightforwardly obtain the thermally averaged DM annihilation cross section via numerical integration. Just as in the flat space case, we specify that $m_{DM}=100 \; \textrm{MeV}$, $x_F = (m_{DM}/T) = 20$, $g_D = 0.3$, and $g_D \epsilon_1 = 10^{-4}$, and consider DM annihilation into an $e^+ e^-$ final state. Our results, depicted in Fig.~\ref{fig16} along with a dashed line marking $<\sigma v> = 7.5 \times 10^{-26} \; \textrm{cm}^3/\textrm{s}$, the approximate necessary cross section to produce the observed DM relic abundance, exhibit substantial similarity with the results for the flat space scenario given in Fig. \ref{fig8}; in particular, in both cases the dependence of the cross section on the BLKT $\tau$ and the brane-localized mass parameter $m_V \propto a_{F,W}$ is extremely limited, and the correct relic abundance is obtained when $m_{DM} \approx 0.36 m_1$ or $m_{DM} \approx 0.53 m_1$. Of course, as we vary the DM mass and 
$g_D \epsilon_1$, other values of $m_1$ will also be allowed. In short, for the annihilation cross section at freeze-out, we observe qualitatively similar behavior in the warped space setup as we do in the flat space scenario: For our choice of parameters resonant enhancement is necessary in order to realize the correct dark matter relic density, and the cross section is largely agnostic to specific selections for the brane-localized kinetic and mass terms for the dark photon field.

\section{Summary and Conclusions}\label{Conclusion}

In this paper, we have discussed a modification to our previous setup in I and II. In lieu of imparting mass to the lightest dark photon Kaluza-Klein (KK) modes via dark photon boundary conditions, which necessitates a bulk DM particle with corresponding KK modes, our current construction simplifies this structure by reinstating the dark Higgs as a scalar localized on the \emph{opposite} brane in the theory from the brane containing the SM, preventing mixing between the SM and dark Higgs scalars. The DM particle can then be placed on the same brane as the dark Higgs, removing the additional complication of a KK tower of DM particles and resulting in substantially simpler phenomenology while still removing the effects of the dark and SM Higgses mixing. 

We then briefly explored the model-building possibilities for this setup in two scenarios, one with a flat extra dimension and the other with a warped Randall-Sundrum metric, in particular considering the behavior of the dark photon tower's mass spectrum, couplings, and mixing parameters with SM fields, as well as briefly touching on the predictions for spin-independent direct detection experiments and thermally averaged annihilation cross sections at freeze-out for various points in parameter space. Exploring the case of a warped extra dimension in addition to that of a flat one affords us significant additional model-building freedom; for example, given the same choice for the lightest dark photon KK mode mass, subsequent KK modes for the warped scenario are approximately $\sim 3$ times heavier than they are in the flat scenario, demonstrating a qualitatively different KK spectrum. The ability for warped extra dimensions to generate hierarchies, meanwhile, can be straightforwardly exploited to naturally explain the mild $O(10^{2-3})$ hierarchy that exists between the SM Higgs scale and the characteristic mass scales of the dark brane, namely the masses of the DM and the lightest dark photon KK modes $\sim 0.1-1$ GeV.

With this model, we find few parameter space restrictions in either the warped or flat space constructions. The requirement that every dark photon KK mode's coupling to DM remain perturbative provides an upper limit on the DM-brane-localized mass term $m_V$, in particular, we find that for the flat construction, $m_V \lsim 1.5 R^{-1}$, where $R$ is the compactification radius of the extra dimension, while for warped space, $m_V \lsim 2 M_{KK}/\sqrt{kR}$, where $M_{KK}$ is the KK mass in the model and $kR \sim 1.5-2.0$. We also find, in agreement with I for the flat space scenario and novelly for the case of warped space, that a positive $O(1)$ value for the SM-brane-localized kinetic term (referred to here as $\tau$) is necessary in order to ensure the validity of our kinetic mixing analysis (in particular to ensure that $O(\epsilon_1^2)$ and higher order terms can in fact be safely neglected). For both the flat and warped space scenarios, however, this constraint is quite mild; requiring $\tau \geq 1/2$ is sufficient to satisfy it.

Regarding possible experimental signals, we explicitly consider that of spin-independent direct detection from scattering with electrons. We find that selecting $g_D \epsilon_1 \sim 10^{-4}$ and $m_1 \sim 100 \; \textrm{MeV}$ still places the spin-independent direct detection cross sections in both the flat and warped space constructions at $\sim 10^{-40} \textrm{cm}^2$, below current experimental constraints. However, we note that such signals are roughly within the order of magnitude of the possible reach of near-term future experiments, and are not especially sensitive to variations in the brane-localized kinetic and mass terms of the particular extra dimensional model (in the flat scenario, we see reasonable variation in these parameters producing at most an approximately 25\% change in the value of the direct detection cross section, while for the warped scenario this variation is approximately 5\%). As such, experiments such as SuperCDMS may place meaningful constraints on dark photon KK mode masses, couplings, and mixings in the near future.

The requirement that the thermally averaged annihilation cross section for the DM gives rise to the correct DM relic density, meanwhile, substantially constrains our selection of the relative DM particle mass $m_{DM}/m_1$. In particular, for natural selections of the other model parameters we see in both the flat and warped scenarios the DM annihilation cross section must enjoy some resonant enhancement of the contribution from the exchange of the lightest dark photon KK mode in order to attain a sufficiently large value. Given the sharpness of the resonance peak, this requirement places a significant constraint on the $m_{DM}$; for the choices $g_D \epsilon_1 = 10^{-4}$, $m_1 = 100 \; \textrm{MeV}$, and $g_D = 0.3$, $m_{DM}$ must lie near 0.36 or 0.54 of $m_1$ for flat space and 0.36 or 0.53 of $m_1$ for warped space. This cross section is also notably largely insensitive to differing choices of the brane-localized dark photon mass $m_V$ and the brane-localized kinetic term $\tau$ provided $m_1$, $g_D$, and $\epsilon_1$ are kept fixed, indicating that the exchange of the lightest KK mode is, somewhat unsurprisingly given its resonant enhancement, of paramount importance as contributors to this process.

Overall, we find that constructing this model within a flat or warped space framework results in little qualitative difference in our results. The most salient potential phenomenological difference lies in the differing relative masses of dark photon KK modes (in particular, the ratio of the  second-lightest dark photon mass to that of the lightest is in general 3-4 times larger in the Randall-Sundrum-like metric we consider than in the flat space case), which would have considerable effect on experimental searches for dark photons in colliders. Otherwise, however, we note that a wide range of natural and currently phenomenologically viable parameter space is available for both constructions.

As we move forward to explore the possibilities of kinetic mixing in theories of extra dimensions, we continue to find alternate constructions that allow for phenomenologically viable models. Here, following the work of I and II, we have presented another, simpler, construction that utilizes the additional model-building freedom afforded by extra dimensions to ameliorate phenomenological concerns that arise in 4-D kinetic mixing theories.

\section*{Acknowledgements}
The authors would like to particularly thank D. Rueter and J.L. Hewett for very valuable discussions related to this work. This work was supported by the Department of Energy, 
Contract DE-AC02-76SF00515.



\end{document}